\documentclass[times,twocolumn,final]{elsarticle}

\usepackage{lineno}
\usepackage{hyperref}
\hypersetup{
    colorlinks=true,
    linkcolor=blue,
    filecolor=magenta,      
    urlcolor=blue,
}
\modulolinenumbers[0]
\let\linenumbers\nolinenumbers\nolinenumbers

\usepackage{medima}
\usepackage{framed,multirow}
\usepackage{amsmath,amssymb,amsfonts}
\usepackage{algorithmic}
\usepackage{graphicx}
\usepackage{textcomp}
\usepackage{comment}
\usepackage{longtable}
\usepackage{afterpage}
\usepackage{caption}
\usepackage{adjustbox}
\usepackage{xltabular}
\usepackage{scalerel}
\usepackage{tikz}
\usetikzlibrary{svg.path}
\usepackage{tabularx} 
\usepackage{amssymb} 
\usepackage{enumitem}
\usepackage{natbib}
\usepackage{pifont}

\usepackage{pgfplots, pgfplotstable}
\usetikzlibrary{plotmarks}

\usepackage [english]{babel}
\usepackage [autostyle, english = british]{csquotes}
\MakeOuterQuote{"}

\usepackage{filecontents}

\definecolor{mycorrect}{rgb}{1, 0, 0} 
\definecolor{mycorrect}{rgb}{0, 0, 0} 

\definecolor{mycorrect2}{rgb}{.8,.349,.1} 
\definecolor{mycorrect2}{rgb}{0, 0, 0} 
\definecolor{done}{rgb}{0, 0.502, 0} 
\definecolor{tbd}{rgb}{0, 0, 1}
\definecolor{newcolor}{rgb}{.8,.349,.1}

\biboptions{authoryear}

\begin{document}

\begin{frontmatter}



\title{\textcolor{mycorrect}{Data Synthesis and Adversarial Networks: A Review and Meta-Analysis in Cancer Imaging}}

\author[mymainaddress]{Richard Osuala\corref{mycorrespondingauthor}}

\cortext[mycorrespondingauthor]{Corresponding author}
\ead{richard.osuala@ub.edu}

\cortext[contrib]{Authors contributed equally}

\author[mymainaddress]{Kaisar Kushibar}
\author[mymainaddress]{Lidia Garrucho}
\author[mymainaddress]{Akis Linardos}
\author[mymainaddress]{Zuzanna Szafranowska}
\author[erasmusmc]{Stefan Klein}
\author[biomedia]{Ben Glocker}
\author[mymainaddress]{Oliver Diaz\corref{contrib}}
\author[mymainaddress]{Karim Lekadir\corref{contrib}}

\address[mymainaddress]{Artificial Intelligence in Medicine Lab (BCN-AIM), Facultat de Matemàtiques i Informàtica, Universitat de Barcelona, Spain
}
\address[erasmusmc]{Biomedical Imaging Group Rotterdam, Department of Radiology \& Nuclear Medicine, Erasmus MC, Rotterdam, The Netherlands
}
\address[biomedia]{Biomedical Image Analysis Group, Department of Computing, Imperial College London, UK
}
\begin{abstract}

Despite technological and medical advances, the detection, interpretation, and treatment of cancer based on imaging data continue to pose significant challenges. These include \textcolor{mycorrect}{ inter-observer variability, class imbalance, dataset shifts, inter- and intra-tumour heterogeneity, malignancy determination, and treatment effect uncertainty}.
\textcolor{mycorrect}{Given the recent advancements in Generative Adversarial Networks (GANs), data synthesis, and adversarial training, we assess the potential of these technologies} to address a number of key challenges of cancer imaging. We \textcolor{mycorrect}{categorise these challenges into (a) data scarcity and imbalance, (b) data access and privacy, (c) data annotation and segmentation, (d) cancer detection and diagnosis, and (e) tumour profiling, treatment planning and monitoring}. 
\textcolor{mycorrect}{Based on our} analysis of \textcolor{mycorrect}{164} publications that apply adversarial training techniques in the context of cancer imaging, \textcolor{mycorrect}{we highlight multiple underexplored solutions with research potential.}
\textcolor{mycorrect}{We further contribute the Synthesis Study Trustworthiness Test (\textit{SynTRUST}), a meta-analysis framework for assessing the validation rigour of medical image synthesis studies. \textit{SynTRUST} is based on 26 concrete measures of thoroughness, reproducibility, usefulness, scalability, and tenability. Based on \textit{SynTRUST}, we analyse 16 of the most promising cancer imaging challenge solutions and observe a high validation rigour in general, but also several desirable improvements.}
With this work, we strive to bridge the gap between the needs of the clinical cancer imaging community and the current and prospective research on \textcolor{mycorrect}{adversarial networks} in the artificial intelligence community.

\end{abstract}


\end{frontmatter}

\linenumbers

\section{Introduction}
\label{sec:introduction}

\begin{figure*}
    \centering
       		\includegraphics[width=1.0\textwidth]{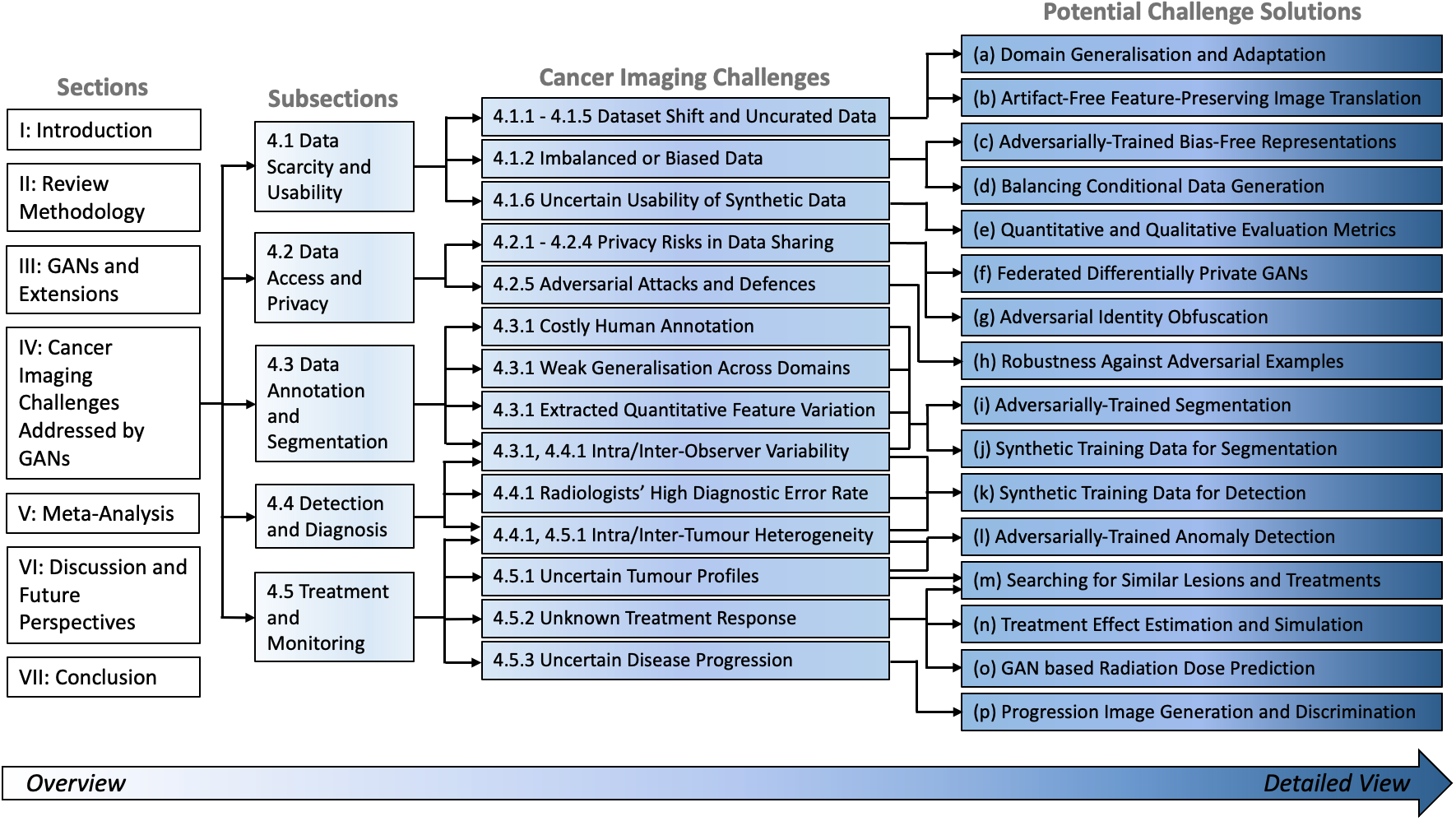}
	\caption[]{\textcolor{mycorrect}{Section organisation:} Illustration of the structure of the present paper starting with paper sections on the left and going into detail towards the right culminating in a selection of cancer imaging challenges and solutions. These solutions (a)-(p) are part of the solutions found in the surveyed GAN literature or are proposed extensions thereof, as is further discussed in section \ref{sec:currentchallenges}.}
  	\label{fig:overview}
\end{figure*}

\subsection{\textcolor{mycorrect}{The Burden of Cancer and Early Detection}}
The evident improvement in global cancer survival in the last decades is arguably attributable not only to health care reforms, but also to advances in clinical research (e.g., targeted therapy based on molecular markers) and diagnostic imaging technology (e.g whole-body magnetic resonance imaging (MRI)~\citep{messiou2019guidelines}, and positron emission tomography–computed tomography (PET-CT)~\citep{arnold2019progress}. Nonetheless, cancers still figure among the leading causes of morbidity and mortality worldwide~\citep{ferlay2015cancer}, with an approximated 9.6 million cancer related deaths in 2018~\citep{WHO2018Cancer}. The most frequent cases of cancer death worldwide in 2018 are lung (1.76 million), colorectal (0.86 million), stomach (0.78 million), liver (0.78 million), and breast (0.63 million)~\citep{WHO2018Cancer}. These figures are prone to continue to increase in consequence of the ageing and growth of the world population~\citep{jemal2011global}.

A large proportion of the global burden of cancer could be prevented due to treatment and early detection~\citep{jemal2011global}. For example, an early detection can provide the possibility to treat a tumour before it acquires critical combinations of genetic alterations (e.g., metastasis with evasion of apoptosis~\citep{hanahan2000hallmarks}). Solid tumours become detectable by medical imaging modalities only at an approximate size of $10^9$ cells ($\approx1cm^3$) after evolving from a single neoplastic cell typically following a Gompertzian~\citep{norton1976predicting} growth pattern~\citep{frangioni2008new}\footnote{In vitro studies reported a \textit{theoretical} detection limit around $10^5$ to $10^6$ for human cancer cell lines using PET. In clinical settings, the \textit{theoretical} detection limit is larger and depends, among others, on background radiation, cancer cell line, and cancer type~\citep{fischer2006few}.}. To detect and diagnose tumours, radiologists inspect, normally by visual assessment, medical imaging modalities such as magnetic resonance imaging (MRI), computed tomography (CT), ultrasound (US), x-ray mammography (MMG), PET~\citep{frangioni2008new, itri2018fundamentals, mccreadie2009eight}. 

Medical imaging data evaluation is time demanding and therefore costly in nature. In addition, volumes of new technologies (e.g., digital breast tomosynthesis~\citep{swiecicki2021generative}) become available and studies generally show an extensive increase in analysable imaging volumes~\citep{mcdonald2015effects}. Also, the diagnostic quality in radiology varies and is very much dependent on the personal experience, skills and invested time of the data examiner~\citep{itri2018fundamentals, elmore1994variability, woo2020intervention}. Hence, to decrease cost and increase quality, automated or semi-automated diagnostic tools can be used to assist radiologists in the decision-making process. Such diagnostic tools comprise traditional machine learning, but also recent deep learning methods, which promise an immense potential for detection performance improvement in radiology.

\subsection{\textcolor{mycorrect}{The Promise of Deep Learning and the Need for Data}}
The rapid increase in graphics processing unit (GPU) processing power has allowed training deep learning algorithms such as convolutional neural networks (CNNs)~\citep{Fukushima1980, lecun1989backpropagation, lecun1998gradient} on large image datasets achieving impressive results in Computer Vision~\citep{cireaan2012multi, krizhevsky2012imagenet}, and Cancer Imaging~\citep{cirecsan2013mitosis}. In particular, the success of AlexNet in the 2012 ImageNet challenge~\citep{krizhevsky2012imagenet} triggered an increased adoption of deep neural networks to a multitude of problems in numerous fields and domains including medical imaging, as reviewed in~\citet{shen2017deep, 9363915, litjens2017survey}.
Despite the increased use of medical imaging in clinical practice, the public availability of medical imaging data remains limited~\citep{mcdonald2015effects}. This represents a key impediment for the training, research, and use of deep learning algorithms in radiology and oncology. Clinical centres refrain from sharing such data for ethical, legal, technical, and financial (e.g., costly annotation) reasons~\citep{bi2019artificial}.\\

Such cancer imaging data not only is necessary to train deep learning models, but also to provide them with sufficient learning possibility to acquire robustness and generalisation capabilities. We define robustness as the property of a predictive model to remain accurate despite of variations in the input data (e.g., noise levels, resolution, contrast, etc). We refer to a model's generalisation capability as its property of preserving predictive accuracy on new data from unseen sites, hospitals, scanners, etc. Both of these properties are in particular desirable in cancer imaging considering the frequent presence of biased or unbalanced data with sparse or noisy labels\footnote{Alongside tumour manifestation heterogeneity, and multi-centre, multi-organ, multi-modality, multi-scanner, and multi-vendor data.}. Both robustness and generalisation are essential to demonstrate the trustworthiness of a deep learning model for usage in a clinical setting, where every edge-case needs to be detected and a false negative can potentially cost the life of a patient. \\

\subsection{\textcolor{mycorrect}{Synthetic Cancer Imaging Data}}
We hypothesise that the variety of data needed to train robust and well-generalising deep learning models for cancer images can be largely synthetically generated using Generative Adversarial Networks (GANs)~\citep{goodfellow2014generative}. The adversarial learning scheme in GANs is based on a generator that generates synthetic (alias "fake") samples of a target distribution trying to fool a discriminator, which classifies these samples as either real or fake. Various papers have provided reviews of GANs in the medical imaging domain~\citet{yi2019generative, kazeminia2020gans, tschuchnig2020generative, sorin2020creating, lan2020generative, singh2020medical}, but they focused on general presentation of the main methods and possible applications. In cancer imaging, however, there are specificities and challenges that call for specific implementations and solutions based on GANs \textcolor{mycorrect}{and the adversarial learning scheme at large}, including:
\begin{enumerate}[label=(\roman*)]
    \item the small size and complexity of cancerous lesions
    \item the high heterogeneity between tumours within as well as between patients and cancer types
    \item the difficulty to annotate, delineate and label cancer imaging studies at large scale
    \item the high data imbalance in particular between healthy and pathological subjects or between benign and malignant cases
    \item the difficulty to gather large consented datasets from highly vulnerable patients undergoing demanding care plans
\end{enumerate}

\begin{figure} [ht]
	\begin{center}
       		\includegraphics[width=0.40\textwidth]{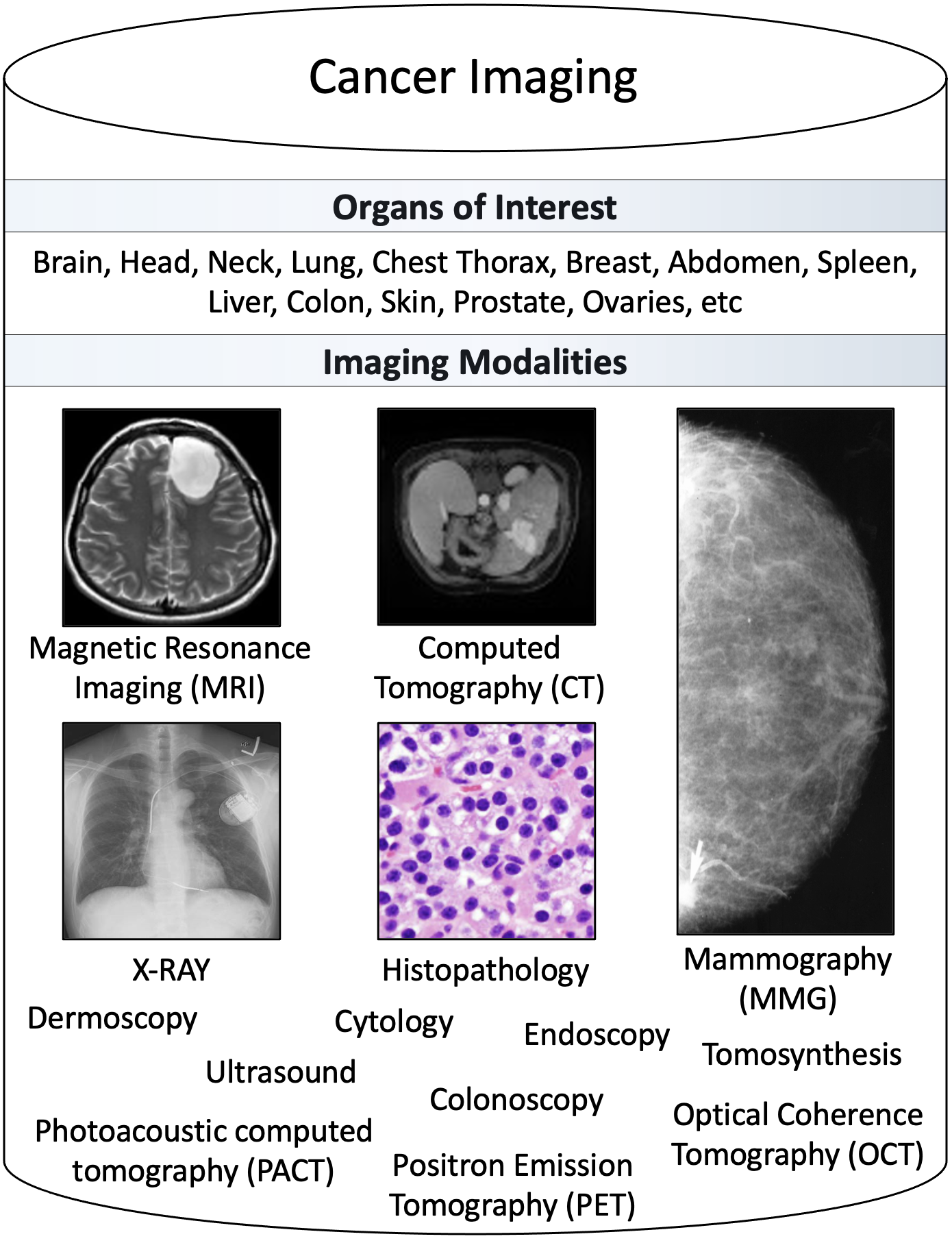}
	\end{center}
	 \caption[]{Overview of the most common organs and modalities targeted by the surveyed cancer imaging publications. A respective histogram that shows the number of papers per modality and per organ can be found in Figure \ref{fig:histogram}.}
  	\label{fig:modalities}
\end{figure}

Hence, the present paper contributes a unique perspective and comprehensive analysis of \textcolor{mycorrect}{adversarial networks} attempting to address the specific challenges in the cancer imaging domain. To the authors' best knowledge, this is the first survey that exclusively focuses on GANs \textcolor{mycorrect}{and adversarial training} in cancer imaging. In this context, we define cancer imaging as the entirety of approaches for research, diagnosis, and treatment of cancer based on medical images. Our survey comprehensively analyses cancer imaging GAN \textcolor{mycorrect}{and adversarial training} applications focusing on radiology modalities. As presented in Figure \ref{fig:modalities}, we recognise that non-radiology modalities are also widely used in cancer imaging. For this reason, we do not restrict the scope of our survey to radiology, but rather also analyse relevant publications in these other modalities including histopathology and cytopathology (e.g., in section \ref{sec:treatment}), and dermatology (e.g., in section \ref{sec:annotation} and \ref{sec:detection}). 

Further, our survey uncovers and highlights promising research directions for \textcolor{mycorrect}{adversarial networks and image synthesis} that can facilitate the sustainable adoption of AI in clinical oncology and radiology. 

\subsection{\textcolor{mycorrect}{Section Organisation}}
\textcolor{mycorrect}{The remainder of this paper is organised as follows. In section \ref{sec:2}, we introduce the methodology of this review. Section \ref{sec:3} provides an overview of GANs and highlights extensions of the adversarial learning framework relevant to cancer imaging. Section \ref{sec:currentchallenges} contains the main contribution that encompasses the systematic review of challenges of cancer imaging and potential solutions based on adversarial networks. This organisation is depicted in more detail in Figure \ref{fig:overview}.} 

\textcolor{mycorrect}{The different challenges are categorised into groups in the subsections \ref{sec:scarcity}, \ref{sec:access}, \ref{sec:annotation}, \ref{sec:annotation}, \ref{sec:detection}, and \ref{sec:treatment}. Each of the challenges categories contains several specific cancer imaging challenges, which we introduce and discuss in \ref{sec:4.1.1}-\ref{sec:4.5.3}. The sections are organised in an independent way allowing the reader to directly jump to a particular cancer imaging category (\ref{sec:scarcity}-\ref{sec:treatment}) of interest without requiring context from previous sections. For each of the specific challenges, we survey and discuss potential solutions, as depicted in Figure \ref{fig:overview}(a)-(p).}

\textcolor{mycorrect}{The subsequent section \ref{sec:trustworthiness} contains our second core contribution, which consists of the \textit{SynTRUST} framework for systematic analysis of trustworthiness criteria of image synthesis and adversarial training publications in medical imaging. Based on this framework, we meta-analyse a set of studies selected based on their strong performance and promising methodology for solving a specific cancer imaging challenge.}

After our literature review in Section \ref{sec:currentchallenges} and our meta-analysis in Section \ref{sec:trustworthiness} to learn how and to what extent GANs \textcolor{mycorrect}{and adversarial training} solutions have addressed the cancer imaging challenges in the past, we highlight and discuss prospective avenues of future research in the Discussion Section \ref{sec:discussion} and point out unexploited potential of \textcolor{mycorrect}{adversarial networks} in cancer imaging.


\section{Review Methodology} \label{sec:2}

Our review comprises two comprehensive literature screening processes. The first screening process surveyed the current challenges in the field of cancer imaging with a focus on radiology imaging modalities. After screening and gaining a deepened understanding of AI-specific and general cancer imaging challenges, we grouped these challenges for further analysis into the following five categories.

\begin{itemize}
    \item \textit{Data scarcity and usability challenges} (section \ref{sec:scarcity}); discussing dataset shifts, class imbalance, fairness, generalisation, domain adaptation and the evaluation of synthetic data.
    \item \textit{Data access and privacy challenges} (section \ref{sec:access}); comprising patient data sharing under privacy constraints, security risks, and adversarial attacks.
    \item \textit{Data annotation and segmentation challenges} (section \ref{sec:annotation}); discussing costly human annotation, high inter and intra-observer variability, and the consistency of extracted quantitative features.
    \item \textit{Detection and diagnosis challenges} (section \ref{sec:detection}); analysing the challenges of high diagnostic error rates among radiologists, early detection, and detection model robustness.
    \item \textit{Treatment and monitoring challenges} (section \ref{sec:treatment}); examining challenges of high inter and intra-tumour heterogeneity, phenotype to genotype mapping, treatment effect estimation and disease progression.
\end{itemize}

The second screening process comprised first of a generic and second a specific literature search to find all papers that apply adversarial learning (i.e. GANs) to cancer imaging. In the generic literature search, generic search queries such as "Cancer Imaging GAN", "Tumour GANs" or "Nodule Generative Adversarial Networks" were used to recall a high number of papers. The specific search focused on answering key questions of interest to the aforesaid challenges such as "Carcinoma Domain Adaptation Adversarial", "Skin Melanoma Detection GAN", "Brain Glioma Segmentation GAN", or "Cancer Treatment Planning GAN".

In Section \ref{sec:currentchallenges}, we map the papers that propose \textcolor{mycorrect}{adversarial training and} GAN applications applied to cancer imaging (second screening) to the surveyed cancer imaging challenges (first screening). The mapping of these GAN-related papers to challenge categories facilitates analysing the extent to which existing solutions solve the current cancer imaging challenges and helps to identify gaps and further potential for \textcolor{mycorrect}{adversarial networks} in this field. The mapping is based on the evaluation criteria used in the GAN-related papers and on the relevance of the reported results to the corresponding section. For example, if a GAN generates synthetic data that is used to train and improve a tumour detection model, then this paper is assigned to the detection and diagnosis challenge section \ref{sec:detection}. If a papers describes a GAN that improves a segmentation model, then this paper is assigned to the segmentation and annotation challenge section \ref{sec:annotation}, and so forth. \\ 
To gather the literature (e.g., first papers describing cancer imaging challenges, second papers proposing GAN solutions), we have searched in medical imaging, computer science and clinical conference proceedings and journals, but also freely on the web using the search engines Google, Google Scholar, and PubMed. After retrieving all papers with a title related to the subject, their abstract was read to filter out non-relevant papers. A full-text analysis was done for the remaining papers to determine whether they were to be included into our manuscript. We analysed the reference sections of the included papers to find additional relevant literature, which also underwent filtering and full-text screening.
Applying this screening process, we reviewed and included a total of \textcolor{mycorrect}{164} GAN \textcolor{mycorrect}{and adversarial training} cancer imaging publications comprising both peer-reviewed articles and conference papers, but also relevant preprints from arXiv and bioRxiv.\\
Details about these \textcolor{mycorrect}{164} cancer imaging applications can be found in tables \ref{table:scarcity-table}-\ref{Table:treatment-table}. The distribution of these publications across challenge category, year, modality, and anatomy is outlined in Figure \ref{fig:histogram}.

\textcolor{mycorrect}{The methodology for deriving and applying the \textit{SynTRUST} meta-analysis framework, which assesses the validity and trustworthiness of medical image synthesis studies, is provided in Section \ref{sec:trustworthiness}.}

\section{GANs and Extensions} \label{sec:3}

\subsection{Introducing the Theoretical Underpinnings of GANs}
\begin{figure*}
	\begin{center}
       		\includegraphics[width=0.85\textwidth]{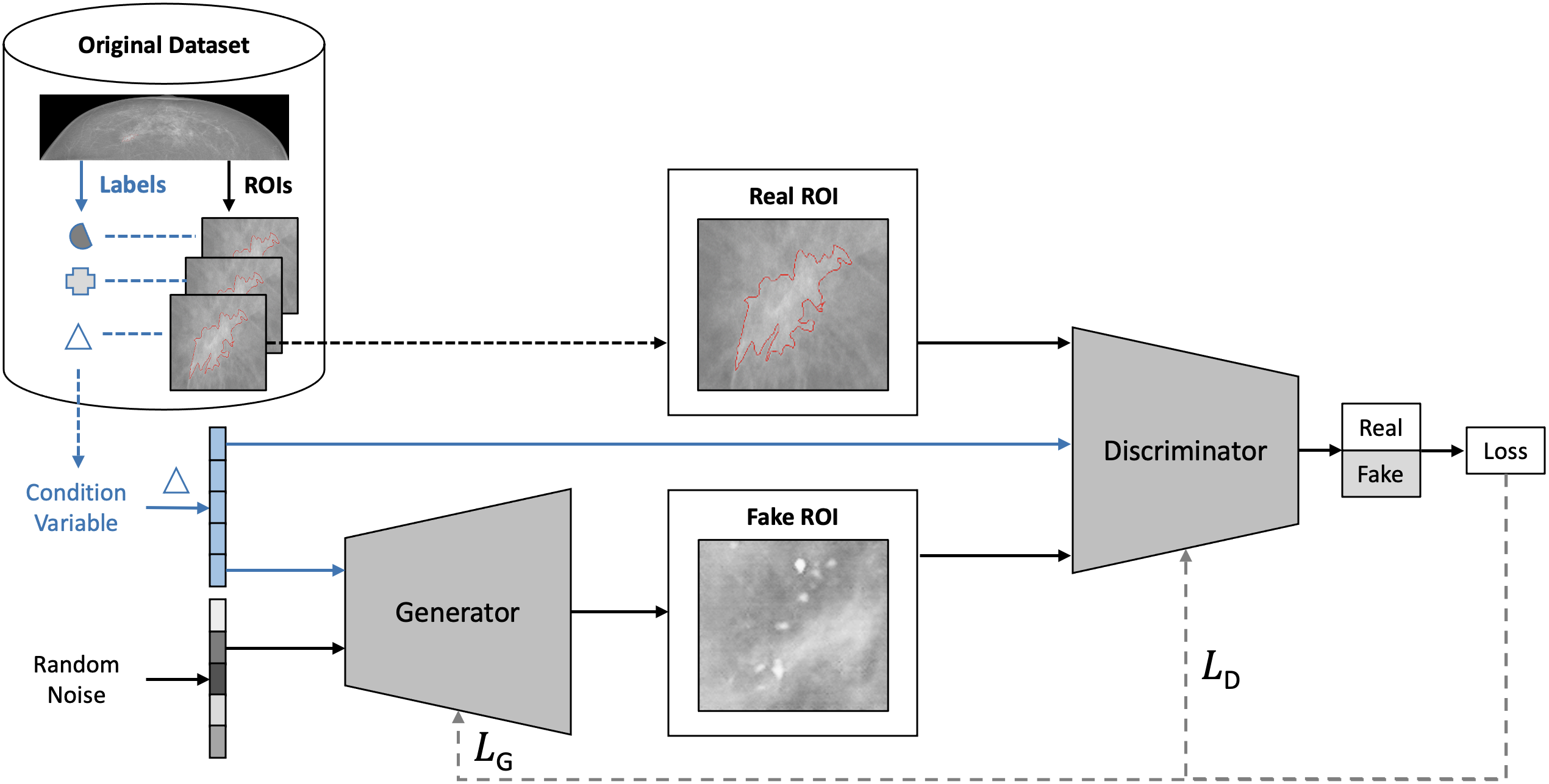}
	\end{center}
	 \caption[]{An example of a generic GAN architecture applied to generation of synthetic mammography region of interest (ROI) images based on the INbreast dataset~\citep{moreira2012inbreast}. Note that including the "Condition" depicted in blue colour extends the vanilla GAN architecture~\citep{goodfellow2014generative} to the cGAN architecture~\citep{mirza2014conditional}.}
  	\label{fig:GAN_architecture_example}
\end{figure*}

Generative Adversarial Networks (GANs)~\citep{goodfellow2014generative} are a type of generative model with a differentiable generator network~\citep{goodfellow2016deep}. GANs are formalised as a minimax two-player game, where the generator network (G) competes against an adversary network called discriminator (D). 
As visualised in Figure \ref{fig:GAN_architecture_example}, given a random noise distribution $z$, G generates samples $x = G(z;\theta(g))$ that D classifies as either real (drawn from training data, i.e. $x\sim p_{data}$) or fake (drawn from G, i.e. $x\sim p_{g}$). $x$ is either sampled from $p_{data}$ or from $p_{g}$ with a probability of 50\%. D outputs a value $p = D(x;\theta(d))$ indicating the probability that $x$ is a real training example rather than one of G's fake samples~\citep{goodfellow2016deep}. As defined by~\citet{goodfellow2014generative}, the task of the discriminator can be characterised as binary classification (CLF) of samples $x$. Hence, the discriminator can be trained using binary-cross entropy resulting in the following loss function $L_{D}$:
\begin{equation}
\begin{aligned}
L_{D} = - \mathbb{E}_{x\sim p_{data}} [log D(x)] + \mathbb{E}_{z\sim p_{z}} [log(1 - D(G(z)))]
\end{aligned}
\end{equation}
D's training objective is to minimise $L_{D}$ (or maximise $-L_{D}$) while the goal of the generator is the opposite (i.e. minimise $-L_{D}$) resulting in the value function $V(G, D)$ of a two-player zero-sum game between D and G:
\begin{equation}
\begin{aligned}
\min_{G} \max_{D} V(D,G) = \min_{G} \max_{D}[\mathbb{E}_{x\sim p_{data}} [log D(x)] \\
+ \mathbb{E}_{z\sim p_{z}} [log(1 - D(G(z)))]]
\end{aligned}
\end{equation}

In theory, in convergence, the generator’s samples become indistinguishable from the real training data ($x\sim p_{data}=p_{g}$) and the discriminator outputs $p = 1/2$ for any given sample $x$~\citep{goodfellow2016deep}. As this is a state where both D and G cannot improve further on their objective by changing only their own strategy, it represents a Nash equilibrium~\citep{farnia2020gans, nash1950equilibrium}. In practice, achieving convergence for this or related adversarial training schemes is an open research problem~\citep{kodali2017convergence, mescheder2018training, farnia2020gans}.

\subsection{Extensions of the Vanilla GAN Methodology}
\textcolor{mycorrect}{As indicated by Figure \ref{fig:chronology},} numerous extensions of GANs have shown to generate synthetic images with high realism~\citep{karras2017progressive, karras2019style, karras2020analyzing, chan2020pi} and under flexible conditions~\citep{mirza2014conditional, odena2017conditional, park2018mc}. GANs have been successfully applied to generate high-dimensional data such as images and, more recently, have also been proposed to generate discrete data~\citep{hjelm2017boundary}. Apart from image generation, GANs have also widely been proposed and applied for paired and unpaired image-to-image translation, domain-adaptation, data augmentation, image inpainting, image perturbation, super-resolution, and image registration and reconstruction~\citep{yi2019generative, kazeminia2020gans, wang2019generative}.\\

Table \ref{Table:gan-table} introduces a selection of common GAN extensions found to be frequently applied to cancer imaging. \textcolor{mycorrect}{For each GAN methodology in this and the following tables \ref{Table:gan-table}-\ref{Table:treatment-table}, we define the 'Task' describing the application of the respective adversarial network. For instance, in 'noise-to-image synthesis' the input into the generator G consists of a noise vector that G translates into an image. A further input into G can be a class label as in 'class-conditional-image-synthesis' based on which an output is generated that corresponds to this class. Paired and unpaired translation refer to the task where the input into G is a sample (e.g. an image in the source domain) based on which G generates another sample (e.g. an image in the target domain). This translation is paired if the training data consists of target and source domain sample pairs.} The key characteristics of each of the GAN extensions of table \ref{Table:gan-table} are described in the following paragraphs.

\subsubsection{Noise-to-Image GAN Extensions}
As depicted in blue in Figure \ref{fig:GAN_architecture_example}, cGAN adds a discrete label as conditional information to the original GAN architecture that is provided as input to both generator and discriminator to generate class conditional samples~\citep{mirza2014conditional}. 

AC-GAN feeds the class label only to the generator while the discriminator is tasked with correctly classifying both the class label and whether the supplied image is real or fake~\citep{odena2017conditional}. 


WGAN is motivated by mathematical rationale and based on the Wasserstein-1 distance (alias "earth mover distance" or "Kantorovich distance") between two distributions. WGAN extends on the theoretic formalisation and optimisation objective of the vanilla GAN to better approximate the distribution of the real data. By applying an alternative loss function (i.e. Wasserstein loss), the discriminator (alias "critic" or "$f_w$") maximises - and the generator minimises - the difference between the critic's scores for generated and real samples. A important benefit of WGAN is the empirically observed correlation of the loss with sample quality, which helps to interpret WGAN training progress and convergence~\citep{arjovsky2017wasserstein}.

In WGAN, the weights of the critic are clipped, which means they have to lie within a compact space $[-c, c]$. This is needed to fulfil that the critic is constraint to be in the space of 1-Lipschitz functions. With clipped weights, however, the critic is biased towards learning simpler functions and prone to have exploding or vanishing gradients if the clipping threshold $c$ is not tuned with care~\citep{gulrajani2017improved, arjovsky2017wasserstein}.

In WGAN-GP, the weight clipping constraint is replaced with a gradient penalty. Gradient penalty of the critic is a tractable and soft version of the following notion: By constraining that the norm of the gradients of a differentiable function is at most 1 everywhere, the function (i.e. the critic) would fulfil the 1-Lipschitz criterion without the need of weight clipping. Compared, among others, to WGAN, WGAN-GP was shown to have improved training stability (i.e. across many different GAN architectures), training speed, and sample quality~\citep{gulrajani2017improved}.

DCGAN generates realistic samples using a convolutional network architecture with batch normalization~\citep{ioffe2015batch} for both generator and discriminator and progressively increases the spatial dimension in the layers of the generator using transposed convolution (alias "fractionally-strided convolution")~\citep{radford2015unsupervised}. 

PGGAN is tested with loss and configurations introduced in WGAN GP. It starts by generating low pixel resolution images, but progressively adds new layers to the generator and discriminator during training resulting in increased pixel resolution and finer image details. It is suggested that after early convergence of initial low-resolution layers, the introduced additional layers enforce the network to only refine the learned representations by increasingly smaller-scale effects and features~\citep{karras2017progressive}. 

In SRGAN, the generator transforms a low-resolution (LR) to a high-resolution (HR, alias "super-resolution") image, while the discriminator learns to distinguish between real high-resolution images and fake super-resolution images. Apart from an adversarial loss, a perceptual loss called 'content loss' measures how well the generator represents higher level image features. This content loss is computed as the euclidean distance between feature representations of the reconstructed image and the reference image based on feature maps of a pretrained 19 layer VGG~\citep{simonyan2014very} network~\citep{ledig2017photo}.

\begin{table*}[t]
\centering
\scriptsize
\caption{A selection of the GAN architectures that we found to be the ones most frequently applied in cancer imaging.}\label{Table:gan-table}
\scalebox{0.85}{
\begin{tabular}{{p{0.33\textwidth}p{0.16\textwidth}p{0.16\textwidth}p{0.20\textwidth}p{0.20\textwidth}}}
    \hline
    \textbf{Publication} &
    \textbf{Input G} &
    \textbf{Input D} &
    \textbf{Losses} &
    \textbf{Task}
    \\
    \hline\hline
    \textbf{Noise to Image}
    \\
    \hline
    GAN~\citep{goodfellow2014generative} &
    Noise & 
    Image &
    Binary cross-entropy based adversarial loss ($L_{ADV}$) &
    \textcolor{mycorrect}{Noise-to-}image synthesis
    \\
    \hline
    conditional GAN (cGAN)~\citep{mirza2014conditional} &
    Noise \& label & 
    Image \& label &
    $L_{ADV}$ &
    \textcolor{mycorrect}{Class-}conditional image synthesis
    \\
    \hline
    Auxiliary Classifier GAN (AC-GAN)~\citep{odena2017conditional} &
    Noise \& label & 
    Image &
    $L_{ADV}$ \& cross-entropy loss (label classification) &
    \textcolor{mycorrect}{Class-}conditional image synthesis
    \\
    \hline
    Deep Convolutional GAN (DCGAN)~\citep{radford2015unsupervised} &
    Noise & 
    Image &
    $L_{ADV}$ &
    \textcolor{mycorrect}{Noise-to-}image synthesis
    \\
    \hline
    Wasserstein GAN (WGAN)~\citep{arjovsky2017wasserstein} &
    Noise & 
    Image &
    Wasserstein loss ($L_{WGAN}$) &
    \textcolor{mycorrect}{Noise-to-}image synthesis
    \\
    \hline
    WGAN Gradient Penalty (WGAN GP)~\citep{gulrajani2017improved} &
    Noise & 
    Image &
    $L_{WGAN}$ with GP ($L_{WGAN-GP}$) &
    \textcolor{mycorrect}{Noise-to-}image synthesis
    \\
    \hline
    Progressively Growing GAN (PGGAN)~\citep{karras2017progressive} &
    Noise & 
    Image &
    $L_{WGAN-GP}$ &
    \textcolor{mycorrect}{Noise-to-}image synthesis
    \\
    \hline\hline
    \textbf{Image to Image}
    \\
    \hline
    Super-Resolution GAN (SRGAN)~\citep{ledig2017photo} &
    Image (LR) & 
    Image (HR) &
    $L_{ADV}$ \& content loss (based on VGG features) & 
    Super-resolution
    \\
    \hline
    CycleGAN~\citep{zhu2017unpaired} &
    Source image & 
    Target image &
    $L_{ADV}$ \& cycle consistency loss \& identity loss &
    Unpaired image-to-image translation
    \\
    \hline
    pix2pix~\citep{isola2017image} &
    Source image & 
    Concatenated source and target images &
    $L_{ADV}$ \& reconstruction loss (i.e. L1) &
    Paired image-to-image translation
    \\
    \hline
    SPatially-Adaptive (DE)normalization (SPADE)~\citep{park2019semantic} &
    Noise or encoded source image \& segmentation map  & 
    Concatenated target image and segmentation map &
    Hinge \& perceptual \& feature matching losses~\citep[from][]{wang2018high} &
    Paired image-to-image translation
    \\
    \hline
    \end{tabular}
}
\end{table*}

\subsubsection{Image-to-Image GAN Extensions}
In image-to-image translation, a mapping is learned from one image distribution to another. For example, images from one domain can be transformed to resemble images from another domain via a mapping function implemented by a GAN generator.

CycleGAN achieves realistic unpaired image-to-image translation using two generators ($G$, $F$) with one traditional adversarial loss each and an additional cycle-consistency loss. Unpaired image-to-image translation transforms images from domain $X$ to another domain $Y$ in the absence of paired training data i.e. corresponding image pairs for both domains. In CycleGAN, the input image $x$ from domain $X$ is translated by generator $G(x)$ to resemble a sample from domain $Y$. Next, the sample is translated back from domain $Y$ to domain $X$ by generator $F(G(x))$. The cycle consistency loss enforces that $F(G(x)) \approx x$ (forward cycle consistency) and that $G(F(y)) \approx y$ (backward cycle consistency)~\citep{zhu2017unpaired}.

Both pix2pix and SPADE are used in paired image-to-image translation where corresponding image pairs for both  domains $X$ and $Y$ are available. pix2pix (alias "condGAN") is a conditional adversarial network that adapts the U-Net architecture\footnote{To reduce information loss in latent space compression, U-Net uses skip connections between corresponding layers (e.g., first to last) in the encoder and decoder.}~\citep{ronneberger2015u} for the generator to facilitate encoding an conditional input image into a latent representation before decoding it back into an output image. pix2pix uses L1 loss to enforce low level (alias "low frequency") image reconstruction and a patch-based discriminator ("PatchGAN") to enforce high level (alias "high frequency") image reconstruction that the authors suggest to interpret as texture/style loss. Note that the input into the PatchGAN discriminator is a concatenation\footnote{Note the concatenation of real\_A and fake\_B before computing the loss in the discriminator backward pass (L93) in the authors' \href{https://github.com/junyanz/pytorch-CycleGAN-and-pix2pix/blob/master/models/pix2pix_model.py\#L93}{pix2pix implementation.}} of the original image (i.e. the generator's input image; e.g. this can be a segmentation map) and the real/generated image (i.e. the generator's output image)~\citep{isola2017image}. 

In SPADE, the generator architecture does not rely on an encoder for downsampling, but uses a conditional normalisation method during upsampling instead: A segmentation mask as conditional input into the SPADE generator is provided to each of its upsampling layers via spatially-adaptive residual blocks. These blocks embed the masks and apply two two-layer convolutions to the embedded mask to get two tensors with spatial dimensions. These two tensors are multiplied/added to each upsampling layer prior to its activation function. The authors demonstrate that this type of normalisation achieves better fidelity and preservation of semantic information in comparison to other normalisation methods that are commonly applied in neural networks (e.g., Batch Normalization). The multi-scale discriminators and the loss functions from pix2pixHD~\citep{wang2018high} are adapted in SPADE, which contains a hinge loss (i.e. as substitute of the adversarial loss), a perceptual loss, and a feature matching loss~\citep{park2019semantic}.

\subsubsection{GAN Network Architectures and Adversarial Loss}
For further methodological detail on the aforementioned GAN methods, loss functions, and architectures, we point the interested reader to the GAN methods review by~\citet{wang2019generative}.
Due to the image processing capabilities of CNNs~\citep{lecun1989backpropagation}, the above-mentioned GAN architectures generally rely on CNN layers internally. Recently, TransGAN~\citep{jiang2021transgan} \textcolor{mycorrect}{and VQGAN \citep{esser2021taming} were} proposed, which diverges from the CNN design pattern to using Transformer Neural Networks~\citep{vaswani2017attention}. Due to the promising performances of these approaches in computer vision tasks, we encourage future studies to investigate the potential of transformer-based GANs for applications in medical and cancer imaging.

Multiple deep learning architectures apply the adversarial loss proposed in~\citet{goodfellow2014generative} together with other loss functions (e.g., segmentation loss functions) for other tasks than image generation (e.g., image segmentation). This adversarial loss is useful for \textcolor{mycorrect}{unsupervised} learning of features and representations that are invariant to some part of the training data. For instance, adversarial learning can be useful to discriminate a domain to learn domain-invariant representations~\citep{ganin2015unsupervised}, as has been successfully demonstrated for medical images~\citep{kamnitsas2017unsupervised}. \textcolor{mycorrect}{Such methods that apply the adversarial loss internally are referred to as 'adversarial training' methods and are included in} the scope of our survey. \textcolor{mycorrect}{That is,} we include and consider all relevant cancer imaging papers that apply or build upon the adversarial learning scheme defined in~\citet{goodfellow2014generative}\textcolor{mycorrect}{, which comprises GANs as well as adversarial training methods}.

\section{Cancer Imaging Challenges Addressed by \textcolor{mycorrect}{Data Synthesis and Adversarial Networks}} \label{sec:currentchallenges}
In this section we follow the structure presented in Figure \ref{fig:overview}, where we categorise cancer imaging challenges into five categories consisting of \textit{data scarcity and usability} (\ref{sec:scarcity}), \textit{data access and privacy} (\ref{sec:access}), \textit{data annotation and segmentation} (\ref{sec:annotation}), \textit{detection and diagnosis} (\ref{sec:detection}), and \textit{treatment and monitoring} (\ref{sec:treatment}). In each subsection, we group and analyse respective cancer imaging challenges and discuss the potential and the limitations of corresponding GAN-based data synthesis \textcolor{mycorrect}{and adversarial training} solutions. In this regard, we also identify and highlight key needs to be addressed by researchers in the field of cancer imaging GANs towards solving the surveyed cancer imaging challenges. 
We provide respective tables \ref{table:scarcity-table}-\ref{Table:treatment-table} for each subsection \ref{sec:scarcity}-\ref{sec:treatment} containing relevant information (publication, method, dataset, modality, task, highlights) for all of the reviewed cancer imaging GAN solutions.

\begin{figure*}
    \centering
       		\includegraphics[width=1.0\textwidth]{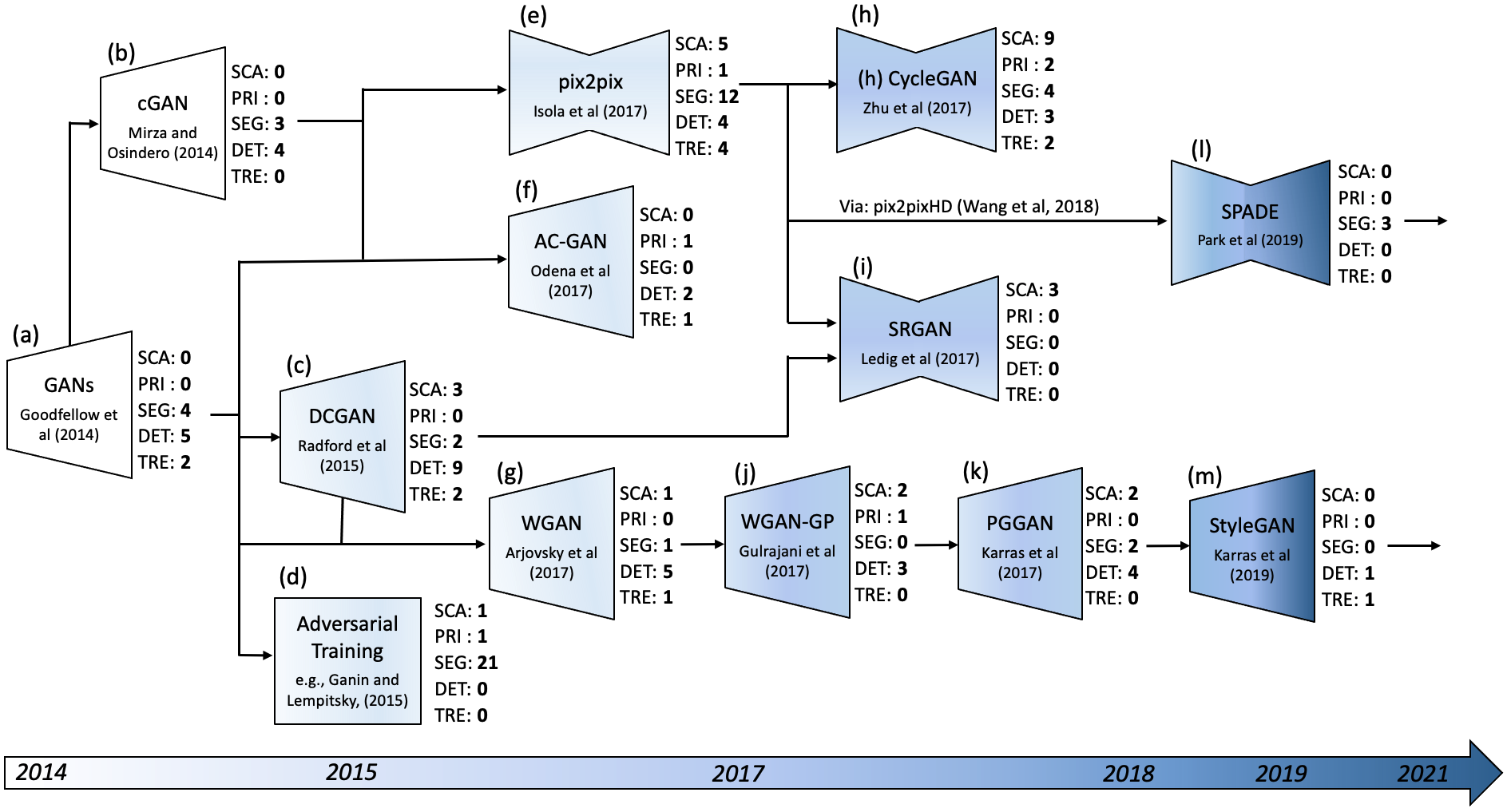}
	\caption[]{\textcolor{mycorrect}{The chronological evolution of adversarial networks in cancer imaging based on the commonly applied GAN types (a)-(m).
	The shape of the respective GAN indicates its mapping function, e.g., a trapezium represents a noise-to-image mapping function. For each cancer imaging challenge category surveyed in \ref{sec:currentchallenges}, the number of occurrences of each GAN type (a)-(m) is counted across reviewed publications. The illustration abbreviates challenge categories as
	SCA: \ref{sec:scarcity} Data scarcity and usability challenges, 
	PRI: \ref{sec:access} Data access and privacy challenges, 
	SEG: \ref{sec:annotation} Data annotation and segmentation challenges, 
	DET: \ref{sec:detection} Detection and diagnosis challenges,
	TRE: \ref{sec:treatment} Treatment and monitoring challenges.}}
  	\label{fig:chronology}
\end{figure*}

\paragraph{\textcolor{mycorrect}{Chronology of key innovations}}
\textcolor{mycorrect}{The most commonly applied adversarial network methodologies in cancer imaging are summarised chronologically in Figure \ref{fig:chronology}. Next to each network (a)-(m), the number of occurrence per cancer imaging challenge category \ref{sec:scarcity}-\ref{sec:treatment} is highlighted.}

\textcolor{mycorrect}{Following Vanilla GANs \ref{fig:chronology}(a), four main lines of innovations have been widely adopted in cancer imaging. These are methods that condition the synthetic data generation e.g. cGAN \ref{fig:chronology}(b), methods that improve upon the network architecture e.g. DCGAN \ref{fig:chronology}(c), methods that improve upon the adversarial loss function e.g. WGAN \ref{fig:chronology}(g), and methods that backpropagate the adversarial loss for representation learning, e.g. domain-invariant representations \ref{fig:chronology}(d).}

\textcolor{mycorrect}{As to conditional methods, further key innovations have been AC-GAN’s \ref{fig:chronology}(f) discriminator classifying the input condition, and methods that conditioning the generation based on an input image using additional reconstruction (e.g., pix2pix \ref{fig:chronology}(e), cycleGAN \ref{fig:chronology}(h)) or perceptual (e.g., SRGAN \ref{fig:chronology}(i)) losses. Recent approaches (e.g., SPADE \ref{fig:chronology}(l)) innovate regarding how the input image is provided to the generator network, e.g., via spatially-adaptive residual blocks in upsampling layers.}

\textcolor{mycorrect}{WGAN’s \ref{fig:chronology}(g) loss based on the discriminator estimating the Wasserstein-1 distance between real and synthetic image distributions is a widely used and extended (e.g., WGAN-GP \ref{fig:chronology}(j)) alternative to the vanilla binary-cross entropy adversarial loss in cancer imaging.}

\textcolor{mycorrect}{The architectural innovation of progressive network growing \ref{fig:chronology}(k) unlocked high-resolution cancer image generation and is adopted by recent approaches such as StyleGAN \ref{fig:chronology}(m), which introduced adaptive instance normalization and pioneered noise (and style condition) input via intermediate activation maps.}

\subsection{Data Scarcity and Usability Challenges}\label{sec:scarcity}

\subsubsection{Challenging Dataset Sizes and Shifts} \label{sec:4.1.1}
Although data repositories such as The Cancer Imaging Archive (TCIA)~\citep{clark2013cancer} have made a wealth of cancer imaging data available for research, the demand is still far from satisfied. As a result, data augmentation techniques are widely used to artificially enlarge the existing datasets, traditionally including simple spatial (e.g., flipping, rotation) or intensity transformations (e.g., noise insertion) of the true data. GANs have shown promise as a more advanced augmentation technique and have already seen use in medical and cancer imaging~\citep{han2018ganBRATS, yi2019generative}.

Aside from the issue of lacking sizeable data, data scarcity often forces studies to be constrained on small-scale single-centre datasets. The resulting findings and models are likely to not generalise well due to diverging distributions between the (synthetic) datasets seen in training and those seen in testing or after deployment, a phenomenon known as dataset shift~\citep{quionero2009dataset}\footnote{More concretely, this describes a case of covariate shift~\citep{quionero2009dataset, shimodaira2000improving} defined by a change of distribution within the independent variables between two datasets}. An example of this in clinical practice are cases where training data is preselected from specific patient sub-populations (e.g., only high-risk patients) resulting in bias and limited generalisability to the broad patient population~\citep{troyanskaya2020artificial, bi2019artificial}.

From a causality perspective, dataset shift can be split into several distinct scenarios~\citep{castro2020causality}:
\begin{itemize}
    \item \textit{Population shift}, caused by differences in age, sex, ethnicities etc.
    \item \textit{Acquisition shift}, caused by differences in scanners, resolution, contrast etc.
    \item \textit{Annotation shift}, caused by differences in annotation policy, annotator experience, segmentation protocols etc.
    \item \textit{Prevalence shift}, caused by differences in the disease prevalence in the population, often resulting from artificial sampling of data
    \item \textit{Manifestation shift}, caused by differences in how the disease is manifested
\end{itemize}

GANs may inadvertently introduce such types of dataset shifts (e.g., due to mode collapse~\citep{goodfellow2014generative}), but it has been shown that this shift can be studied, measured and avoided~\citep{santurkar2018classification, arora2018gans}. GANs can be a sophisticated tool for data augmentation or curation~\citep{diaz2021data} and by calibrating the type of shift introduced, they have the potential to turn it into an advantage, generating diverse training data that can help models generalise better to unseen target domains.
The research line studying this problem is called \textit{domain generalisation}, and it has presented promising results for harnessing adversarial models towards learning of domain-invariant features~\citep{zhou2021domain}. GANs \textcolor{mycorrect}{and adversarial training} have been used in various ways in this context, using multi-source data to generalise to unseen targets~\citep{rahman2019multi, li2018domain} or in \textcolor{mycorrect}{unsupervised domain generalisation} using adaptive data augmentation to append adversarial examples iteratively~\citep{volpi2018generalizing}. As indicated in Figure \ref{fig:overview}(a), the domain generalisation research line has recently been further extended to cancer imaging~\citep{lafarge2019learning, chen2021generative}.

\begin{figure*}
    \centering
       		\includegraphics[width=0.95\textwidth]{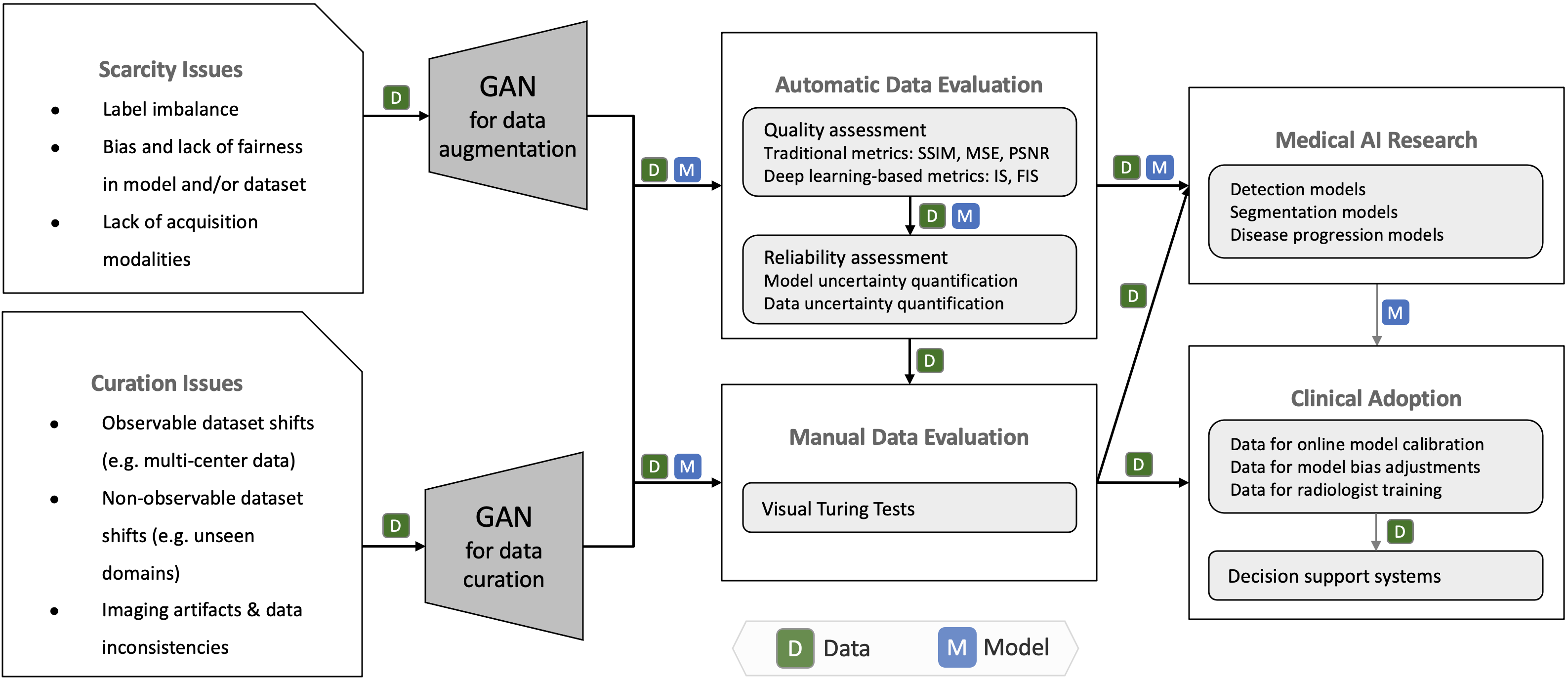}
	\caption[]{Illustration of a workflow that applies GANs to the challenges of data scarcity and data curation. After the GAN generates synthetic data specific to the issue at hand, the data is automatically and manually evaluated before further used in medical AI research. Ultimately, both synthetic data and medical AI models are integrated as decision support tools into clinical practice.}
  	\label{fig:data-scarcity-workflow}
\end{figure*}

In the following, further cancer imaging challenges in the realm of data scarcity and usability are described and related GAN solutions are referenced. Given these challenges and solutions, we derive a workflow for clinical adoption of (synthetic) cancer imaging data, which is illustrated in Figure \ref{fig:data-scarcity-workflow}.

\subsubsection{Imbalanced Data and Fairness} \label{sec:imbalance_fairness}
Apart from the rise of data-hungry deep learning solutions and the need to cover the different organs and data acquisition modalities, a major problem that arises from data scarcity is that of imbalance---i.e. the overrepresentation of a certain type of data over others~\citep{bi2019artificial}. In its more common form, imbalance of diagnostic labels can hurt a model's specificity or sensitivity, as a prior bias from the data distribution may be learned. The Lung Screening Study (LSS) Feasibility Phase exemplifies the common class imbalance in cancer imaging data: 325 (20.5\%) suspicious lung nodules were detected in the 1586 first low-dose CT screening, of which only 30 (1.89\%) were lung cancers~\citep{gohagan2004baseline, gohagan2005final, national2011national}. This problem directly translates to multi-task classification (CLF), with imbalance between different types of cancer leading to worse sensitivity on the underrepresented categories~\citep{yu2013recognition}. It is important to note that by solving the imbalance with augmentation techniques, bias is introduced as the prior distribution is manipulated, causing prevalence shift. As such, the test set should preserve the population statistics.
Aside from imbalance of labels, more insidious forms of imbalance such as that of race/ethnicity~\citep{adamson2018machineRACEdisparity} or gender~\citep{larrazabal2020gender} of patients are easily omitted in studies. This leads to fairness problems in real world applications as underrepresenting such categories in the training set will hurt performance on these categories in the real world (population shift)~\citep{li2021estimating}.
Because of their potential to generate synthetic data, GANs are a promising solution to the aforementioned problems and have already been thoroughly explored in this regard in Computer Vision~\citep{sampath2021surveyIMBALANCE, mullick2019generative}. Concretely, the discriminator and generator can be conditioned on underrepresented labels, forcing the generator to create images for a specific class\footnote{The class can be something as simple as "malignant" or "benign", or a more complex score for risk assessment of a tumour such as the BiRADs scoring system for breast tumours~\citep{liberman2002breast}}, as indicated in Figure \ref{fig:overview}(d). 
Many lesions classifiable by complex scoring systems such as RADS reporting are rare and, hence, effective conditional data augmentation is needed to improve the recognition of such lesions by ML detection models~\citep{kazuhiro2018generative}. GANs have already been used to adjust label distributions in imbalanced cancer imaging datasets, e.g. by generating underrepresented grades in a risk assessment scoring system~\citep{hu2018prostategan} for prostate cancer. A further promising applicable method is to enrich the data using a related domain as proxy input~\citep{addepalli2020degan}.
Towards the goal of a more diverse distribution of data with respect to gender and ethnicity, similar principles can be applied. For instance,~\citet{li2021estimating} proposed an adversarial training scheme to improve fairness in classification of skin lesions for underrepresented groups (age, sex, skin tone) by learning a neutral representation using an adversarial bias discrimination loss.
Fairness imposing GANs can also generate synthetic data with a preference for underrepresented groups, so that models may ingest a more balanced dataset, improving demographic parity without excluding data from the training pipeline. Such models have been trained in computer vision tasks~\citep{sattigeri2018fairnessGAN, wang2019balanced, zhang2018mitigating, xu2018fairgan, beutel2017data}, but corresponding research on medical and cancer imaging denoted by Figure \ref{fig:overview}(c) has been limited~\citep{li2021estimating, ghorbani2020dermgan}.

\subsubsection{Cross-modal Data Generation} \label{sec:cross_modal}
In cancer, multiple acquisition modalities are enlisted in clinical practice~\citep{kim2016predictive, chen2017direct, barbaro2017potential, chang2020multi, chang2020synthetic}; thus automated diagnostic models should ideally learn to interpret various modalities as well or learn a shared representation of these modalities. Conditional GANs offer the possibility to generate one \textcolor{mycorrect}{or multiple \citep{yurt2019mustgan, li2019diamondgan, zhou2020hi}} modalities from another, alleviating the need to actually perform the potentially more harmful screenings---i.e. high-dose CT, PET---that expose patients to radiation, or require invasive contrast agents such as intravenous iodine-based contrast media (ICM) in CT~\citep{haubold2021contrast}, gadolinium-based contrast agents in MRI~\citep{zhao2020tripartite}(in Table \ref{table:detection-table}) or radioactive tracers in PET~\citep{wang20183dMRItoPET, zhao2020study}. Furthermore, extending the acquisition modalities used in a given task would also enhance the performance and generalisability of AI models, allowing them to learn shared representations among these imaging modalities~\citep{bi2019artificial, hosny2018artificial}. Towards this goal, multiple GAN domain-adaptation solutions have been proposed to generate CT using MRI~\citep{wolterink2017deep, kearney2020attention, tanner2018generative, kaiser2019mriMRItoCT, nie2017medical, kazemifar2020dosimetricMRItoCT, prokopenko2019unpairedMRItoCT}, PET from MRI~\citep{wang20183dMRItoPET}, PET from CT~\citep{ben2017virtual},~\citep{bi2017synthesisCTtoPET}  (in Table \ref{table:detection-table}), and CT from PET as in~\citet{armanious2020medgan}, where also GAN-based PET denoising and MR motion correction are demonstrated. If not indicated otherwise, these image-to-image translation studies are outlined in Table \ref{table:scarcity-table}.
Because of its complexity, clinical cancer diagnosis is based not only on imaging but also non-imaging data (genomic, molecular, clinical, radiological, demographic, etc). In cases where this data is readily available, it can serve as conditional input to GANs towards the generation of images with the corresponding phenotype-genotype mapping, as is also elaborated in regard to tumour profiling for treatment in Section \ref{sec:treatment-profiling}. A multimodal cGAN was recently developed, conditioned on both images and gene expression code~\citep{xu2020correlation}; however, research along this line is otherwise limited. 

\subsubsection{Feature Hallucinations in Synthetic Data}\label{sec:feature-hallucation}
\begin{figure} [ht]
	\begin{center}
       		\includegraphics[width=0.48\textwidth]{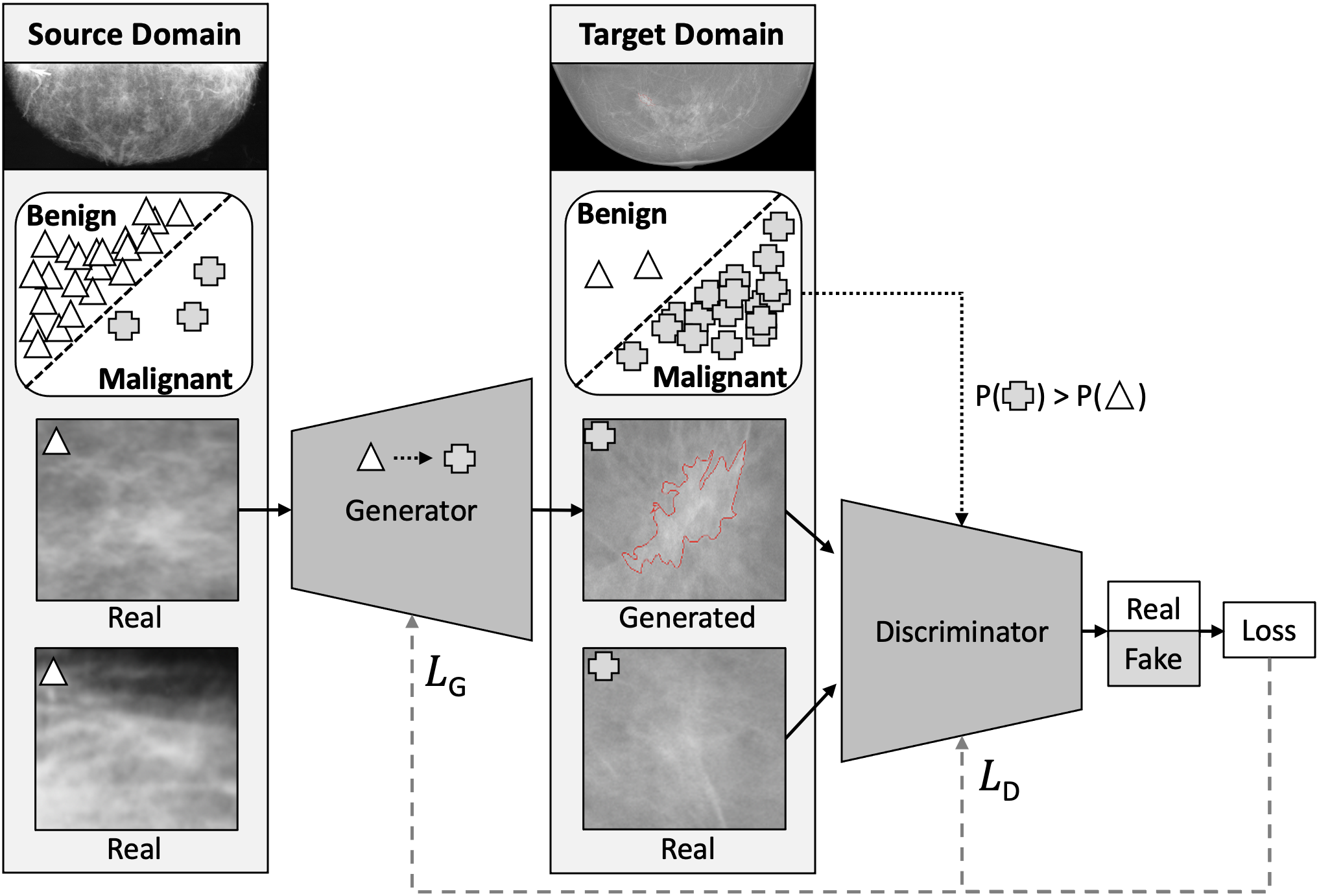}
	\end{center}
	 \caption[]{Example of a GAN that translates Film Scanned MMG (source) to Full-Field Digital MMG (target). The generator transforms 'benign' source images (triangles) into 'malignant' target images (plus symbols). As opposed to source, the target domain contains more malignant MMGs than benign ones. If the discriminator thus learns to associate malignancy with realness, this incentivises the generator to inject malignant features (depicted by dotted arrows).
	 For simplicity additional losses (e.g., reconstruction losses) are omitted.}
  	\label{fig:pathology-feature-translation}
\end{figure}

As displayed in Figure \ref{fig:pathology-feature-translation} and denoted in Figure \ref{fig:overview}(b), conditional GANs can unintentionally\footnote{Intentional feature injection or removal is discussed in \ref{paragraph:tampering}} hallucinate non-existent artifacts into a patient image. This is particularly likely to occur in cross-modal data augmentation, especially but not exclusively if the underlying dataset is imbalanced. For instance,~\citet{cohen2018distribution} describe GAN image feature hallucinations embodied by added and removed brain tumours in cranial MRI. The authors tested the relationship between the ratio of tumour images in the GAN target distribution and the ratio of images diagnosed with tumours by a classifier. The classifier was trained on the GAN generated target dataset, but tested on a balanced holdout test set.
It was thereby shown that the generator of CycleGAN effectively learned to hide source domain image features in target domain images, which arguably helped it to fool its discriminator. Paired image-to-image translation with pix2pix~\citep{isola2017image} was more stable, but still some hallucinations were shown to likely have occurred. A cause for this can be a biased discriminator that has learned to discriminating specific image features (e.g., tumours) that are more present in one domain.~\citet{cohen2018distribution, cohen2018cure} and~\citet{ wolterink2018generative} warn that models that map source to target images, have an incentive to add/remove features during translation if the feature distribution in the target domain is distinct from the feature distribution in the source domain\footnote{For example, if one domain contains mainly healthy images, while the other domain contains mainly pathological images.}.

Domain-adaptation with unpaired image-to-image translation GANs such as CyleGAN has become increasingly popular in cancer imaging~\citep{wolterink2017deep, tanner2018generative, modanwal2019normalization, fossen2020synthesizing, zhao2020study, hognon2019standardization, mathew2020augmenting, kearney2020attention, peng2020magnetic, jiang2018tumor, sandfort2019data}. As described, these methods are hallucination-prone and, thus, can put patients at risk when used in clinical settings. More research is needed on how to robustly avoid or detect and eliminate hallucinations in generated data. To this end, we highlight the potential of investigating feature preserving image translation techniques and methods for evaluating whether features have been accurately translated. For instance, in the presence of feature masks or annotations, an additional local reconstruction loss can be introduced in GANs that enforces feature translation in specific image areas.

\subsubsection{Data Curation and Harmonisation} \label{sec:curation}

Aside from the limited availability of cancer imaging datasets, a major problem is that the ones available are often not readily usable and require further curation~\citep{hosny2018artificial}. 
Curation includes dataset formatting, normalising, structuring, de-identification, quality assessment and other methods to facilitate subsequent data processing steps, one of which is the ingestion of the data into AI models~\citep{diaz2021data}.
In the past, GANs have been proposed for curation of data labelling, segmentation and annotation of images (details in Section \ref{sec:annotation}) and de-identification of facial features, EHRs, etc (details in Section \ref{sec:access}). Particular to cancer imaging datasets and of significant importance is the correction of artifacts, such as patient motion, metallic objects, chemical shifts and others caused by the image processing pipeline~\citep{pusey1986magnetic, nehmeh2002effect}, which run the risk of confusing models with spurious information. Towards the principled removal of artifacts, several GAN solutions have been proposed~\citep{vu2020generativeARTIFACTREMOVAL, koike2020deepARTIFACTREMOVAL, armanious2020medgan}. As for the task of reconstruction of compressed data (e.g., compressed sensing MRI~\citep{mardani2017deep}), markedly,~\citet{yang2018dagan} proposed DAGAN, which is based on U-Net~\citep{ronneberger2015u}, reduces aliasing artifacts, and faithfully preserves texture, boundaries and edges (of brain tumours) in the reconstructed images.~\citet{kim2018improving} feed down-sampled high-resolution brain tumour MRI into a GAN framework similar to pix2pix to reconstruct high-resolution images with different contrast. The authors highlight the possible acceleration of MR imagery collection while retaining high-resolution images in multiple contrasts, necessary for further clinical decision-making. As relevant to the context of data quality curation, GANs have also been proposed for image super-resolution in cancer imaging (e.g., for lung nodule detection~\citep{gu2020medsrgan}, abdominal CT~\citep{you2019ct}, and breast histopathology~\citep{shahidi2021breast}).

Beyond the lack of curation, a problem particular to multi-centre studies is that of inconsistent curation between data derived in different centres. These discontinuities arise from different scanners, segmentation protocols, demographics, etc, and can cause significant problems to subsequent ML algorithms that may overfit or bias towards one configuration over another (i.e. acquisition and annotation shifts). GANs have the potential to contribute in this domain as well by bringing the distributions of images across different centres closer together. In this context recent work by~\citet{li2021normalization} and~\citet{wei2020using} used GAN-based volumetric normalisation to reduce the variability of heterogeneous 3D chest CT scans of different slice thickness and dose levels. The authors showed that features in subsequent radiomics analysis exhibit increased alignment. Other works in this domain include a framework that could standardise heterogeneous datasets with a single reference image and obtained promising results on an MRI dataset~\citep{hognon2019standardization}, and GANs that learn bidirectional mappings between different vendors to normalise dynamic contrast enhanced (DCE) breast MRI~\citep{modanwal2019normalization}. An interesting research direction to be explored in the future is synthetic multi-centre data generation using GANs, simulating the distribution of various scanners/centres.

\subsubsection{Synthetic Data Assessment}
As indicated in Figure \ref{fig:overview}(e), a condition of paramount importance is proper evaluation of GAN-generated or GAN-curated data. This evaluation is to verify that synthetic data is usable for a desired downstream task (e.g., segmentation, classification) and/or indistinguishable from real data while ensuring that no private information is leaked. 
GANs are commonly evaluated based on \textit{fidelity} (realism of generated
samples) and \textit{diversity} (variation of generated samples compared to real samples)~\citep{borji2021pros}. Different quantitative measures exist to assess GANs based on the fidelity and diversity of its generated synthetic medical images~\citep{yi2019generative, borji2021pros}.

Visual Turing tests (otherwise referred to as Visual Assessment, Mean Opinion Score (MOS) Test, and sometimes used interchangeably with In-Silico Clinical Trials) are arguably the most reliable approach, where clinical experts are presented with samples from real and generated data and are tasked to identify which one is generated.~\citet{korkinof2020perceived} showed that their PGGAN-generated~\citep{karras2017progressive} 1280x1024 mammograms were inseparable by the majority of participants, including trained breast radiologists. A similar visual Turing test was successfully done in the case of skin disease~\citep{ghorbani2020dermgan}, super-resolution of CT~\citep{you2019ct}, brain MRI~\citep{kazuhiro2018generative, han2018ganBRATS}, lung cancer CT scans~\citep{chuquicusma2018fool}, and histopathology images~\citep{levine2020synthesis}. For instance,~\citet{chuquicusma2018fool} trained a DCGAN~\citep{radford2015unsupervised} on the LIDC-IDRI dataset\citep{armato2011lung} to generate 2D (56x56 pixel) pulmonary lung nodule scans that were realistic enough to deceive 2 radiologists with 11 and 4 years of experience.
In contrast to computer vision techniques where synthetic data can often be easily evaluated by any non-expert, the requirement of clinical experts makes Visual Turing Tests in this domain much more costly. Furthermore, a lack of scalability and consistency in medical judgement needs to be taken into account as well~\citep{brennan1992statistical} and visual Turing tests should in the ideal case engage a range of experts to address inter-observer variation in the assessments. Also, iterating over the same observer addresses intra-observer variation---i.e. repeating the process within a certain amount of intervals that could be days or weeks. These problems are further magnified by the shortage of radiology experts~\citep{mahajan2020auditSHORTAGE, rimmer2017radiologistSHORTAGE}
which brings up the necessity for supplementary metrics that can automate the evaluation of generative models. Such metrics allow for preliminary evaluation and can enable research to progress without the logistical hurdle of enlisting experts.

Furthermore, in cases where the sole purpose of the generated data is to improve a downstream task---i.e. classification or segmentation---then the prediction success of the downstream task would be the metric of interest. The latter can reasonably be prioritised over other metrics given that the underlying reasons why the synthetic data alters downstream task performance are examined and clarified\footnote{For example, synthetic data may balance imbalanced datasets, reduce overfitting on limited training data, or improve model robustness to better capture domain shifts in the test dataset.}.

\paragraph{Image Quality Assessment Metrics} 
~\citet{wang2004imageAUTOMATICEVALUATION} have thoroughly investigated image quality assessment metrics. The most commonly applied metrics include structural similarity index measure (SSIM)\footnote{SSIM predicts perceived quality and considers image statistics to assess structural information based on luminance, contrast, and structure.} between generated image and reference image~\citep{wang2004imageAUTOMATICEVALUATION}, mean squared error (MSE)\footnote{MSE is computed by averaging the squared intensity differences between corresponding pixels of the generated image and the reference image.} and peak signal-to-noise ratio (PSNR)\footnote{PSNR is an adjustment to the MSE score, commonly used to measure reconstruction quality in lossy compression.}.
In a recent example that followed this framework of evaluation, synthetic brain MRI with tumours generated by edge-aware EA-GAN~\citep{yu2019eaEVALUATION} was assessed using three such metrics: PSNR, SSIM, and normalised mean squared error (NMSE). The authors integrated an end-to-end sobel edge detector to create edge maps from real/synthetic images that are input into the discriminator in the dEa-GAN variant to enforce improved textural structure and object boundaries. Interestingly, aside from evaluating on the whole image, the authors demonstrated evaluation results focused on the tumour regions, which were overall significantly lower than the whole image. 
Other works that have evaluated their synthetic images in an automatic manner have focused primarily on the SSIM and PSNR metrics and include generation of CT~\citep{kearney2020attention, mathew2020augmenting} and PET scans~\citep{zhao2020study}. While indicative of image quality, these similarity-based metrics might not generalise well to human judgement of image similarity, the latter depending on high-order image structure and context~\citep{zhang2018unreasonable}. Finding evaluation metrics that are strong correlates of human judgement of perceptual image similarity is a promising line of research. In the context of cancer and medical imaging, we highlight the need for evaluation metrics for synthetic images that correlate with the perceptual image similarity judged by medical experts. Apart from perceptual image similarity, further evaluation metrics in cancer and medical imaging are to be investigated that are able to estimate the diagnostic value of (synthetic) images and, in the presence of reference images, the diagnostic value proportion between target and reference image.

\paragraph{Deep Generative Model-specific Assessment Metrics} In recent years, the Inception score (IS)~\citep{salimans2016improved} and Fréchet Inception distance (FID)~\citep{heusel2017gans} have emerged, offering a more sophisticated alternative for the assessment of synthetic data. The IS uses a classifier to generate a probability distribution of labels given a synthetic image.
If the probability distribution is highly skewed, it is indicative that a specific object is present in the image (resulting in a higher IS), while in the case where it is uniform, the image contains a jumble of objects and that is more likely to be non-sense (resulting in a lower IS).\footnote{Not only a low label entropy within an image is desired, but also a high label entropy across images: IS also assesses the variety of peaks in the probability distributions generated from the synthetic images, so that a higher variety is indicative of more diverse objects being generated by the GAN (resulting in a higher IS).}
The FID metric compares the distance between the synthetic image distribution to that of the real image distribution by comparing extracted high-level features from one of the layers of a classifier (e.g., Inception v3 as in IS). Both metrics have shown promise in the evaluation of GAN-generated data; however, they come with several bias issues that need to be taken into account during evaluation~\citep{chong2020effectively, devries2019evaluation, borji2019pros}. As these metrics have not been widely used in cancer imaging yet, their applicability on GAN-synthesised cancer images remains to be investigated. In contrast to computer vision datasets containing diverse objects, medical imaging datasets commonly only contain images of one specific organ. In this regard, we promote further research as to how object diversity based methods such as IS can be applied to medical and cancer imaging, which requires, among others, meaningful adjustments of the dataset-specific pretrained classifications models (i.e. Inception v3) that IS and FID rely upon.

\paragraph{Uncertainty Quantification as GAN Evaluation Metric?} \label{sec:uncertainty} 
A general problem facing the adoption of deep learning methods in clinical tasks is their inherent unreliability exemplified by high prediction variation caused by minimal input variation (e.g., one pixel attack~\citep{korpihalkola2020one}). This is further exacerbated by the nontransparent decision making process inside deep neural networks thus often described as "black box models"~\citep{bi2019artificial}. Also, the performance of deep learning methods in out-of-domain datasets has been assessed as unreliable~\citep{lim2019buildingTRUST}. To eventually achieve beneficial clinical adoption and trust, examining and reporting the inherent uncertainty of these models on each prediction becomes a necessity. Besides classification, segmentation~\citep{hu2020coarse, alshehhi2021quantification}, etc, uncertainty estimation is applicable to models in the context of data generation as well~\citep{lim2019buildingTRUST, abdar2020reviewUNCERTAINTY, hu2020coarse}.
~\citet{edupuganti2019uncertainty} studied a GAN architecture based on variational autoencoders (VAE)~\citep{kingma2013auto} on the task of MRI reconstruction, with emphasis on uncertainty studies. Due to their probabilistic nature, VAEs allowed for a Monte Carlo sampling approach which enables quantification of pixel-variance and the generation of uncertainty maps. Furthermore, they used Stein’s Unbiased Risk Estimator (SURE)~\citep{stein1981estimation} as a measure of uncertainty that serves as surrogate of MSE even in the absence of ground truth. Their results indicated that adversarial losses introduce more uncertainty. 
Parallel to image reconstruction, uncertainty has also been studied in the context of brain tumours (glioma) in MRI enhancement~\citep{tanno2021uncertainty}. In this study, a probabilistic deep learning framework for model uncertainty quantification was proposed, decomposing the problem into two uncertainty types: \textit{intrinsic uncertainty} (particular to image enhancement and pertaining to the one-to-many nature of the super-resolution mapping) and \textit{parameter uncertainty} (a general challenge, it pertains to the choice of the optimal model parameters). 
The overall model uncertainty in this case is a combination of the two and was evaluated for image super-resolution. Through a series of systematic studies the utility of this approach was highlighted, as it resulted in improved overall prediction performance of the evaluated models even for out-of-distribution data. It was further shown that predictive uncertainty highly correlated with reconstruction error, which not only enabled spotting unrealistic synthetic images, but also highlights the potential in further exploring uncertainty as an evaluation metric for GAN-generated data.
A further use-case of interest for GAN evaluation via uncertainty estimation is the "adherence" to provided conditional inputs. As elaborated in \ref{sec:feature-hallucation} for image-to-image translation, conditional GANs are likely to introduce features that do not correspond to the conditional class label or source image. After training a classification model on image features of interest (say, tumour vs non-tumour features), we can examine the classifier's prediction and estimated uncertainty\footnote{The uncertainty can be estimated using methods such as \textcolor{mycorrect}{Bayesian Neural Networks~\citep{mackay1992practical, neal2012bayesian}}, Monte-Carlo Dropout~\citep{gal2016dropout} or Deep Ensembles~\citep{lakshminarayanan2016simple}.} for the generated images. Given the expected features in the generated images are known beforehand, the classifier's uncertainty of the presence of these features can be used to estimate not only image fidelity (e.g., image features are not generated realistic enough), but also "condition adherence" (e.g., expected image features are altered during generation).

\paragraph{Outlook on Clinical Adoption}
Alongside GAN-specific and standard image assessment metrics, uncertainty-based evaluation schemes can further automate the analysis of generative models. To this end, the challenge of clinical validation for predictive uncertainty as a reliability metric for synthetic data assessment remains~\citep{tanno2021uncertainty}. In practice, building clinical trust in AI models is a non-trivial endeavour and will require rigorous performance monitoring and calibration especially in the early stages~\citep{kelly2019key, duran2021afraid}. This is particularly the case when CADe and CADx models are trained on entirely (or partially) synthetic data given that the data itself was not first assessed by clinicians. Until a certain level of trust is built in these pipelines, automatic metrics will be a preliminary evaluation step that is inevitably followed by diligent clinical evaluation for deployment. A research direction of interest in this context would be "gatekeeper" GANs---i.e. GANs that simulate common data (and/or difficult edge cases) of the target hospital, on which deployment-ready candidate models (e.g., segmentation, classification, etc) are then tested to ensure they are sufficiently generalisable. If the candidate model performance on such test data satisfies a predefined threshold, it has passed this quality gate for clinical deployment.

\begin{table*}[htbp] 
\centering
\scriptsize
\caption{Overview of adversarially-trained models applied to cancer imaging \textbf{data scarcity and usability} challenges. Publications are clustered by section and ordered by year in ascending order.}\label{table:scarcity-table}
\scalebox{0.85}{
\begin{tabular}{{p{0.17\textwidth}p{0.13\textwidth}p{0.24\textwidth}p{0.13\textwidth}p{0.13\textwidth}p{0.25\textwidth}}}
    \hline
    \textbf{Publication} &
    \textbf{Method} &
    \textbf{Dataset} &
    \textbf{Modality} &
    \textbf{Task} &
    \textbf{Highlights} 
    \\
    \hline\hline
    \textbf{Imbalanced Data \& Fairness}
    \\
    \hline
    \citet{hu2018prostategan} &
    ProstateGAN & 
    Private &
    Prostate MRI &
    \textcolor{mycorrect}{Class-}conditional synthesis & 
    Gleason score (cancer grade) class conditions.
    \\
    \hline
    \citet{ghorbani2020dermgan} &
    DermGAN & 
    Private & 
    Dermoscopy & 
    Paired translation &
    Adapted pix2pix evaluated via Turing Tests and rare skin condition CLF.
    \\
    \hline
    \citet{li2021estimating} &
    Encoder & 
    ISIC 2018~\citep{codella2018skin} & 
    Dermoscopy & 
    \textcolor{mycorrect}{Adversarial training,} Representation learning & 
    Fair Encoder with bias discriminator and skin lesion CLF.
    \\
    \hline\hline
    \textbf{Cross-Modal Data Generation}
    \\
    \hline
    \citet{wolterink2017deep} &
    CycleGAN & 
    Private &
    Cranial MRI/CT  & 
    Unpaired translation & 
    First CNN for unpaired MR-to-CT translation. Evaluated via PSNR and MAE. 
    \\
    \hline
    \citet{ben2017virtual} &
    pix2pix & 
    Private & 
    Liver PET/CT & 
    Paired translation & 
    Paired CT-to-PET translation with focus on hepatic malignant tumours.
    \\
    \hline
    \citet{nie2017medical} &
    context-aware GAN & 
    ADNI~\citep{wyman2013standardization, weiner2017alzheimer}
    & 
    Cranial/pelvic MRI/CT  & 
    Paired translation & 
    Supervised 3D GAN for MR-to-CT translation with "Auto-Context Model" (ACM).
    \\
    \hline
    \citet{wang20183dMRItoPET} &
    Locality Adaptive GAN (LA-GAN) & 
    BrainWeb phantom~\citep{cocosco1997brainweb} & 
    Cranial MRI, PET phantom  & 
    Paired translation &
    3D auto-context, synthesising PET from low-dose PET and multimodal MRI. 
    \\
    \hline
    \citet{tanner2018generative} &
    CycleGAN & 
    VISCERAL~\citep{jimenez2016cloud} & 
    Lung/abdominal MRI/CT &
    Image registration & 
    MR-CT CycleGAN for registration.
    \\
    \hline
    \citet{kaiser2019mriMRItoCT} &
    pix2pix, context-aware GAN~\citep{nie2017medical} & 
    RIRE~\citep{rire1998} & 
    Cranial MRI/CT  & 
    Paired translation & 
    Detailed preprocessing description, MR-to-CT translation, comparison with U-Net.
    \\
    \hline
    \citet{prokopenko2019unpairedMRItoCT} &
    DualGAN, SRGAN & 
    CPTAC3~\citep{national2018radiology} \& Head-Neck-PET-CT~\citep{vallieres9data}
    &
    Cranial MRI/CT &
    Unpaired translation &
    DualGAN for unpaired MR-to-CT, visual Turing tests. 
    \\
    \hline
    \textcolor{mycorrect}{\citet{yurt2019mustgan}} 
    
    \href{http://github.com/icon-lab/mrirecon}{mrirecon} & 
    \textcolor{mycorrect}{mustGAN} & 
    \textcolor{mycorrect}{IXI~\citep{ixidataset} \& ISLES~\citep{maier2017isles}} & 
    \textcolor{mycorrect}{Cranial MRI} & 
    \textcolor{mycorrect}{Paired translation} & 
    \textcolor{mycorrect}{FLAIR, T1, T2 synthesis via feature fusion of one-to-one and many-to-one pix2pix networks.}
    \\
    \hline
    \textcolor{mycorrect}{\citet{li2019diamondgan}} & 
    \textcolor{mycorrect}{diamondGAN} & 
    \textcolor{mycorrect}{Private \& MICCAI-WMH~\citep{kuijf2019standardized}} & 
    \textcolor{mycorrect}{Cranial MRI} & 
    \textcolor{mycorrect}{Unpaired translation} & 
    \textcolor{mycorrect}{Target modality synthesis from flexible set of non-aligned source modalities.}
    \\
    \hline
    \textcolor{mycorrect}{\citet{zhou2020hi}} & 
    \textcolor{mycorrect}{hi-Net} & 
    \textcolor{mycorrect}{BRATS 2018~\citep{menze2014multimodal, bakas2018identifying}} & 
    \textcolor{mycorrect}{Cranial MRI} & 
    \textcolor{mycorrect}{Paired translation} & 
    \textcolor{mycorrect}{Domain-specific encoder network features fused into layers of pix2pix-based network.}
    \\
    \hline
    \citet{zhao2020study} &
    S-CycleGAN & 
    Private & 
    Cranial low/full dose PET & 
    Paired translation &
    Low (LDPET) to full dose (FDPET) translation with supervised loss for paired images. 
    \\
    \hline
    \citet{kearney2020attention} &
    VAE-enhanced A-CycleGAN & 
    Private & 
    Cranial MRI/CT & 
    Unpaired translation & 
    MR-to-CT evaluated via PSNR, SSIM, MAE. Superior to paired alternatives.
    \\
    \hline
    \citet{kazemifar2020dosimetricMRItoCT} &
    context-aware GAN & 
    Private & 
    Cranial MRI/CT & 
    Paired translation & 
    Feasibility of generated CT from MRI for dose calculation for radiation treatment.
    \\
    \hline
    \citet{armanious2020medgan} &
    MedGAN &  
    Private & 
    Liver PET/CT & 
    Paired translation & 
    CasNet architecture, PET-to-CT, MRI motion artifact correction, PET denoising.
    \\
    \hline
    \citet{xu2020correlation} &
    multi-conditional GAN &
    NSCLC~\citep{zhou2018non}& 
    Lung CT, gene expression &
    \textcolor{mycorrect}{Multi-input} conditional synthesis &  
    Image-gene data fusion, nodule generator input: background, segmentation, gene code.
    \\
    \hline
    \citet{haubold2021contrast}  &
    Pix2PixHD~\citep{wang2018high} &
    Private & 
    Arterial phase CT & 
    Paired translation & 
    Low-to-full ICM CT (thorax, liver, abdomen), 50\% reduction in intravenous ICM dose.
    \\
    \hline\hline
    \textbf{Feature Hallucinations}
    \\
    \hline
    \citet{cohen2018distribution, cohen2018cure}
    
    \href{https://github.com/ieee8023/dist-bias}{dist-bias} &
    CycleGAN, pix2pix &
    BRATS2013~\citep{menze2014multimodal} &
    Cranial MRI & 
    Paired/unpaired translation & 
    Removed/added tumours during image translation can lead to misdiagnosis.
    \\
    \hline\hline
    \textbf{Data Curation}
    \\
    \hline
    \citet{yang2018dagan} &
    DAGAN &
    MICCAI 2013 grand challenge dataset & 
    Cranial MRI & 
    Image reconstruction & 
    Fast GAN compressed sensing MRI reconstruction outperformed conventional methods.
    \\
    \hline
    \citet{kim2018improving} &
    pix2pix-based &  
    BRATS~\citep{menze2014multimodal} &
    Cranial MRI & 
    Reconstruction/ super-resolution &
    Information transfer between different contrast MRI, effective pretraining/fine-tuning.
    \\
    \hline
    \citet{hognon2019standardization} &
    CycleGAN, pix2pix &  
    BRATS~\citep{menze2014multimodal}, BrainWeb phantom~\citep{cocosco1997brainweb} &
    Cranial MRI & 
    Paired/unpaired translation, normalisation  &
    CycleGAN translation to BrainWeb reference image, pix2pix back-translation to source.
    \\
    \hline
    \citet{modanwal2019normalization} &
    CycleGAN &  
    Private & 
    Breast MRI & 
    Unpaired translation & 
    Standardising DCE-MRI across scanners, anatomy preserving mutual information loss.
    \\
    \hline
    \citet{you2019ct} & 
    CycleGAN-based & 
    Mayo Low Dose CT~\citep{LowDoseCTGrandChallenge} & 
    Abdominal CT & 
    Super-resolution & 
    Joint constraints to Wasserstein loss for structural preservation. Evaluated by 3 radiologists.
    \\
    \hline
    \citet{gu2020medsrgan} &
    MedSRGAN &
    LUNA16~\citep{setio2017validation} & 
    MRI/thoracic CT & 
    Super-resolution & 
     Residual Whole Map Attention Network (RWMAN) in G. Evaluated by 5 radiologists.
    \\
    \hline
    \citet{vu2020generativeARTIFACTREMOVAL} &
    WGAN-GP-based &
    k-Wave toolbox~\citep{treeby2010k}& 
    Photoacoustic CT (PACT) & 
    Paired translation & 
    U-Net \& WGAN-GP based generator for artifact removal. Evaluated via SSIM, PSNR.
    \\
    \hline
    \citet{koike2020deepARTIFACTREMOVAL} &
    CycleGAN &
    Private & 
    Head/neck CT & 
    Unpaired translation & 
    Metal artifact reduction via CT-to-CT translation, evaluated via radiotherapy dose accuracy. 
    \\
    \hline
    \citet{wei2020using} &
    WGAN-GP-inspired &
    Private &
    Chest CT & 
    Paired translation &
    CT normalisation of dose/slice thickness. Evaluated via Radiomics Feature Variability.
    \\
    \hline
    \citet{shahidi2021breast} &
    WA-SRGAN &  
    BreakHis~\citep{benhammou2020breakhis}, Camelyon~\citep{litjens20181399} & 
    Breast/lymph node histopathology & 
    Super-resolution &
    Wide residual blocks, self-attention SRGAN for improved robustness \& resolution
    \\
    \hline
    \citet{li2021normalization} &
    SingleGAN-based~\citep{yu2018singlegan} &
    Private &
    Spleen/colorectal CT &
    Unpaired translation & 
    Multi-centre (4) CT normalisation. Evaluated via cross-centre radiomics features variation. Short/long-term survivor CLF improvement.
    \\
    \hline\hline
    \textbf{Synthetic Data Assessment}
    \\
    \hline
    \citet{kazuhiro2018generative} &
    DCGAN &
    Private & 
    Cranial MRI & 
    \textcolor{mycorrect}{Noise-to-}image synthesis &
    Feasibility study for brain MRI synthesis evaluated by 7 radiologists.
    \\
    \hline
    \citet{han2018ganBRATS} &
    DCGAN, WGAN &
    BRATS 2016~\citep{menze2014multimodal} & 
    Cranial MRI & 
    \textcolor{mycorrect}{Noise-to-}image synthesis &
    128x128 brain MRI synthesis evaluated by one expert physician.
    \\
    \hline
    \citet{chuquicusma2018fool} &
    DCGAN &
    LIDC-IDRI\citep{armato2011lung} & 
    Thoracic CT & 
    \textcolor{mycorrect}{Noise-to-}image synthesis &
    Malignant/benign lung nodule ROI generation evaluated by two radiologists.
    \\
    \hline
    \citet{yu2019eaEVALUATION} &
    Ea-GAN &
    BRATS 2015~\citep{menze2014multimodal}, IXI~\citep{ixidataset} &
    Cranial MRI & 
    Paired image-to-image translation &
    Loss based on edge maps of synthetic images. Evaluated via PSNR, NMSE, SSIM.
    \\
    \hline
    \citet{korkinof2020perceived} &
    PGGAN &
    Private &
    Full-field digital MMG & 
    \textcolor{mycorrect}{Noise-to-}image synthesis &
    1280×1024 MMG synthesis from $>10^6$ image dataset. Evaluated by 55 radiologists.
    \\
    \hline
    \citet{levine2020synthesis} &
    PGGAN, VAE, ESRGAN &
    TCGA~\citep{grossman2016toward}, OVCARE archive & 
    Ovarian Histopathology & 
    \textcolor{mycorrect}{Noise-to-}image synthesis &
    1024×1024 whole slide synthesis. Evaluated via FID and by 15 pathologists (9 certified)
    \\
    \hline
    \end{tabular}
    }
\end{table*}

\subsection{Data Access and Privacy Challenges}\label{sec:access}
Access to sufficiently large and labelled data resources is the main constraint for the development of deep learning models for medical imaging tasks~\citep{esteva2019guide}. In cancer imaging, the practice of sharing validated data to aid the development of AI algorithms is restricted due to technical, ethical, and legal concerns~\citep{bi2019artificial}. The latter is exemplified by regulations such as the Health Insurance Portability and Accountability Act~\citep{act1996health} in the United States of America (USA) or the European Union's General Data Protection Regulation~\citep{gdpr2016} with which respective clinical centres must comply with. Alongside the need and numerous benefits of patient privacy preservation, it can also limit data sharing initiatives and restrict the availability, size and usability of public cancer imaging datasets.~\citet{bi2019artificial} assess the absence of such datasets as a noteworthy challenge for AI in cancer imaging.

The published GANs \textcolor{mycorrect}{and adversarial training methods} that are suggested for or applied to cancer imaging challenges within this section \ref{sec:access} are summarise below in Table \ref{Table:privacy-table1}.

\subsubsection{Decentralised Data Generation} \label{sec:federated}
As AI systems are often developed and trained outside of medical institutions, prior approval to transfer data out of their respective data silos is required, adding significant hurdles to the logistics of setting up a training pipeline or rendering it entirely impossible. In addition, medical institutions can often not guarantee a secured connection to systems deployed outside their centres~\citep{hosny2018artificial}, which further limits their options to share valuable training data. 

One privacy preserving approach solving this problem is federated learning~\citep{mcmahan2017communication}, where copies of an AI model are trained in a distributed fashion inside each clinical centre in parallel and are aggregated to a global model in a central server. This eliminates the need for sensitive patient data to leave any of the clinical centres~\citep{Kaissis2020, sheller2020federated}. However, it is to be noted that federated learning cannot guarantee full patient privacy.~\citet{hitaj2017deep} demonstrated that any malicious user can train a GAN to violate the privacy of the other users in a federated learning system. While difficult to avoid, the risk of such GAN-based attacks can be minimised, e.g., by using a combination of selective parameter updates~\citep{shokri2015privacy} (sharing only a small selected part of the model parameters across centres) and the sparse vector technique\footnote{Sparse Vector Technique (SVT)~\citep{lyu2016understanding} is a Differential Privacy (DP)~\citep{10.1007/11787006_1} method that introduces noise into a deep learning model's gradients.} as shown by~\citet{li2019privacy}.

To solve the challenge of privacy assurance of clinical medical imaging data, a distributed GAN~\citep{hardy2019md, xin2020private, guerraoui2020fegan, rasouli2020fedgan, zhang2021training} can be trained on sensitive patient data to generate synthetic training data. The technical, legal, and ethical constraints for sharing de-identified synthetic data are typically less restrictive than for real patient data. Such generated data can be used instead of the real patient data to train models on disease detection, segmentation, or prognosis.

\begin{figure*}
    \centering
       		\includegraphics[width=0.85\textwidth]{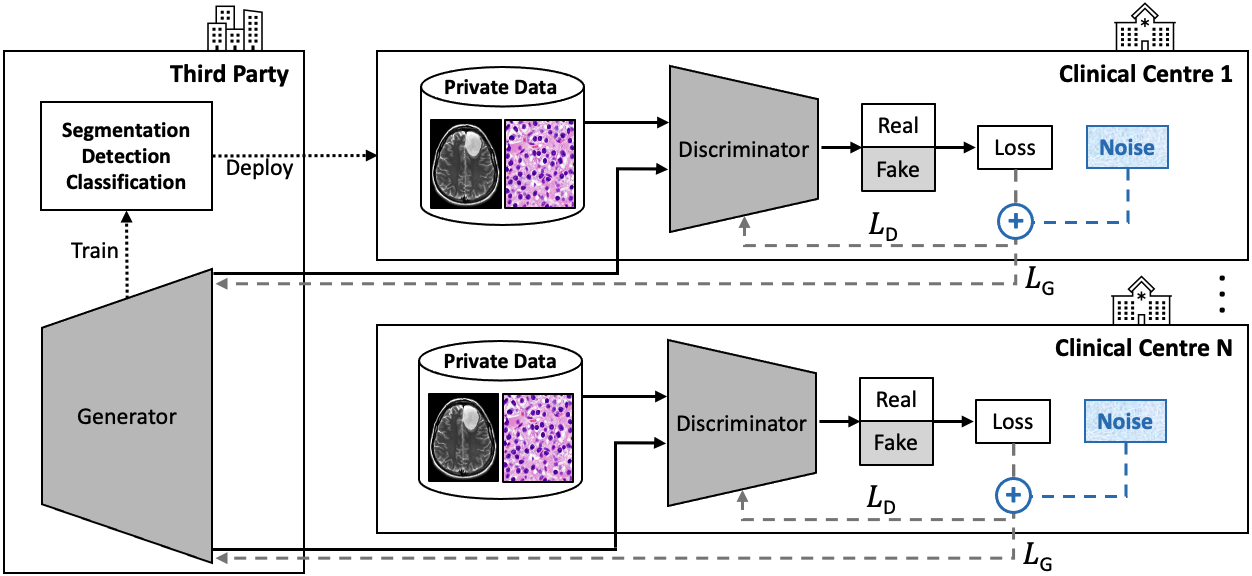}
	\caption[]{Visual example of a GAN in a federated learning setup with a central generator trying to generate realistic samples that fool all of the discriminators, which are distributed across clinical centres as in ~\citet{chang2020multi,chang2020synthetic}. Once trained, the generator can produce training data for a downstream task model (e.g., segmentation, detection, classification). As depicted in blue colour, we suggest to extend the federated learning setup by adding "Noise" to the gradients providing a differential privacy guarantee. This reduces the possibility of reconstruction of specific records of the training data (i.e. images of a specific patient) by someone with access to the trained GAN model (i.e. to the generator) or by someone intercepting the synthetic images while they are transferred from the central generator to the centres during training.
	}
  	\label{fig:federated-dp-gan}
\end{figure*}

For instance,~\citet{chang2020multi,chang2020synthetic} proposed the Distributed Asynchronized Discriminator GAN (AsynDGAN), which consists of multiple discriminators deployed inside various medical centres and one central generator deployed outside the medical centres. The generator never needs to see the private patient data, as it learns by receiving the gradient updates of each of the discriminators. The discriminators are trained to differentiate images of their medical centre from synthetic images received from the central generator. After training AsynDGAN, its generator is used and evaluated based on its ability to provide a rich training set of images to successfully train a segmentation model. AsynDGAN is evaluated on MRI brain tumour segmentation and cell nuclei segmentation. The segmentation models trained only on AsynDGAN-generated data achieves a competitive performance when compared to segmentation models trained on the entire dataset of real data. Notably, models trained on AsynDGAN-generated data outperform models trained on local data from only one of the medical centres. To our best knowledge, AsynDGAN is the only distributed GAN applied to cancer imaging to date. Therefore, we promote further research in this line to fully exploit the potential of privacy-preservation using distributed GANs. As demonstrated in Figure \ref{fig:federated-dp-gan} and suggested in Figure \ref{fig:overview}(f), for maximal privacy preservation we recommend exploring methods that combine privacy during training (e.g., federated GANs) with privacy after training (e.g., differentially-private GANs), the latter being described in the following section.

\subsubsection{Differentially-Private Data Generation}
\citet{shin2018medical} train a GAN to generate brain tumour images and highlight the usefulness of their method for anonymisation, as their synthetic data cannot be attributed to a single patient but rather only to an instantiation of the training population. However, it is to be scrutinised whether such synthetic samples are indeed fully private, as, given a careful analysis of the GAN model and/or its generated samples, a risk of possible reconstruction of part of the GAN training data exists~\citep{papernot2016semi}. For example,~\citet{chen2020improved} propose a GAN for model inversion (MI) attacks, which aim at reconstructing the training data from a target model's parameters. A potential solution to avoid training data reconstruction is highlighted by~\citet{xie2018differentially}, who propose the Differentially Private Generative Adversarial Network (DPGAN). In Differential Privacy (DP)~\citep{10.1007/11787006_1} the parameters ($\varepsilon, \delta$) denote the privacy budget~\citep{torfi2020differentially}, where $\varepsilon$ measures the privacy loss and $\delta$ represents the probability that a range of outputs with a privacy loss $>\varepsilon$ exists\footnote{For example, if an identical model $M$ is trained two times, once with training data $D$ resulting in $M_{D}$ and once with marginally different training data $D\textquotesingle$ resulting in $M_{D\textquotesingle}$, it is ($\varepsilon$)-DP if the following holds true: For any possible output $x$, the output probability for $x$ of model $M_{D}$ differs no more than $exp(\varepsilon)$ from the output probability for $x$ of $M_{D\textquotesingle}$.}. Hence, the smaller the parameters ($\varepsilon, \delta$) for a given model, the less effect a single sample in the training data has on model output. The less effect of such a single sample, the stronger is the confidence in the privacy of the model to not reveal samples of the training data.

\paragraph{Examples of GANs with Differential Privacy Guarantees}
In DPGAN noise is added to the model's gradients during training to ensure training data privacy. Extending on the concept of DPGAN,~\citet{jordon2018pate} train a GAN coined PATE-GAN based on the Private Aggregation of Teacher Ensembles (PATE) framework~\citep{papernot2016semi,papernot2018scalable}. In the PATE framework, a student model learns from various unpublished teacher models each trained on data subsets. The student model cannot access an individual teacher model nor its training data. PATE-GAN consists of $k$ discriminator teachers, $T_{1}, ..., T_{k}$, and a student discriminator $S$ that backpropagates its loss back into the generator. This limits the effect of any individual sample in PATE-GAN's training. In a ($\varepsilon=1, \delta=10^{-5}$)-DP setting, classification models trained on PATE-GAN's synthetic data achieves competitive performances e.g. on a \textit{non-imaging} cervical cancer dataset~\citep{fernandes2017transfer} compared to an upper bound vanilla GAN baseline without DP.

On the same dataset,~\citet{torfi2020differentially} achieve competitive results using a Rényi Differential Privacy and Convolutional Generative Adversarial Networks (RDP-CGAN) under an equally strong ($\varepsilon=1, \delta=10^{-5}$)-DP setting. 

For the generation of biomedical participant data in clinical trials,~\citet{beaulieu2019privacy} apply an AC-GAN under a ($\varepsilon=3.5, \delta=10^{-5}$)-DP setting based on Gaussian noise added to AC-GAN's gradients during training. 

\citet{bae2020anomigan} propose AnomiGAN to anonymise private medical data via some degree of output randomness during inference. This randomness of the generator is achieved by randomly adding, for each layer, one of its separately stored training variances. AnomiGAN achieves competitive results on a \textit{non-imaging} breast cancer dataset and a \textit{non-imaging} prostate cancer for any of the reported privacy parameter values $\varepsilon \in [0.0, 0.5]$ compared to DP, where Laplacian noise is added to samples.

\paragraph{Outlook on Synthetic Cancer Image Privacy}
Despite the above efforts, DP in GANs has only been applied to non-imaging cancer data which indicates research potential for extending these methods reliably to cancer imaging data.
According to~\citet{stadler2021synthetic}, using synthetic data generated under DP can protect outliers in the original data from linkage attacks, but likely also reduces the statistical signal of these original data points, which can result in lower utility of the synthetic data. Apart from this privacy-utility tradeoff, it may not be readily controllable/predictable which original data features are preserved and which omitted in the synthetic datasets~\citep{stadler2021synthetic}.
In fields such as cancer imaging where patient privacy is critical, desirable privacy-utility tradeoffs need to be defined and thoroughly evaluated to enable trust, shareability, and usefulness of synthetic data. Consensus is yet to be found as to how privacy preservation in GAN-generated data can be evaluated and verified in the research community and in clinical practice. Promising approaches include methods that define a privacy gain/loss for synthetic samples~\citep{stadler2021synthetic, yoon2020anonymization}.~\citet{yoon2020anonymization}, for instance, define and backpropagate an identifiability loss to the generator to synthesis anonymised electronic health records (EHRs). The identifiability loss is based on the notion that the minimum weighted euclidean distance between two patient records from two different patients can serve as a desirable anonymisation target for synthetic data.
Designing or extending reliable methods and metrics for standardised quantitative evaluation of patient privacy preservation in synthetic medical images is a line of research that we call attention to.

\subsubsection{Obfuscation of Identifying Patient Features in Images}
If the removal of all sensitive patient information within a cancer imaging dataset allows for sharing such datasets, then GANs can be used to obfuscate such sensitive data. As indicated by Figure \ref{fig:overview}(g), GANs can learn to remove the features from the imaging data that could reveal a patient's identity, e.g. by learning to apply image inpainting to pixel or voxel data of burned in image annotations or of identifying body parts. Such identifying body parts could be the facial features of a patient, as was shown by~\citet{schwarz2019identification} on the example of cranial MRI. Numerous studies exist where GANs accomplish facial feature de-identification on non-medical imaging modalities~\citep{wu2018privacy, hukkelaas2019deepprivacy,li2019anonymousnet,maximov2020ciagan}. For medical imaging modalities, GANs have yet to prove themselves as tool of choice for anatomical and facial feature de-identification against common standards~\citep{segonne2004hybrid,bischoff2007technique,schimke2011preserving, milchenko2013obscuring} with solid baselines. These standards, however, have shown to be susceptible to reconstruction achieved by unpaired image-to-image GANs on MRI volumes with high reversibility for blurred faces and partial reversibility for removed facial features~\citep{abramian2019refacing}.~\citet{goten2021adversarial} provide a first proof-of-concept for GAN-based facial feature de-identification in 3D ($128^3$ voxel) cranial MRI. The generator of their conditional de-identification GAN (C-DeID-GAN) receives brain mask, brain intensities and a convex hull of the brain MRI as input and generates de-identified MRI slices. C-DeID-GAN generates the entire de-identified brain MRI scan and, hence, may not be able to guarantee that the generation process does not alter any of the original brain features. A solution to this can be to only generate and replace the 2D MRI slices or parts thereof that do contain non-pathological facial features while retaining all other original 2D MRI slices. Presuming preservation of medically relevant features and robustness of de-identification, GAN-based approaches can allow for subsequent medical analysis, privacy preserving data sharing and provision of de-identified training data. Hence, we highlight the research potential of GANs for robust medical image de-identification e.g. via image inpainting GANs that have already been successful applied to other tasks in cancer imaging such as synthetic lesion inpainting into mammograms~\citep{wu2018conditional, becker2019injecting} and lung CT scans~\citep{mirsky2019ct}. Also, GAN-based patient feature de-identification methods that are adjustable and trainable to remain quantifiably robust against adversarial image reconstruction are a research line of interest.

\begin{figure*}
    \centering
       		\includegraphics[width=0.85\textwidth]{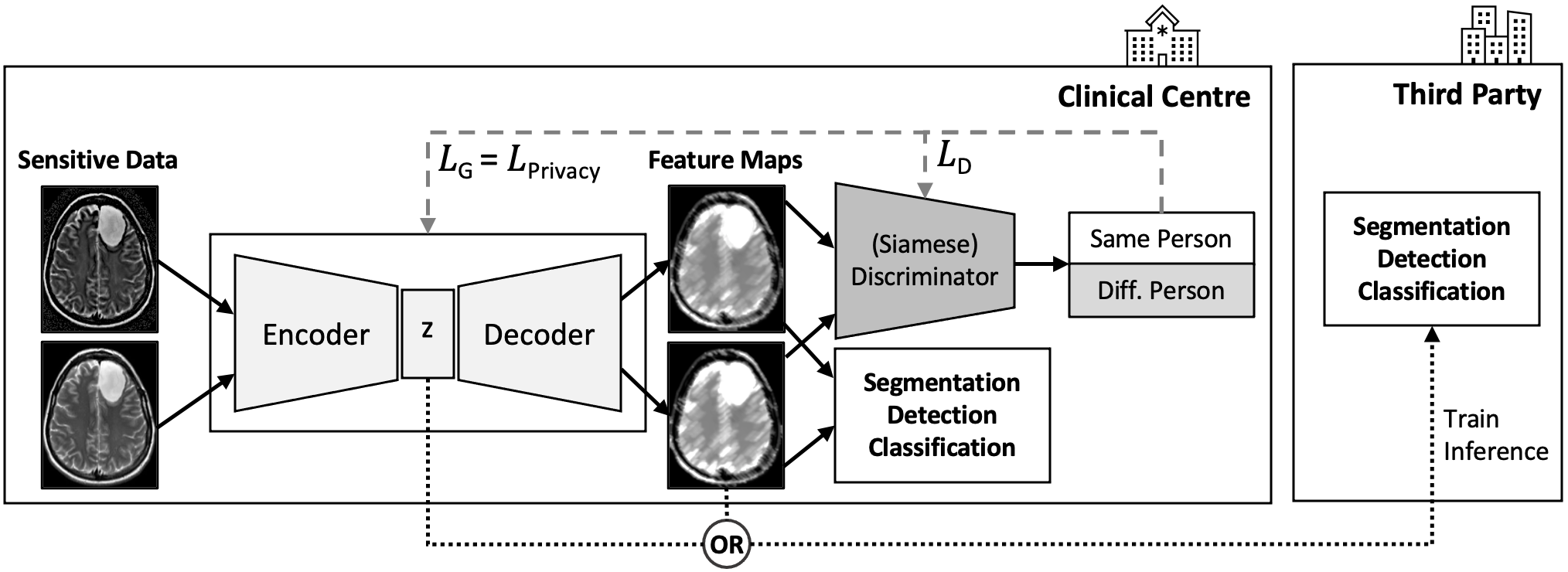}
	\caption[]{Example of an autoencoder architecture trained via adversarial loss to learn privacy-preserving feature maps as in~\citet{kim2019privacy} and/or a privacy-preserving latent representation $z$. Once trained and after thorough periodic manual verification of its ability to preserve privacy, the representation $z$ and/or the feature maps can be sent to third parties outside the clinical centre for model training or inference requests.
	}
  	\label{fig:privacy-preserving-encoder}
\end{figure*}

\subsubsection{Identifying Patient Features in Latent Representations}
In line with Figure \ref{fig:overview}(g), a further example for privacy preserving methods are autoencoders\footnote{For example adversarial autoencoders~\citep{makhzani2015adversarial, creswell2018denoising}, which adversarially learn latent space representations that match a chosen prior distribution.} that learn patient identity-specific features and obfuscate such features when encoding input images into latent space representation. Such an identity-obfuscated representation can be used as input into further models (classification, segmentation, etc) or decoded back into a de-identified image. Adversarial training has been shown to be effective for learning a privacy-preserving encoding function, where a discriminator tries to succeed at classifying the private attribute from the encoded data~\citep{raval2017protecting,wu2018towards,yang2018learning,pittaluga2019learning}. Apart from being trained via the backpropagated adversarial loss, the encoder needs at least one further utility training objective to learn to generate useful representations, such as denoising~\citep{vincent2008extracting} or classification of a second attribute (e.g., facial expressions~\citep{chen2018vgan, oleszkiewicz2018siamese}). Siamese Neural Networks~\citep{bromley1994signature} such as the Siamese Generative Adversarial Privatizer (SGAP)~\citep{oleszkiewicz2018siamese} have been used effectively for adversarial training of an identity-obfuscated representation encoder. In SGAP, two weight-sharing Siamese Discriminators are trained using a distance based loss function to learn to classify whether a pair of images belongs to the same person. As visualised in Figure \ref{fig:privacy-preserving-encoder},~\citet{kim2019privacy} follow a similar approach with the goal of de-identifying and segmenting brain MRI data. Two feature maps are encoded from a pair of MRI scans and fed into a Siamese Discriminator that evaluates via binary classification whether the two feature maps are from the same patient. The generated feature maps are also fed into a segmentation model that backpropagates a Dice loss~\citep{sudre2017generalised} to train the encoder.
Figure \ref{fig:privacy-preserving-encoder} illustrates the scenario where the encoder is deployed in a trusted setting after training, e.g. in a clinical centre, and the segmentation model is deployed in an untrusted setting, e.g. outside the clinical centre at a third party. The encoder shares the identity-obfuscated feature maps with the external segmentation model without the need of transferring the sensitive patient data outside the clinical centre. This motivates further research into adversarial identity-obfuscated encoding methods e.g., to allow sharing and usage of cancer imaging data representations and models across clinical centres.

\subsubsection{Adversarial Attacks Putting Patients at Risk} \label{sec:adversarial_attacks}

\paragraph{Examples of GAN-based Tampering with Cancer Imaging Data}\label{paragraph:tampering}
For instance,~\citet{mirsky2019ct} added and removed evidence of cancer in lung CT scans. Of two identical deep 3D convolutional cGANs (based on pix2pix), one was used to inject (diameter $\geq10mm$) and the other to remove (diameter $<3mm$) multiple solitary pulmonary nodules indicating lung cancer. The GANs were trained on 888 CT scans from the Lung Image Database Consortium image collection (LIDC-IDRI) dataset~\citep{armato2011lung} and inpainted on an extracted region of interest of $32^3$ voxel cuboid shape. The trained GANs can be autonomously executed by malware and are capable of ingesting nodules into standard CT scans that are realistic enough to deceive both radiologists and AI disease detection systems. Three radiologists with 2, 5 and 7 years of experience analysed 70 tampered and 30 authentic CT scans. Spending on average 10 minutes on each scan, the radiologists diagnosed 99\% of the scans with added nodules as malignant and 94\% of the scans with removed nodules as healthy. After disclosing the presence of the attack to the radiologists, the percentages dropped to 60\% and 87\%, respectively~\citep{mirsky2019ct}.

\citet{becker2019injecting} trained a CycleGAN~\citep{zhu2017unpaired} on 680 down-scaled mammograms from the Breast Cancer Digital Repository (BCDR)~\citep{lopez2012bcdr} and the INbreast~\citep{moreira2012inbreast} datasets to generate suspicious features and was able to remove or inject them into existing mammograms. They showed that their approach can fool radiologists at lower pixel dimensions (i.e. 256×256) demonstrating that alterations in patient images by a malicious attacker can remain undetected by clinicians, influence the diagnosis, and potentially harm the patient~\citep{becker2019injecting}.

\paragraph{Defending Adversarial Attacks}
In regard to fooling diagnostic models, one measure to circumvent adversarial attacks is to increase model robustness against adversarial examples~\citep{madry2017towards}, as suggested by Figure \ref{fig:overview}(h). Augmenting the robustness has been shown to be effective for medical imaging segmentation models~\citep{he2019non, park2020robustification}, lung nodule detection models~\citep{liu2020no, paul2020mitigating}, skin cancer recognition~\citep{huq2020analysis, hirano2021universal}, and classification of histopathology images of lymph node sections with metastatic tissue~\citep{wetstein2020adversarial}.~\citet{liu2020no} provide model robustness by adding adversarial chest CT examples to the training data. These adversarial examples are composed of synthetic nodules that are generated by a 3D convolutional variational encoder trained in conjunction with a WGAN-GP~\citep{gulrajani2017improved} discriminator. To further enhance robustness, Projected Gradient Descent (PGD)~\citep{madry2017towards} is applied to find and protect against noise patterns for which the detector network is prone to produce over-confident false predictions~\citep{liu2020no}.

Apart from being the adversary, GANs can also detect adversarial attacks and thus are applicable as security counter-measure enabling attack anticipation, early warning, monitoring and mitigation. Defense-GAN, for example, learns the distribution of non-tampered images and can generate a close output to an inference input image that does not contain adversarial modifications~\citep{samangouei2018defense}. 

We highlight the research potential in adversarial attacks and examples, alongside prospective GAN detection and defence mechanisms that can, as elaborated, highly impact the field of cancer imaging. Apart from the image injection of entire tumours and the generation of adversarial radiomics examples, a further attack vector to consider in future studies is the perturbation of the specific imaging features within an image that are used to compute radiomics features.

\begin{table*}[h]
\centering
\scriptsize
\caption{Overview of adversarially-trained models applied/applicable to \textbf{data access and privacy} cancer imaging challenges. Publications are clustered by section and ordered by year in ascending order.}\label{Table:privacy-table1}
\scalebox{0.81}{
\begin{tabular}{{p{0.17\textwidth}p{0.14\textwidth}p{0.23\textwidth}p{0.15\textwidth}p{0.13\textwidth}p{0.27\textwidth}}}
    \hline
    \textbf{Publication} &
    \textbf{Method} &
    \textbf{Dataset} &
    \textbf{Modality} &
    \textbf{Task} &
    \textbf{Highlights}
    \\
    \hline\hline
    \textbf{Decentralised GANs}
    \\
    \hline
    \citet{chang2020multi,chang2020synthetic} 
    
    \href{https://github.com/tommy-qichang/AsynDGAN}{AsynDGAN}&
    AsynDGAN, PatchGAN~\citep{isola2017image}  & 
    BRATS 2018~\citep{bakas2018identifying}, Multi-Organ~\citep{kumar2017dataset} &
    Cranial MRI, nuclei images & 
    Paired translation & 
    Mask-to-image, central G gets distributed Ds' gradients, synthetic only-trained segmentation.
    \\
    \hline\hline
    \textbf{Differential-Privacy GANs}
    \\
    \hline
    \citet{xie2018differentially} &
    DPGAN &
    MNIST~\citep{lecun1998gradient}, MIMIC-III~\citep{johnson2016mimic} &
    MNIST images, EHRs & 
    \textcolor{mycorrect}{Noise-to-}image synthesis  & 
    Noisy gradients during training ensure DP guarantee.
    \\
    \hline
    \citet{jordon2018pate} &
    PATE-GAN &
    Cervical cancer~\citep{fernandes2017transfer}  &
    [non-imaging] Medical records & 
    Data synthesis &
    DP via PATE framework. G gradient from student D that learns from teacher Ds.
    \\
    \hline
    \citet{beaulieu2019privacy} &
    AC-GAN &
    MIMIC-III~\citep{johnson2016mimic}  &
    [non-imaging] EHRs, clinical trial data & 
    \textcolor{mycorrect}{Class-}conditional synthesis & 
    DP via Gaussian noise added to AC-GAN gradient. Treatment arm (standard/intensive) as condition.
    \\
    \hline
    \citet{bae2020anomigan} &
    AnomiGAN &
    UCI b \& Prostate~\citep{blake1998uci} &
    [non-imaging] Cell nuclei tabular data  & 
    \textcolor{mycorrect}{Multi-class-}conditional synthesis, classification & 
    DP via training variances added to G's layers in inference. Real data row as G's condition.
    \\
    \hline
    \citet{torfi2020differentially} &
    RDP-CGAN &
    Cervical cancer~\citep{fernandes2017transfer}, MIMIC-III~\citep{johnson2016mimic} &
    [non-imaging] Medical records, EHRs & 
    Data synthesis & 
    DP GAN based on Rényi divergence. Allows to track a DP loss.
    \\
    \hline\hline
    \textbf{Patient De-Identification}
    \\
    \hline
    \citet{abramian2019refacing} &
    CycleGAN &
    IXI~\citep{ixidataset} &
    Cranial MRI & 
    Unpaired translation &
    Reconstruction of blurring/removed faces in MRI shows privacy vulnerability.
    \\
    \hline
    \citet{kim2019privacy} &
    PrivacyNet &
    PPMI~\citep{marek2011parkinson} &
    Cranial MRI &
    \textcolor{mycorrect}{Adversarial training}, segmentation & 
    Segmenting de-identified representations learned via same-person CLF by Siamese Ds.
    \\
    \hline
    \citet{goten2021adversarial} &
    C-DeID-GAN &
    ADNI~\citep{wyman2013standardization, weiner2017alzheimer}, OASIS-3~\citep{lamontagne2019oasis} &
    Cranial MRI & 
    Paired translation & 
    Face de-id. Concatenated convex hull, brain mask \& brain volumes as G \& D inputs.
    \\
    \hline\hline
    \textbf{Adversarial Data Tampering}
    \\
    \hline
    \citet{mirsky2019ct} &
    pix2pix-based CT-GAN &
    LIDC-IDRI~\citep{armato2011lung} &
    Lung CT & 
    Image inpainting  & 
    Injected/removed lung nodules in CT fool radiologists and AI models. 
    \\
    \hline
    \citet{becker2019injecting} &
    CycleGAN &
    BCDR~\citep{lopez2012bcdr}, INbreast~\citep{moreira2012inbreast} &
    Digital/Film MMG & 
    Unpaired \textcolor{mycorrect}{image-to-image} translation & 
    Suspicious features can be learned and injected/removed from MMG.
    \\
    \hline
    \citet{liu2020no} &
    Variational Encoder, WGAN-GP &
    LUNA~\citep{setio2017validation}, NLST~\citep{national2011national} &
    Lung CT & 
    \textcolor{mycorrect}{Noise-to-}image synthesis & 
    Robustness via adversarial data augmentation, reduce false positives in nodule detection.
    \\
    \hline
    \end{tabular}
}
\end{table*} 

\subsection{Data Annotation and Segmentation Challenges}\label{sec:annotation}
\subsubsection{Annotation-Specific Issues in Cancer Imaging} \label{sec:annotation_issues}
\paragraph{Missing Annotations in Datasets}
In cancer imaging, not only the availability of large datasets is rare, but also the availability of labels, annotations, and segmentation masks within such datasets. The generation and evaluation of such labels, annotations, and segmentation masks is a task for which trained health professionals (radiologists, pathologists) are needed to ensure validity and credibility~\citep{hosny2018artificial, bi2019artificial}. Nonetheless, radiologist annotations of large datasets can take years to generate~\citep{bi2019artificial}. The tasks of labelling and annotating (e.g., bounding boxes, segmentation masks, textual comments) cancer imaging data is, hence, expensive both in time and cost, especially considering the large amount of data needed to train deep learning models.

\paragraph{Intra/Inter-Observer Annotation Variability} \label{sec:annotation:variability}
This cancer imaging challenge is further exacerbated by the high intra- and inter-observer variability between both pathologists~\citep{gilles2008pathologist, dimitriou2018principled, martin2018interobserver, klaver2020interobserver} and radiologists~\citep{elmore1994variability, hopper1996analysis, hadjiiski2012inter, teh2017inter, wilson2018inter, woo2020intervention, brady2017error} in interpreting cancer images across imaging modalities, affected organs, and cancer types. Automated annotation processes based on deep learning models allow to produce reproducible and standardised results in each image analysis. In one of most common case where the annotations consist of a segmentation mask, reliably segmenting both tumour and non-tumour tissues is crucial for disease analysis, biopsy, and subsequent intervention and treatment~\citep{hosny2018artificial, huynh2020artificial}, the latter being further discussed in Section~\ref{sec:treatment}. For example, automatic tumour segmentation models are useful in the context of radiotherapy treatment planning~\citep{cuocolo2020machine}.

\paragraph{Human Biases in Cancer Image Annotation} \label{sec:annotation:human-bias}
During routine tasks, such as medical image analysis, humans are prone to account for only a few of many relevant qualitative image features. On the contrary, the strength of GANs and deep learning models is the evaluation of large numbers of multi-dimensional image features alongside their (non-linear) inter-relationships and combined importance~\citep{hosny2018artificial}. Deep learning models are likely to react to unexpected and subtle patterns in the imaging data (e.g., anomalies, hidden comorbidities, etc.) that medical practitioners are prone to overlook for instance due to any of multiple existing cognitive biases (e.g., anchoring bias, framing bias, availability bias)~\citep{brady2017error} or inattentional blindness~\citep{drew2013invisible}. Inattentional blindness occurs when radiologists (or pathologist) have a substantial amount of their attention drawn to a specific task, such as finding an expected pattern (e.g., a lung nodule) in the imaging data, that they become blind to other patterns in that data.

\paragraph{Implications of Low Segmentation Model Robustness}
As for the common annotation task of segmentation mask delineation, automated segmentation models can minimise the risk of the aforesaid human biases. However, to date, segmentation models have difficulties when confronted with intricate segmentation problems including domain shifts, rare diseases with limited sample size, or small lesion and metastasis segmentation. In this sense, the performance of many automated and semi-automated clinical segmentation models has been sub-optimal~\citep{sharma2010automated}. This emphasises the need for expensive manual verification of segmentation model results by human experts~\citep{hosny2018artificial}. The challenge of training automated models for difficult segmentation problems can be approached by applying methods for learning discriminative features without explicit labels. Such methods include GANs and variational autoencoders~\citep{kingma2013auto} capable of automating robust segmentation~\citep{hosny2018artificial}.

In addition, segmented regions of interest (ROI) are commonly used to extract quantitative imaging features with diagnostic value such as radiomics features. The latter are used to detect and monitor tumours (e.g., lymphoma~\citep{kang2018diffusion}), biomarkers, and tumour-specific phenotypic attributes~\citep{lambin2012radiomics, parmar2015machine}. The accuracy and success of such commonly applied diagnostic image feature quantification methods, hence, depends on accurate and robust ROI segmentations. Segmentation models need to be able to provide reproducibility of extracted quantitative features and biomarkers~\citep{bi2019artificial} with reliably-low variation, among others, across different scanners, CT slice thicknesses, and reconstruction kernels~\citep{balagurunathan2014test, zhao2016reproducibility}. To this end, we promote lines of research that use adversarial training schemes to target the robustification of segmentation models. Progress in this open research challenge can beneficially unlock trust, usability, and clinical adoption of biomarker quantification methods in clinical practice.

\begin{figure*} [tb]
	\begin{center}
       		\includegraphics[width=0.72\textwidth]{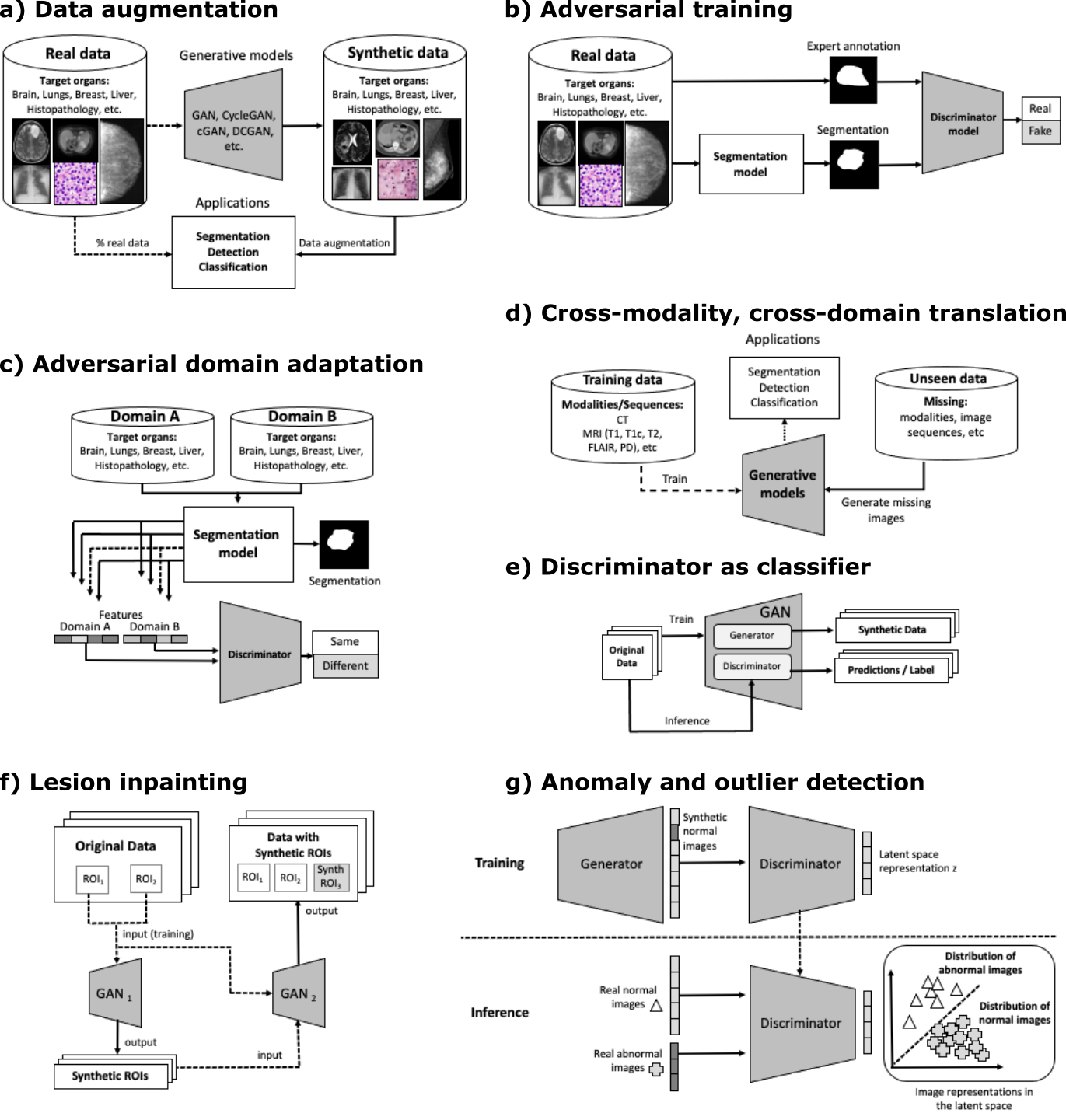}
	\end{center}
	 \caption[]{Overview of cancer imaging GAN applications for detection and segmentation. (a) describes training data augmentation of downstream task models (e.g., segmentation, detection, classification, etc). In (b) a discriminator scrutinises the segmentations created by a segmentation model, while in (c) the discriminator enforces the model to create domain-agnostic latent representations. (d) illustrates domain-adaptation, where the translated target domain images are used for downstream model training. In (e), the AC-GAN~\citep{odena2017conditional} discriminator classifies original data. In (f), one GAN generates ROIs while another inpaints them into full-sized images. (g) uses the discriminator's latent space to find abnormal/outlier representations.}
  	\label{fig:Figure3_GAN_application_pipelines}
\end{figure*}

\subsubsection{\textcolor{mycorrect}{GAN Applications for Cancer Image Segmentation}}\label{sec:annotation:examples}

Table~\ref{Table:segmentation-table} summarises the collection of segmentation publications that utilise such adversarial training approaches and GAN-based data synthesis for cancer imaging. 

\textcolor{mycorrect}{
In Table~\ref{Table:segmentation-table}, we further report the baseline performance alongside the performance increase attributable to applying GANs or adversarial training for each surveyed publication. For the common Dice Score segmentation performance metric, Figure \ref{fig:ScatterSeg1} visualises these differences. Comparing the figure's black identity line and the red trend line over publications, we observe a general improvement of approximately 5 percentage points of adversarial learning methods compared to their baselines. Figure \ref{fig:ScatterSeg2} further displays the variation in performance between baselines and adversarial network methods for the years 2017 to 2021. Based on visual analysis, performance gains seem to be both anatomy-invariant and invariant to the strength of the baseline, where similar gains are achieved for initially low (e.g., $<0.7$) and high (e.g., $>=0.7$) baseline Dice scores.
While Figure \ref{fig:ScatterSeg1} and Figure \ref{fig:ScatterSeg2} offer interesting quantitative insights, we recommend taking potential publication bias\footnote{Publication bias likely influences the trends observable in Figures \ref{fig:ScatterSeg1} and \ref{fig:ScatterSeg2}: Only papers that show an improvement attributable to adversarial networks were published and therefore only such studies could be included.} into account when drawing conclusions from these figures. Trends in the presented data in these plots can be analysed holistically, however, they are not intended to benchmark and compare individual publications against each other. This is due to multiple limiting factors of such comparisons including the differences in (a) the used baselines, (b) organs, (c) modalities, (d) the segmentation task and its associated difficulty, (e) the amount of training and testing data, (f) data and annotation quality, (g) pre- and post-processing methods, or (h) the study's objectives. In regard to (g), some studies may focus on other benefits of adversarial learning methods instead of or apart from Dice Score improvement, such as, reducing the needed training dataset size, domain adaptation in general, protecting patient privacy with synthetic data, 
or simply improving other metrics (e.g. Hausdorff distance, FID).}

In the following sections, we provide a summary of the commonly used techniques and trends in the GAN literature that address the challenges in cancer image segmentation.

\begin{figure*}[tb]
\begin{tikzpicture}[scale=1.0,transform shape]
\centering
  \pgfplotsset{
      scale only axis,
  }
  
\begin{axis}[enlargelimits=0.05, 
        xlabel=Dice Score: After Introducing Adversarial Network,
        ylabel=Dice Score: Baseline,
        grid=both,
        scale only axis=true,
        width=0.875\textwidth, 
        height=5.5cm,
        xtick={0.35,0.40,...,1.0},
        ytick={0.35,0.40,...,1.0},
        xmin=0.35,
        xmax=1,
        ymin=0.35,
        ymax=1,
        legend style={at={(0.01, 0.975)}, anchor=north west},
        legend cell align={left},
      ]
    \addplot[
        scatter/classes={a={mark=pentagon*, mark options={line width=0.35pt}, draw=black, fill=blue}, b={mark=diamond*, mark options={line width=0.35pt}, draw=black, fill=red}, c={mark=*, mark options={line width=0.35pt}, draw=black, fill=yellow}, d={mark=square*, mark options={line width=0.35pt}, draw=black, fill=orange}, e={mark=triangle*, mark options={line width=0.35pt}, draw=black, fill=green}, f={mark=otimes*, mark options={line width=0.35pt}, draw=black, fill=pink}, g={mark=halfcircle*, mark options={line width=0.35pt}, draw=black, fill=black}, h={mark=oplus*, mark options={line width=0.35pt}, draw=black, fill=gray}},
        scatter, mark=*, only marks, 
        scatter src=explicit symbolic,
        nodes near coords*={\Label},
        visualization depends on={value \thisrow{label} \as \Label}
    ] table [meta=class] {
        x y class label hidden
            0.63 0.60 a \footnotesize{\citeauthor{kamnitsas2017unsupervised}} \phantom{\tiny{a}}
        0.84 0.79 a \phantom{\tiny{a}} \small{\citeauthor{mok2018learning}} 
            0.68 0.67 a \small{\citeauthor{yu20183d}} \phantom{\tiny{a}}
        0.81 0.81 a \phantom{\tiny{a}} \small{\citeauthor{shin2018medical}}
        0.59 0.57 a \phantom{\tiny{a}} \small{\citeauthor{kim2020synthesis}}
        0.71 0.69 a \phantom{\tiny{a}} \small{\citeauthor{hu2020coarse}}
        0.93 0.87 a \phantom{\tiny{a}} \small{\citeauthor{cirillo2020vox2vox}} 
            0.67 0.48 a \small{\citeauthor{han2021deep}} \phantom{\tiny{a}}
        0.85 0.80 a \phantom{\tiny{a}} \small{\citeauthor{xue2018segan}} 
            0.94 0.86 b \small{\citeauthor{singh2018conditional}} \phantom{\tiny{a}}
        0.93 0.70 b \phantom{\tiny{a}} \footnotesize{\citeauthor{caballo2020deep}} 
        0.88 0.85 b \phantom{\tiny{a}} \small{\citeauthor{negi2020rda}} 
            0.94 0.93 c \small{\citeauthor{huo2018splenomegaly}} \phantom{\tiny{a}}
        0.68 0.62 c \phantom{\tiny{a}} \small{\citeauthor{chen2019liver}} 
        0.93 0.92 c \phantom{\tiny{a}} \small{\citeauthor{sandfort2019data}} 
            0.92 0.72 c \scriptsize{\citeauthor{xiao2019radiomics}} \phantom{\tiny{a}}
        0.61 0.58 c \phantom{\tiny{a}} \small{\citeauthor{oliveira2020implanting}} 
            0.80 0.66 d \small{\citeauthor{jiang2018tumor}} \phantom{\tiny{a}}
        0.99 0.96 d \phantom{\tiny{a}} \small{\citeauthor{jin2018ct}} 
        0.87 0.84 d \phantom{\tiny{a}} \small{\citeauthor{dong2019automatic}} 
        0.98 0.97 d \small{\citeauthor{tang2019xlsor}} \phantom{\tiny{a}} 
        0.85 0.82 d \phantom{\tiny{a}} \small{\citeauthor{shi2020automatic}} 
        0.97 0.96 d \phantom{\tiny{a}} \small{\citeauthor{munawar2020segmentation}} 
            0.41 0.35 e \small{\citeauthor{kohl2017adversarial}} \phantom{\tiny{a}}
        0.75 0.69 e \phantom{\tiny{a}} \small{\citeauthor{grall2019using}}
        0.90 0.88 e \phantom{\tiny{a}} \small{\citeauthor{nie2020adversarial}}
        0.90 0.85 e \phantom{\tiny{a}} \small{\citeauthor{zhang2020semi}} 
            0.76 0.73 e \footnotesize{\citeauthor{cem2020investigating}} \phantom{\tiny{a}}
        0.92 0.87 f \phantom{\tiny{a}} \small{\citeauthor{liu2019accurate}} 
                0.73 0.66 f \phantom{\tiny{a}} \small{\citeauthor{xie2020mi}}
        0.83 0.79 g \phantom{\tiny{a}} \small{\citeauthor{pandey2020image}} 
            0.84 0.85 h \small{\citeauthor{chi2018controlled}} \phantom{\tiny{a}}
        0.81 0.77 h \phantom{\tiny{a}} \small{\citeauthor{abhishek2019mask2lesion}} 
            0.91 0.76 h \footnotesize{\citeauthor{sarker2019mobilegan}} \phantom{\tiny{a}}
            0.53 0.40 h \small{\citeauthor{chaitanya2021semi}} \phantom{\tiny{a}}
    };
    \addplot[black,samples at={0,1}] {x};
    \addplot [thick, smooth, red, dashed] table[y={create col/linear regression={y=Y}}]{
        X Y class label hidden
            0.63 0.60 a \footnotesize{\citeauthor{kamnitsas2017unsupervised}} \phantom{\tiny{a}}
        0.84 0.79 a \phantom{\tiny{a}} \small{\citeauthor{mok2018learning}} 
            0.68 0.67 a \small{\citeauthor{yu20183d}} \phantom{\tiny{a}}
        0.81 0.81 a \phantom{\tiny{a}} \small{\citeauthor{shin2018medical}}
        0.59 0.57 a \phantom{\tiny{a}} \small{\citeauthor{kim2020synthesis}}
        0.71 0.69 a \phantom{\tiny{a}} \small{\citeauthor{hu2020coarse}}
        0.93 0.87 a \phantom{\tiny{a}} \small{\citeauthor{cirillo2020vox2vox}} 
            0.67 0.48 a \small{\citeauthor{han2021deep}} \phantom{\tiny{a}}
        0.85 0.80 a \phantom{\tiny{a}} \small{\citeauthor{xue2018segan}} 
            0.94 0.86 b \small{\citeauthor{singh2018conditional}} \phantom{\tiny{a}}
        0.93 0.70 b \phantom{\tiny{a}} \footnotesize{\citeauthor{caballo2020deep}} 
        0.88 0.85 b \phantom{\tiny{a}} \small{\citeauthor{negi2020rda}} 
            0.94 0.93 c \small{\citeauthor{huo2018splenomegaly}} \phantom{\tiny{a}}
        0.68 0.62 c \phantom{\tiny{a}} \small{\citeauthor{chen2019liver}} 
        0.93 0.92 c \phantom{\tiny{a}} \small{\citeauthor{sandfort2019data}} 
            0.92 0.72 c \scriptsize{\citeauthor{xiao2019radiomics}} \phantom{\tiny{a}}
        0.61 0.58 c \phantom{\tiny{a}} \small{\citeauthor{oliveira2020implanting}} 
            0.80 0.66 d \small{\citeauthor{jiang2018tumor}} \phantom{\tiny{a}}
        0.99 0.96 d \phantom{\tiny{a}} \small{\citeauthor{jin2018ct}} 
        0.87 0.84 d \phantom{\tiny{a}} \small{\citeauthor{dong2019automatic}} 
        0.98 0.97 d \small{\citeauthor{tang2019xlsor}} \phantom{\tiny{a}} 
        0.85 0.82 d \phantom{\tiny{a}} \small{\citeauthor{shi2020automatic}} 
        0.97 0.96 d \phantom{\tiny{a}} \small{\citeauthor{munawar2020segmentation}} 
            0.41 0.35 e \small{\citeauthor{kohl2017adversarial}} \phantom{\tiny{a}}
        0.75 0.69 e \phantom{\tiny{a}} \small{\citeauthor{grall2019using}}
        0.90 0.88 e \phantom{\tiny{a}} \small{\citeauthor{nie2020adversarial}}
        0.90 0.85 e \phantom{\tiny{a}} \small{\citeauthor{zhang2020semi}} 
            0.76 0.73 e \footnotesize{\citeauthor{cem2020investigating}} \phantom{\tiny{a}}
        0.92 0.87 f \phantom{\tiny{a}} \small{\citeauthor{liu2019accurate}} 
                0.73 0.66 f \phantom{\tiny{a}} \small{\citeauthor{xie2020mi}}
        0.83 0.79 g \phantom{\tiny{a}} \small{\citeauthor{pandey2020image}} 
            0.84 0.85 h \small{\citeauthor{chi2018controlled}} \phantom{\tiny{a}}
        0.81 0.77 h \phantom{\tiny{a}} \small{\citeauthor{abhishek2019mask2lesion}} 
            0.91 0.76 h \footnotesize{\citeauthor{sarker2019mobilegan}} \phantom{\tiny{a}}
            0.53 0.40 h \small{\citeauthor{chaitanya2021semi}} \phantom{\tiny{a}}
    };

\legend{\footnotesize{Head/Neck},\footnotesize{Breast},\footnotesize{Abdomen}, \footnotesize{Chest}, \footnotesize{Prostate}, \footnotesize{Colorectal}, \footnotesize{Pathology}, \footnotesize{Other}}
\end{axis}
\end{tikzpicture}
    \caption[]{\textcolor{mycorrect}{Scatter plot illustrating the segmentation performance improvement attributable to adversarial networks for the surveyed publications. 
    Each publication is represented by a \textcolor{mycorrect2}{marker} with a colour \textcolor{mycorrect2}{and shape} encoding depicting the publication's anatomical category.
    Only the publications are included that measure performance via Dice Score and compare against a baseline, as reported in Table \ref{Table:segmentation-table}. For publications reporting multiple Dice Scores, their mean was computed and included herein. The black identity line indicates no change between baseline and adversarial network intervention, while dots below this line represent an improvement. The red regression line depicts the trend of improvement across publications. The author names of a few publications have been manually selected for highlighting based on the distance to the trend line.}} 
  	\label{fig:ScatterSeg1}
\end{figure*}
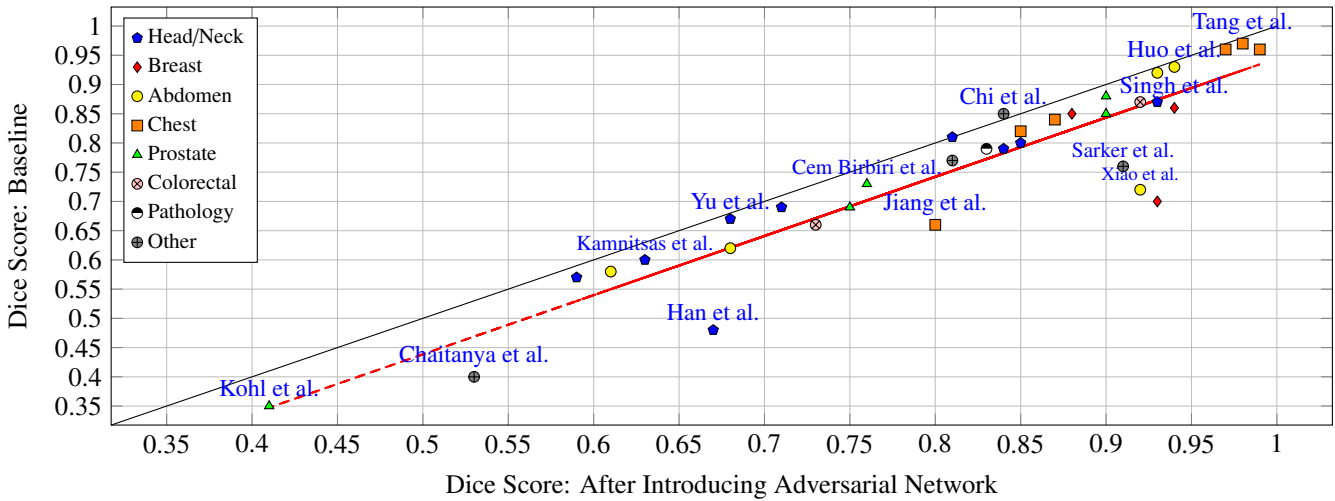

\begin{figure*}[tb]
\begin{tikzpicture}[scale=1.0,transform shape]
\centering
  \pgfplotsset{
      scale only axis,
  }
 
\begin{axis}[enlargelimits=0.025,  
        ylabel=Year of Publication,
        xlabel=Baseline+$\Delta$: Dice Score Change After Introducing Adversarial Network (in \textit{\%}), 
        grid=both,
        scale only axis=true,
        width= 0.8\textwidth, 
        height=3.5cm,
        ytick={2017,2018,...,2021},
        xtick={-1,0,...,24},
        xmin=-1,
        xmax=24,
        ymin=2017,
        ymax=2021,
        legend style={at={(1.0075,0.95)},anchor=north west},
        legend cell align={left},
        /pgf/number format/.cd,
        use comma,
        1000 sep={}]
      ]
    \addplot[
        scatter/classes={a={mark=pentagon*, mark options={line width=0.35pt}, draw=black, fill=blue}, b={mark=diamond*, mark options={line width=0.35pt}, draw=black, fill=red}, c={mark=*, mark options={line width=0.35pt}, draw=black, fill=yellow}, d={mark=square*, mark options={line width=0.35pt}, draw=black, fill=orange}, e={mark=triangle*, mark options={line width=0.35pt}, draw=black, fill=green}, f={mark=otimes*, mark options={line width=0.35pt}, draw=black, fill=pink}, g={mark=halfcircle*, mark options={line width=0.35pt}, draw=black, fill=black}, h={mark=oplus*, mark options={line width=0.35pt}, draw=black, fill=gray}},
        scatter, mark=*, only marks, 
        scatter src=explicit symbolic,
        nodes near coords*={\Label},
        visualization depends on={value \thisrow{label} \as \Label}
    ] 
        table [meta=class] {
        x y class label hidden
            3 2017 a \scriptsize{\citeauthor{kamnitsas2017unsupervised}} \phantom{\tiny{a}}
            5 2018 a \tiny{\citeauthor{mok2018learning}} \phantom{\tiny{a}}
        1 2018 a \phantom{\tiny{a}} \small{\citeauthor{yu20183d}}
            0 2018 a \small{\citeauthor{shin2018medical}} \phantom{\tiny{a}}
        2 2020 a \phantom{\tiny{a}} \small{\citeauthor{kim2020synthesis}}
        2 2020 a \phantom{\tiny{a}} \small{\citeauthor{hu2020coarse}}
        5 2020 a \phantom{\tiny{a}} \small{\citeauthor{cirillo2020vox2vox}}
            19 2021 a \small{\citeauthor{han2021deep}} \phantom{\tiny{a}}
        5 2018 a \phantom{\tiny{a}} \small{\citeauthor{xue2018segan}}
            8 2018 b \small{\citeauthor{singh2018conditional}} \phantom{\tiny{a}}
            23 2020 b \small{\citeauthor{caballo2020deep}} \phantom{\tiny{a}}
        3 2020 b \phantom{\tiny{a}} \small{\citeauthor{negi2020rda}}
        1 2018 c \phantom{\tiny{a}} \small{\citeauthor{huo2018splenomegaly}}
        6 2019 c \phantom{\tiny{a}} \small{\citeauthor{chen2019liver}}
        1 2019 c \phantom{\tiny{a}} \small{\citeauthor{sandfort2019data}}
            20 2019 c \small{\citeauthor{xiao2019radiomics}} \phantom{\tiny{a}}
        3 2020 c \phantom{\tiny{a}} \small{\citeauthor{oliveira2020implanting}}
            14 2018 d \small{\citeauthor{jiang2018tumor}} \phantom{\tiny{a}}
            3 2018 d \scriptsize{\citeauthor{jin2018ct}} \phantom{\tiny{a}}
            3 2019 d \small{\citeauthor{dong2019automatic}} \phantom{\tiny{a}}
        1 2019 d \phantom{\tiny{a}} \small{\citeauthor{tang2019xlsor}}
        3 2020 d \phantom{\tiny{a}} \small{\citeauthor{shi2020automatic}} 
            1 2020 d  \scriptsize{\citeauthor{munawar2020segmentation}} \phantom{\tiny{a}}
            6 2017 e \small{\citeauthor{kohl2017adversarial}} \phantom{\tiny{a}}
            6 2019 e \small{\citeauthor{grall2019using}} \phantom{\tiny{a}}
        2 2020 e \phantom{\tiny{a}} \small{\citeauthor{nie2020adversarial}}
        5 2020 e \phantom{\tiny{a}} \footnotesize{\citeauthor{zhang2020semi}} 
        3 2020 e  \phantom{\tiny{a}} \small{\citeauthor{cem2020investigating}}
        5 2019 f \phantom{\tiny{a}} \small{\citeauthor{liu2019accurate}}
            7 2020 f \small{\citeauthor{xie2020mi}} \phantom{\tiny{a}}
            4 2020 g \small{\citeauthor{pandey2020image}} \phantom{\tiny{a}}
        -1 2018 h \phantom{\tiny{a}} \small{\citeauthor{chi2018controlled}}
        4 2019 h \phantom{\tiny{a}} \small{\citeauthor{abhishek2019mask2lesion}}
            15 2019 h \small{\citeauthor{sarker2019mobilegan}} \phantom{\tiny{a}}
            13 2021 h \small{\citeauthor{chaitanya2021semi}} \phantom{\tiny{a}}
    };

\legend{\footnotesize{Head/Neck},\footnotesize{Breast},\footnotesize{Abdomen}, \footnotesize{Chest}, \footnotesize{Prostate}, \footnotesize{Colorectal}, \footnotesize{Pathology}, \footnotesize{Other}}
\end{axis}
\end{tikzpicture}
     \caption[]{\textcolor{mycorrect}{Scatter plot displaying year of publication and Dice Score improvement (in \%). Each  \textcolor{mycorrect2}{marker} represents a publication and its colour \textcolor{mycorrect2}{and shape} encoding represents its corresponding anatomical category.
     Only the publications are included that report Dice Score alongside a baseline comparison. For publications reporting multiple Dice Scores, their mean was computed and included herein. 
     Author names have been manually selected at random for highlighting.}} 
  	\label{fig:ScatterSeg2}
\end{figure*}
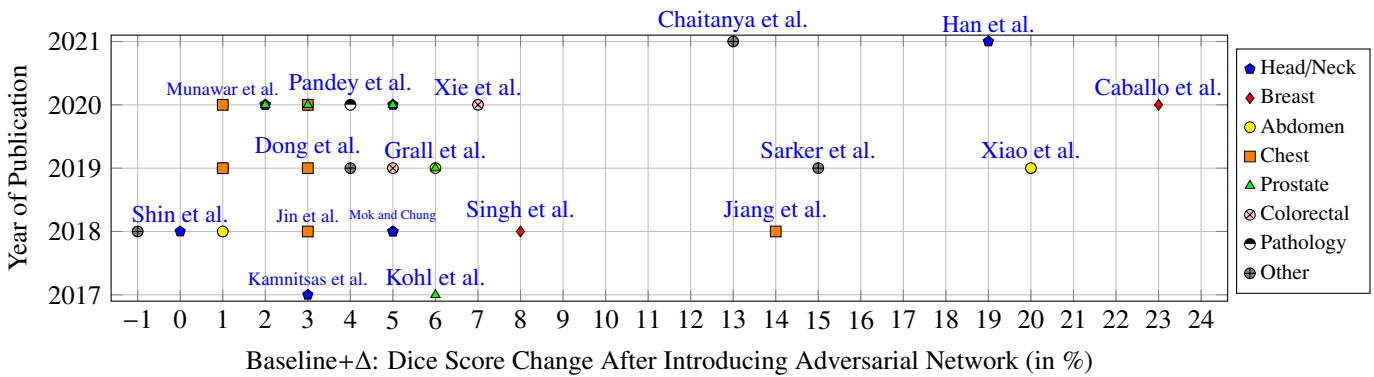

\paragraph{Robust quantitative imaging feature extraction} \label{sec:feature_extraction}
For example,~\citet{xiao2019radiomics} addressed the challenge of robustification of segmentation models and reliable biomarker quantification.~\citet{xiao2019radiomics} provide radiomics features as conditional input to the discriminator of their adversarially trained liver tumour segmentation model. Their learning procedure strives to inform the generator to create segmentations that are specifically suitable for subsequent radiomics feature computation. Apart from adversarially training segmentation models, we also highlight the research potential of adversarially training quantitative imaging feature extraction models (e.g., deep learning radiomics) for reliable application in multi-centre and multi-domain settings.


\paragraph{Synthetic Segmentation Model Training Data}
By augmenting and varying the training data of segmentation models, it is possible to substantially decrease the amount of manually annotated images during training while maintaining the performance~\citep{foroozandeh2020synthesizing}. A general pipeline of such usage of GAN based generative models is demonstrated in Figure~\ref{fig:Figure3_GAN_application_pipelines}(a) and mentioned in Figure \ref{fig:overview}(j).

Over the past few years, CycleGAN~\citep{zhu2017unpaired} based approaches have been widely used for synthetic data generation due to the possibility of using unpaired image sets in training, as compared to paired image translation methods like pix2pix~\citep{isola2017image} or SPADE~\citep{park2019semantic}. CycleGAN based data augmentation has been shown to be useful for segmentation model training, in particular, for generating images with different acquisition characteristics such as contrast enhanced MRI from non-contrast MRI~\citep{wang2021contrast}, cross-modality image translation between different modalities such as CT and MRI images~\citep{huo2018splenomegaly}, and domain adaptation tasks~\citep{jiang2018tumor}. The popularity of the CycleGAN based methods lies not only in image synthesis or domain adaptation, but also in the inclusion of simultaneous image segmentation in its pipeline~\citep{lee2020study}.

Although pix2pix methods require paired samples, it is also a widely used type of GAN in data augmentation for medical image segmentation (see Table~\ref{Table:segmentation-table}). Several works on segmentation have demonstrated its effectiveness in generating synthetic medical images. By manipulating its input, the variability of the training dataset for image segmentation could be remarkably increased in a controlled manner~\citep{abhishek2019mask2lesion, oliveira2020implanting}. Similarly, the conditional GAN methods have also been used for controllable data augmentation for improving lesion segmentation~\citep{oliveira2020controllable}. Providing a condition as an input to generate a mask is particularly useful to specify the location, size, shape, and heterogeneity of the synthetic lesions. One of the recent examples, proposed by~\citet{kim2020synthesis}, demonstrates this in brain MRI tumour synthesis by conditioning an input with simplified controllable concentric circles to specify lesion location and characteristics. A further method for data augmentation is the inpainting of generated lesions into healthy real images or into other synthetic images, as depicted by Figure~\ref{fig:Figure3_GAN_application_pipelines}(f).
Overall, the described data augmentation techniques have shown to improve generalisability and performance of segmentation models by increasing both the number and the variability of training samples~\citep{qasim2020red, foroozandeh2020synthesizing, lee2020study}.

\paragraph{Segmentation Models with Integrated Adversarial Loss} \label{sec:annotation_mask_discriminator}
As stated in Figure \ref{fig:overview}(i), GANs can also be used as the algorithm that generates robust segmentation masks, where the generator is used as a segmenter and the discriminator scrutinises the segmentation masks given an input image. One intuition behind this approach is the detection and correction of higher-order inconsistencies between the ground truth segmentation maps and the ones created by the segmenter via adversarial learning~\citep{luc2016semantic, hu2020coarse, cirillo2020vox2vox}. This approach is demonstrated in Figure~\ref{fig:Figure3_GAN_application_pipelines}(b). With the additional adversarial loss when training a segmentation model, this approach has been shown to improve semantic segmentation accuracy~\citep{hung2019adversarial, sarker2019mobilegan, shi2020automatic}.
Using adversarial training, similarity of a generated mask to manual segmentation given an input image is taken under consideration by the discriminator allowing a global assessment of the segmentation quality. \textcolor{mycorrect}{This approach further offers a practical solution towards handling intra- and inter-observer annotation variability, as the mask discriminator learns an average over observers, which is backpropagated to the segmenter via adversarial loss.}

A unique way of incorporating the adversarial loss from the discriminator has been recently proposed in \citet{nie2020adversarial}. In their work, the authors utilise a fully-convolutional network as a discriminator, unlike its counterparts that use binary, single neuron output networks. In doing so, a dense confidence map is produced by the discriminator, which is further used to train the segmenter with an attention mechanism.

Overall, using an adversarial loss as an additional global segmentation assessment is likely to be a helpful further signal for segmentation models, in particular, for heterogeneously structured datasets of limited size~\citep{kohl2017adversarial}, as is common for cancer imaging datasets. We highlight potential further research in GAN-based segmentation models to learn to segment increasingly fine radiologic distinctions. These models can help to solve further cancer imaging challenges, for example, accurate differentiation between neoplasms and tissue response to injury in the regions surrounding a tumour after treatment~\citep{bi2019artificial}.

\paragraph{Segmentation Models with Integrated Adversarial Domain Discrimination} \label{sec:annotation_domain_discriminator}
Moreover, a similar adversarial loss can also be performed internally on the segmentation model features as illustrated in Figure~\ref{fig:Figure3_GAN_application_pipelines}(c). Such an approach can benefit  \textcolor{mycorrect}{unsupervised} domain adaptation and domain generalisation by enforcing the segmentation model to learn to base its prediction on domain-invariant feature representations~\citep{kamnitsas2017unsupervised}.







\begin{table*}[htbp] 
\centering
\scriptsize
\caption{Overview of adversarial training and GAN-based approaches applied to \textbf{segmentation} in cancer imaging tasks. Publications are clustered by organ type and ordered by year in ascending order.\textcolor{mycorrect}{ `*' indicates that the metrics are only available in figures and the baseline numbers are lower than using GANs in the corresponding paper.} \textcolor{mycorrect}{`n.a.' indicates that there was no comparison with a specific baseline with the reason for this being indicated in the `Highlights' column.}} \label{Table:segmentation-table}
\scalebox{0.65}{
\begin{tabular}{{p{0.18\textwidth}p{0.16\textwidth}p{0.24\textwidth}p{0.10\textwidth}p{0.12\textwidth}p{0.13\textwidth}p{0.07\textwidth}p{0.37\textwidth}}}
\hline
    \textbf{Publication} &
    \textbf{Method} &
    \textbf{Dataset} &
    \textbf{Modality} &
    \textbf{Task} &
    \textbf{Metric w/o GAN (Baseline)} &
    \textbf{Metric with GAN (Baseline+$\Delta$)} &
    \textbf{Highlights} \\
    \hline
    \hline
    \textbf{Head/Brain/Neck}
    \\
    \hline
    \textcolor{mycorrect}{\citet{kamnitsas2017unsupervised}} 
    
    \href{https://github.com/deepmedic/deepmedic}{deepmedic}& 
    \textcolor{mycorrect}{deepmedic} & 
    \textcolor{mycorrect}{Private} & 
    \textcolor{mycorrect}{MRI} & 
    \textcolor{mycorrect}{Adversarial training} & 
    \textcolor{mycorrect}{Dice: 0.60 Recall: 0.56 Precision: 0.70} &
    
    \textcolor{mycorrect}{0.63, 0.59, 0.72} &
    \textcolor{mycorrect}{Multi-domain (MPRAGE, FLAIR, T2, Proton Density, and Gradient-Echo) segmentation with domain discriminator loss.}
    \\
    \hline
    \citet{rezaei2017conditional} & 
    pix2pix, MarkovianGAN~\citep{li2016precomputed} & 
    BRATS 2017~\citep{menze2014multimodal, bakas2018identifying, bakas2017advancing} & 
    MRI & 
    Adversarial training & 
    Dice: n.a. &
    0.80 & 
    D detects G-generated masks for high/low grade glioma segmentation. 
    \textcolor{mycorrect}{Benchmarked in a challenge, thus missing metric w/o GAN.}
    \\
    \hline
    \citet{mok2018learning} & 
    CB-GAN &  
    BRATS \citep{menze2014multimodal} & 
    MRI & 
    Data augmentation & 
    Dice: 0.79 & 
    0.84 &
    Coarse-to-fine G captures training set manifold, generates generic samples in HGG \& LGG segmentation.
    \\
    \hline
    \citet{yu20183d}& 
    pix2pix-based & 
    BRATS \citep{menze2014multimodal} & 
    MRI & 
    Data augmentation & 
    Dice: 0.67 & 
    0.68 &
    Cross-modal paired FLAIR to T1 translation, training tumour segmentation with T1+real/synthetic FLAIR. Baseline: only-T1 training
    \\
    \hline
    \citet{shin2018medical} & 
    pix2pix & 
    BRATS \citep{menze2014multimodal} & 
    MRI & 
    Data augmentation & 
    Dice: 0.81 & 
    0.81 &
    Training on synthetic, before fine-tuning on 10\% of the real data.
    \\
    \hline
    \citet{xue2018segan}&
    SegAN & 
    BRATS \citep{menze2014multimodal} & 
    MRI &
    Adversarial training &
    Dice: \textcolor{mycorrect}{0.80} & 
    0.85 &
    Paired image-to-mask. New multi-scale loss: L1 of D representations between GT- \& prediction-masked input MRI. \textcolor{mycorrect}{U-Net baseline.}
    \\
    \hline
    \citet{giacomello2020brain} &
    SegAN-CAT & 
    BRATS \citep{menze2014multimodal} & 
    MRI &
    Adversarial training &
    Dice: \textcolor{mycorrect}{0.71} & 
    0.86 &
    Paired image-to-mask. Combined dice loss \& multi-scale loss using concatenation on channel axis instead of masking. \textcolor{mycorrect}{SegAN baseline}.
    \\
    \hline
    \citet{kim2020synthesis}
    
    \href{https://github.com/KSH0660/BrainTumor}{BrainTumor} & 
    GAN & 
    BRATS \citep{bakas2018identifying, menze2014multimodal} & 
    MRI & 
    Image inpainting, data augmentation & 
    Dice: 0.57 & 
    0.59 &
    Simplifying tumour features into concentric circle \& grade mask to inpaint.
    \\
    \hline
    \citet{hu2020coarse} & 
    UNet-based GAN segmenter & 
    Private & 
    CT, PET & 
    Adversarial training & 
    Dice: 0.69 & 
    0.71 &
    Spatial context information \& hierarchical features. Nasal-type lymphoma segmentation with uncertainty estimation.
    \\
    \hline
    \citet{qasim2020red} & 
    SPADE-based & 
    BRATS \citep{bakas2018identifying}, ISIC \citep{codella2019skin} &
    MRI, dermoscopy & 
    Cross-domain translation & 
    Dice: \textcolor{mycorrect}{*} & 
    B:0.66 S:0.62 &
    Brain and skin segmentation. Frozen segmenter as 3rd player in GAN to condition on local apart from global information. 
    \\
    \hline
    \citet{foroozandeh2020synthesizing} & 
    PGGAN, SPADE & 
    BRATS \citep{menze2014multimodal} & 
    MRI & 
    Data augmentation &
    Av. dice error(\%): 16.80 & 
    16.18 &
    Sequential noise-to-mask  and paired mask-to-image translation to synthesise labelled tumour images.
    \\
    \hline
    \citet{cirillo2020vox2vox} & 
    vox2vox: 3D pix2pix & 
    BRATS \citep{menze2014multimodal} & 
    MRI &
    Adversarial training & 
    Dice: 0.87 & 
    0.93 &
    3D adversarial training to enforce segmentation results to look realistic.
    \\
    \hline
    \citet{han2021deep} &
    Symmetric adaptation network & 
    BRATS \citep{menze2014multimodal} & 
    MRI &
    Cross-domain translation &
    Dice: 0.48 & 
    0.67 &
    Simultaneous source/target (T2 to other sequences) translation and segmentation. Compared to CycleGAN baseline.
    \\
    \hline
    \citet{alshehhi2021quantification} &
    U-Net based GAN segmenter & 
    BRATS \citep{menze2014multimodal} & 
    MRI &
    Adversarial training &
    Dice: n.a. & 
    n.a. &
    \textcolor{mycorrect}{Quantitative comparison of 7 active learning acquisition functions using existing adversarial networks (including SegAN, SegAN-CAT)}
    \\
    \hline
    \hline
    \textbf{Breast}
    \\
    \hline
    \citet{singh2018conditional} & 
    pix2pix & 
    DDSM \citep{heath2001digital} & 
    Film MMG & 
    Adversarial training & 
    Dice: 0.86 &
    0.94 & 
    Adversarial loss to make automated segmentation close to manual masks for breast mass segmentation.
    \\
    \hline
    \citet{caballo2020deep} & 
    DCGAN \citep{radford2015unsupervised} & 
    Private & 
    CT & 
    Data augmentation & 
    Dice: 0.70 & 
    0.93 &
    GAN based data augmentation; Validated by matching extracted radiomics features. \\
    \hline
    \citet{negi2020rda}&
    GAN, WGAN-RDA-UNET & 
    Private & 
    Ultrasound & 
    Adversarial training & 
    Dice: 0.85 & 
    0.88  &
    Outperforms state-of-the-art using Residual-Dilated-Attention-Gate-UNet and WGAN for lesion segmentation.
    \\
    \hline
    \hline
    \textbf{Abdominal}
    \\
    \hline
    \citet{huo2018splenomegaly} & 
    Conditional PatchGAN & 
    Private & 
    MRI & 
    Adversarial training & 
    Dice: \textcolor{mycorrect}{0.93} & 
    0.94 &
    Adversarial loss as segmentation post-processing for spleen segmentation. \textcolor{mycorrect}{ResNet baseline.}
    \\
    \hline
    \citet{chen2019liver}& 
    DC-FCN-based GAN segmenter  & 
    LiTS \citep{bilic2019liver} & 
    CT & 
    Adversarial training & 
    Dice: \textcolor{mycorrect}{0.62} & 
    0.68 &
    Cascaded framework with densely connected adversarial training. \textcolor{mycorrect}{DC-Fully connected network baseline.} 
    \\
    \hline
    \citet{sandfort2019data} & 
    CycleGAN &
    NIH \citep{prior2017public}, Decathlon \citep{simpson2019large}, DeepLesion \citep{yan2018deeplesion} & 
    CT & 
    Cross-domain translation & 
    Dice (od): 0.92, 0.10 & 
    0.93, 0.75 &
    Contrast enhanced to non-enhanced translation to improve out-of-distribution (od) segmentation.
    \\
    \hline
    \citet{xiao2019radiomics} & 
    Radiomics-guided GAN & 
    Private & 
    MRI &
    Adversarial training & 
    Dice: \textcolor{mycorrect}{0.72} & 
    0.92 &
    Radiomics-guided adversarial mechanism to map relationship between contrast and non-contrast images. \textcolor{mycorrect}{U-Net baseline comparison.} \\
    \hline
    \citet{oliveira2020implanting}& 
    pix2pix, SPADE & 
    LiTS \citep{bilic2019liver} & 
    CT & 
    Image inpainting & 
    Dice: 0.58 & 
    0.61  &
    Realistic lesion inpainting in CT slices to provide controllable set of training samples.
    \\
    \hline
    \hline
    \textbf{Chest and lungs} 
    \\
    \hline
    \citet{jiang2018tumor} & 
    CycleGAN-based & 
    NSCLC \citep{prior2017public} & 
    CT, MRI & 
    Cross-domain translation &
    Dice: \textcolor{mycorrect}{0.66} & 
    0.80 &
    Tumour-aware loss for unsupervised cross-domain adaptation \textcolor{mycorrect}{compared with standard cycleGAN benchmark.}
    \\
    \hline
    \citet{jin2018ct} & 
    cGAN & 
    LIDC \citep{armato2011lung} & 
    CT & 
    Image inpainting & 
    Dice: 0.96 & 
    0.99 &
    Generated lung nodules to improve segmentation robustness; A novel multi-mask reconstruction loss.
    \\
    \hline
    \citet{dong2019automatic} & 
    UNet-based GAN segmenter & 
    AAPM Challenge \citep{yang2018autosegmentation} & 
    CT & 
    Adversarial training & 
    Dice: 0.97 (l),  0.83 (sc), 0.71 (o), 0.85 (h) & 
    0.97, 0.90, 0.75, 0.87 &
    Adversarial training to discriminate manual and automated segmentation of lungs, spinal cord, oesophagus, heart.
    \\
    \hline
    \citet{tang2019xlsor} 
    
    \href{https://github.com/rsummers11/CADLab/tree/master/Lung_Segmentation_XLSor.}{XLSor}& 
    MUNIT \citep{huang2018multimodal} & 
    JSRT \citep{shiraishi2000development}, Montgomery \citep{jaeger2013automatic} & 
    Chest X-Ray & 
    Lung segmentation & 
    Dice: 0.97 & 
    0.98 &
    Unpaired normal-to-abnormal (pathological) translation. Synthetic data augmentation for lung segmentation. 
    \\
    \hline
    \citet{shi2020automatic} & 
    AUGAN & 
    LIDC-IDRI\citep{armato2011lung} & 
    CT & 
    Adversarial training & 
    Dice: 0.82 & 
    0.85 &
    A deep layer aggregation based on U-Net++.
    \\
    \hline
    \citet{munawar2020segmentation}& 
    Unet-based GAN segmenter & 
    JSRT \citep{shiraishi2000development}, Montgomery \& Shenzhen \citep{jaeger2013automatic} & 
    Chest X-Ray & 
    Adversarial training & 
    Dice: 0.96 & 
    0.97 &
    Adversarial training to discriminate manual and automated segmentation.
    \\
    \hline
    \hline
    \textbf{Prostate} 
    \\
    \hline
    \citet{kohl2017adversarial} & 
    UNet-based GAN segmenter & 
    Private & 
    MRI & 
    Adversarial training & 
    Dice: 0.35 & 
    0.41 &
    Adversarial loss to discriminate manual and automated segmentation.
    \\
    \hline
    \citet{grall2019using} 
    
    \href{https://github.com/AzmHmd/Analysing-MRI-Prostate-scans-with-cGANs.git}{Prostate cGAN}& 
    pix2pix & 
    Private &
    MRI &
    Adversarial training & 
    Dice: 0.67 (ADC), 0.74 (DWI), 0.67 (T2) & 
    0.73, 0.79, 0.74 &
    Paired prostate image-to-mask translation. Investigated the robustness of the pix2pix against noise.
    \\
    \hline
    \citet{nie2020adversarial} & 
    GAN & 
    PROMISE12 \citep{litjens2014evaluation} & 
    MRI & 
    Adversarial confidence learning & 
    Dice: 88.25 (b), 90.11 (m), 86.67 (a) & 
    89.52, 90.97, 88.20 &
    Difficulty-aware mechanism to alleviate the effect of easy samples during training. b = base, m = middle, a = apex.
    \\
    \hline
    \citet{zhang2020semi}& 
    PGGAN & 
    Private & 
    CT & 
    Data augmentation & 
    Dice: 0.85 & 
    0.90 &
    Semi-supervised training using both annotated and un-annotated data. Synthetic data augmentation using PGGAN.
    \\
    \hline
    \citet{cem2020investigating}& 
    pix2pix, CycleGAN & 
    PROMISE12 \citep{litjens2014evaluation} & 
    MRI & 
    Data augmentation & 
    Dice: 0.72 & 
    0.76 &
    Compared the performance of pix2pix, U-Net, and CycleGAN. 
    \\
    \hline
    \citet{chaitanya2021semi}& 
    cGAN, DCGAN-based D  & 
    ACDC \citep{bernard2018deep}, Decathlon \citep{simpson2019large} & 
    MRI, CT & 
    Data augmentation & 
    Dice: 0.40 & 
    0.53 &
    GAN data augmentation for intensity and shape variations. Tested on cardiac, prostate, and pancreas datasets.
    \\
    \hline
    \hline
    \textbf{Colorectal} 
    \\
    \hline
    \citet{liu2019accurate}& 
    GAN, LAGAN & 
    Private & 
    CT & 
    Adversarial training & 
    Dice: 0.87 & 
    0.92 &
    Automatic post-processing to refine the segmentation of deep networks. \\
    \hline
    \citet{poorneshwaran2019polyp}& 
    pix2pix & 
    CVC-clinic \citep{vazquez2017benchmark} & 
    Endoscopy & 
    Adversarial training & 
    Dice: \textcolor{mycorrect}{n.a.} & 
    0.88 &
    Adversarial learning to make automatic segmentation close to manual. \textcolor{mycorrect}{Ablations were compared instead of baselines.}
    \\
    \hline
    \citet{xie2020mi}& 
    MI$^2$GAN, CycleGAN & 
    CVC-clinic \citep{vazquez2017benchmark}, ETIS-Larib \citep{silva2014toward} & 
    Endoscopy & 
    Cross-domain translation & 
    Dice: 0.66 & 
    0.73 &
    Content features and domain information decoupling and maximising the mutual information. 
    \\
    \hline
    \hline
    \textbf{Pathology} 
    \\
    \hline
    \citet{pandey2020image} & 
    GAN \& cGAN & 
    Kaggle \citep{ljosa2012annotated} & 
    Histopathology & 
    Data augmentation & 
    Dice: 0.79 & 
    0.83 &
    Two-stage GAN to generate masks and conditioned synthetic images.
    \\
    \hline
    \hline
    \textbf{Other} 
    \\
    \hline
    \citet{chi2018controlled}& 
    pix2pix & 
    ISBI ISIC \citep{codella2018skin} & 
    Dermoscopy & 
    Data augmentation & 
    Dice: 0.85 & 
    0.84 &
    Similar performance replacing half of real with synthetic data.
    Colour labels as lesion specific characteristics.
    \\
    \hline
    \citet{abhishek2019mask2lesion} & 
    pix2pix & 
    ISBI ISIC \citep{codella2018skin} & 
    Dermoscopy & 
    Data augmentation & 
    Dice: 0.77 & 
    0.81 &
    Generate new lesion images given any arbitrary mask.
    \\
    \hline
    \citet{sarker2019mobilegan}& 
    MobileGAN & 
    ISBI ISIC \citep{codella2018skin} & 
    Dermoscopy & 
    Adversarial training & 
    Dice: \textcolor{mycorrect}{0.76} & 
    0.91 &
    Lightweight and efficient GAN model with position and channel attention. \textcolor{mycorrect}{Dice was compared to a U-Net baseline.}
    \\
    \hline
    \citet{zaman2020generative} & 
    pix2pix & 
    Private & 
    Ultrasound & 
    Data augmentation & 
    Dice: \textcolor{mycorrect}{*} & 
    $\approx\Delta$+0.05 &
    Recommendations on standard data augmentation approaches. Bone surface segmentation. pix2pix based discriminator. 
    \\
    \hline
\end{tabular}
}
\end{table*}

\subsubsection{Limitations and Future Prospects for Cancer Imaging Segmentation}
As shown in Table~\ref{Table:segmentation-table}, the applications of GANs in cancer image segmentation cover a variety of clinical requirements. Remarkable steps have been taken to advance this field of research over the past few years. However, the following limitations and future prospects can be considered for further investigation:
\begin{itemize}
    \item Although the data augmentation using GANs could increase the number of training samples for segmentation, the variability of the synthetic data is limited to the training data. Hence, it may limit the potential of improving the performance in terms of segmentation accuracy. Moreover, training a GAN that produce high sample variability requires a large dataset also with a high variability, and, in most of the cases, with corresponding annotations. Considering the data scarcity challenge in the cancer imaging domain, this can be difficult to achieve.
    \item In some cases, using GANs could be excessive, considering the difficulties related to convergence of competing generator and discriminator parts of the GAN architectures. For example, the recently proposed SynthSeg model~\citep{billot2020partial} is based on Gaussian Mixture Models to generate images and train a contrast agnostic segmentation model. Such approaches can be considered as an alternative to avoid common pitfalls of the GAN training process (e.g., mode collapse). However, this approach needs to be further investigated for cancer imaging tasks where the heterogeneity of tumours is challenging.
    \item A great potential for using synthetic cancer images is to generate common shareable datasets as benchmarks for automated segmentation methods~\citep{bi2019artificial}. Although this benchmark dataset needs its own validation, it can be beneficial in testing the limits of automated methods with systematically controlled test cases. Such benchmark datasets can be generated by controlling the shape, location, size, intensities of tumours, and can simulating diverse images of different domains that reflect the distributions from real institutions. To avoid learning patterns that are only available in synthetic datasets (e.g., checkerboard artifacts), it is a prospect to investigate further metrics that measure the distance of such synthetic datasets to real-world datasets and the generalisation and extrapolation capabilities of models trained on synthetic benchmarks to real-world data.
\end{itemize}

\subsection{Detection and Diagnosis Challenges}\label{sec:detection}
\subsubsection{Common Issues in Diagnosing Malignancies}

\paragraph{Clinicians' High Diagnostic Error Rates}\label{sec:detection_radiologist_error}
Studies of radiological error report high ranges of diagnostic error rates (e.g., discordant interpretations in 31–37\% in Oncologic CT, 13\% major discrepancies in Neuro CT and MRI)~\citep{brady2017error}. 
After~\citet{mccreadie2009eight} critically reviewed the radiology cases of the last 30 months in their clinical centre, they found that from 256 detected errors (62\% CT, 12\% Ultrasound, 11\% MRI, 9\% Radiography, 5\% Fluoroscopy) in 222 patients, 225 errors (88\%) were attributable to poor image interpretation (14 false positive, 155 false negative, 56 misclassifications).
A recent literature review on diagnostic errors by~\citet{newman2021rate} estimated a false negative rate\footnote{In~\citet{newman2021rate}, the false negative rates includes both missed (patient encounters at which the diagnosis might have been made but was not) and delayed diagnosis (diagnostic delay relative to urgency of illness detection).} of 22.5\% for lung cancer, 8.9\% for breast cancer, 9.6\% for colorectal cancer, 2.4\% for prostate and 13.6\% for melanoma.
These findings exemplify the uncomfortably high diagnostic and image interpretation error rates that persist in the field of radiology despite decades of interventions and research~\citep{itri2018fundamentals}. 

\paragraph{The Challenge of Reducing Clinicians' High Workload}
In some settings, radiologists must interpret one CT or MRI image every 3–4 seconds in an average 8-hour workday~\citep{mcdonald2015effects}. 
Automated CADe and CADx systems can provide a more balanced quality-focused workload for radiologists, where radiologists focus on scrutinising the automated detected lesions (false positive reduction) and areas/patches with high predictive uncertainty (false negative reduction). A benefit of CADe/CADx deep learning models are their real-time inference and strong pattern recognition capabilities that are not readily susceptible to cognitive bias (discussed in \ref{sec:annotation:human-bias}), environmental factors~\citep{itri2018fundamentals}, or inter-observer variability (discussed in \ref{sec:annotation:variability}).

\paragraph{Detection Model Performance on Critical Edge Cases}
Challenging cancer imaging problems are the high intra- and inter-tumour heterogeneity~\citep{bi2019artificial}, the detection of small lesions and metastasis across the body (e.g., lymph node involvement and distant metastasis~\citep{hosny2018artificial}) and the accurate distinction between malignant and benign tumours (e.g., for detected lung nodules that seem similar on CT scans~\citep{hosny2018artificial}). Methods are needed to extend on and further increase the current performance of deep learning detection models~\citep{bi2019artificial}.

\subsubsection{\textcolor{mycorrect}{GAN Applications for Cancer Detection and Diagnosis}}
As we detail in the following, the capability of adversarial learning to improve malignancy detection has been demonstrated for multiple tumour types and imaging modalities. To this end, Table~\ref{table:detection-table} summarises the collection of recent publications that utilise GANs and adversarial training for cancer detection, classification, and diagnosis.

\textcolor{mycorrect}{
Figures \ref{fig:ScatterDia1} and \ref{fig:ScatterDia2} visualise the publications' performance metric values reported in Table~\ref{table:detection-table}. Figure \ref{fig:ScatterDia1} provides visual estimate of the effectiveness of GANs and adversarial training in increasing downstream task performance. A performance increase of approximately 5 percentage points can be observed by comparing the figure's black identity line with the red trend line over publications. We note that no visual pattern seems to be observable that indicates a difference in performance gain between anatomical categories. Across all publications, the performance gains seem not to be a function of the strength of the baseline, as they remain approximately constant with increasing baseline performance. This is indicated by the minimal change of distance between the black identity line and the red trend line throughout the graph. Figure \ref{fig:ScatterDia2} shows the GAN-induced variation in performance for the years 2017 to 2021 with multiple adversarial models achieving a performance increase of over 10\% and most models over 3\% on their respective diagnostic downstream task. As emphasised in Section \ref{sec:annotation:examples}, conclusion drawn from Figures \ref{fig:ScatterDia1} and \ref{fig:ScatterDia2} have to take publication bias into account. Further, benchmarking and comparison of individual publications based on the presented data in these figures is not part of their intended use due to the differences in baselines, modalities, organs, train and test datasets, and publication objectives.}

\begin{figure*}[tb]
\begin{tikzpicture}[scale=1.0,transform shape]
\centering
  \pgfplotsset{
      scale only axis,
  }
\begin{axis}[enlargelimits=0.025, 
        xlabel=Performance: After Introducing Adversarial Network (in \%),
        ylabel=Performance: Baseline (in \%),
        grid=both,
        scale only axis=true,
        width=0.875\textwidth, 
        height=6cm,
        xtick={0, 5,..., 100},
        ytick={0, 5,..., 100},
        xmin=54,
        xmax=100,
        ymin=54,
        ymax=100,
        legend style={at={(0.01, 0.975)}, anchor=north west},
        legend cell align={left},
      ]
    \addplot[
        scatter/classes={a={mark=pentagon*, mark options={line width=0.35pt}, draw=black, fill=blue}, b={mark=diamond*, mark options={line width=0.35pt}, draw=black, fill=red}, c={mark=*, mark options={line width=0.35pt}, draw=black, fill=yellow}, d={mark=square*, mark options={line width=0.35pt}, draw=black, fill=orange}, e={mark=triangle*, mark options={line width=0.35pt}, draw=black, fill=green}, f={mark=otimes*, mark options={line width=0.35pt}, draw=black, fill=pink}, 
        g={mark=oplus*, mark options={line width=0.35pt}, draw=black, fill=gray}},
        scatter, mark=*, only marks, 
        scatter src=explicit symbolic,
        nodes near coords*={\Label},
        visualization depends on={value \thisrow{label} \as \Label}
    ] table [meta=class] {
        x y class label hidden
        80.00 70.00 a \phantom{\tiny{a}} \footnotesize{\citeauthor{chen2018deep}(AUC)} 
        91.08 90.06 a \phantom{\tiny{a}} \footnotesize{\citeauthor{han2020infinite}(ACC)}
        97.48 93.67 a \phantom{\tiny{a}} \footnotesize{\citeauthor{han2019combining}(SEN)}
        74.96 73.48 a \phantom{\tiny{a}} \footnotesize{\citeauthor{benson2020gan}(ACC)}
        77.00 67.00 a \footnotesize{\citeauthor{han1904h2019}(SEN)} \phantom{\tiny{a}}
            34.80 26.10 a \footnotesize{\citeauthor{siddiquee2019learning}(IoU)} \phantom{\tiny{a}}
            91.70 89.60 a \tiny{\citeauthor{sun2020adversarial}(F1)} \phantom{\tiny{a}}
        89.60 88.70 b \phantom{\tiny{a}} \footnotesize{\citeauthor{wu2018conditional}(AUC)}
        74.96 73.48 b \phantom{\tiny{a}} \footnotesize{\citeauthor{guan2019breast}(ACC)} 
        63.81 62.53 b \phantom{\tiny{a}} \footnotesize{\citeauthor{jendele2019adversarial}(F1)}  
            67.00 57.00 b \tiny{\citeauthor{lee2020simulating}(AUC)} \phantom{\tiny{a}}
        84.60 82.90 b \phantom{\tiny{a}} \footnotesize{\citeauthor{wu2020synthesizing}(AUC)}
        96.00 90.00 b \phantom{\tiny{a}} \footnotesize{\citeauthor{alyafi2020dcgans}(F1)} 
        87.00 78.23 b \phantom{\tiny{a}} \footnotesize{\citeauthor{desai2020breast}(ACC)}
            81.40 79.20 b \tiny{\citeauthor{muramatsu2020improving}(ACC)} \phantom{\tiny{a}}
        94.10 92.50 b \phantom{\tiny{a}} \footnotesize{\citeauthor{swiderski2021deep}(AUC)}
        93.7 92.0 b \phantom{\tiny{a}} \footnotesize{\citeauthor{kansal2020generative}(ACC)} 
            17.2 15.1 b \footnotesize{\citeauthor{shen2021mass}(FROC-AUC)} \phantom{\tiny{a}}
        87.94 86.60 b \phantom{\tiny{a}} \footnotesize{\citeauthor{pang2021semi}(SEN)}
        85.70 78.60 c \phantom{\tiny{a}} \footnotesize{\citeauthor{frid2018gan}(SEN)}
            89.4 79.2 c \footnotesize{\citeauthor{zhao2020tripartite}(ACC)} \phantom{\tiny{a}}
            95.0 65.0 c \footnotesize{\citeauthor{doman2020lesion}(DR)} \phantom{\tiny{a}}
        63.2 59.6 d \phantom{\tiny{a}} \footnotesize{\citeauthor{kanayama2019gastric}(AP)}
        85.3 81.9 d \phantom{\tiny{a}} \footnotesize{\citeauthor{shin2018abnormal}(PRE)}
            33.9 24.7 d \tiny{\citeauthor{rau2019implicit}(L1)} \phantom{\tiny{a}}
        88.5 85.1 d \phantom{\tiny{a}} \footnotesize{\citeauthor{yu2020synthesis}(AUC)}
        77.7 75.7 d  \scriptsize{\citeauthor{krause2021deep}(AUC)} \phantom{\tiny{a}}
        84.7 83.4 e \phantom{\tiny{a}} \footnotesize{\citeauthor{bissoto2018skin}(AUC)}
        89.0 89.0 e \phantom{\tiny{a}} \footnotesize{\citeauthor{creswell2018denoising}(AUC)} 
        74.0 71.6 e \phantom{\tiny{a}} \footnotesize{\citeauthor{baur2018melanogans}(ACC)}
        86.1 81.5 e \phantom{\tiny{a}} \footnotesize{\citeauthor{rashid2019skin}(ACC)}
        76.3 72.1 e \phantom{\tiny{a}} \footnotesize{\citeauthor{fossen2020synthesizing}(REC)}
        76.9 71.8 e \phantom{\tiny{a}} \footnotesize{\citeauthor{qin2020gan}(PRE)} 
            63.84 66.38 f \footnotesize{\citeauthor{bi2017synthesisCTtoPET}(F1)} \phantom{\tiny{a}}
            92.1 70.87 f  \tiny{\citeauthor{salehinejad2018generalization}(ACC)} \phantom{\tiny{a}}
        95.24 92.86 f \phantom{\tiny{a}} \footnotesize{\citeauthor{zhao2018synthetic}(ACC)}
        80.5 72.5 f \phantom{\tiny{a}} \footnotesize{\citeauthor{onishi2019automated}(ACC)}
            95.0 84.8 f \footnotesize{\citeauthor{gao2019augmenting}(SEN)} \phantom{\tiny{a}}  
            55.0 51.8 f  \footnotesize{\citeauthor{han2019synthesizing}(CPM)} \phantom{\tiny{a}}
        88.12 87.56 f \phantom{\tiny{a}} \footnotesize{\citeauthor{yang2019class}(AUC)}
        89.03 85.88 f \phantom{\tiny{a}} \footnotesize{\citeauthor{wang2020class}(F1)}
        95.32 91.6 f \phantom{\tiny{a}} \footnotesize{\citeauthor{kuang2020unsupervised}(ACC)} 
        97.0 95.0 f \phantom{\tiny{a}} \footnotesize{\citeauthor{ghosal2020lung}(AUC)}
            94.5 93.8 f \tiny{\citeauthor{sun2020classification}(ACC)} \phantom{\tiny{a}} 
            60.5 53.2 f \scriptsize{\citeauthor{wang2020combination}(ACC)} \phantom{\tiny{a}}
            98.57 97.81 f \footnotesize{\citeauthor{bu20203d}(SEN)} \phantom{\tiny{a}} 
            85.0 85.0 f \tiny{\citeauthor{nishio2020attribute}(ACC)} \phantom{\tiny{a}}
        77.8 66.7 f \phantom{\tiny{a}} \footnotesize{\citeauthor{onishi2020multiplanar}(SPE)}
        85.3 81.0 f \phantom{\tiny{a}} \footnotesize{\citeauthor{teramoto2020deep}(ACC)}
            89.0 73.0 g \footnotesize{\citeauthor{schlegl2017unsupervised}(AUC)} \phantom{\tiny{a}}
        98.83 95.67 g \phantom{\tiny{a}} \footnotesize{\citeauthor{zhang2018medical}(ACC)}
        90.925 69.2575 g \phantom{\tiny{a}} \scriptsize{\citeauthor{chaudhari2019data}(ACC)}
            70.97 64.52 g \footnotesize{\citeauthor{liu2019wasserstein}(ACC)} \phantom{\tiny{a}}
        94.7 89.2 g \phantom{\tiny{a}} \footnotesize{\citeauthor{rubin2019top}(AUC)} 
    };
    \addplot[black,samples at={0,100}] {x};
    \addplot [thick, smooth, red, dashed] table[y={create col/linear regression={y=Y}}]{
        x Y class label hidden
        80.00 70.00 a \phantom{\tiny{a}} \footnotesize{\citeauthor{chen2018deep}(AUC)} 
        91.08 90.06 a \phantom{\tiny{a}} \footnotesize{\citeauthor{han2020infinite}(ACC)}
        97.48 93.67 a \phantom{\tiny{a}} \footnotesize{\citeauthor{han2019combining}(SEN)}
        74.96 73.48 a \phantom{\tiny{a}} \footnotesize{\citeauthor{benson2020gan}(ACC)}
            77.00 67.00 a \footnotesize{\citeauthor{han1904h2019}(SEN)} \phantom{\tiny{a}}
            34.80 26.10 a \footnotesize{\citeauthor{siddiquee2019learning}(IoU)} \phantom{\tiny{a}}
            91.70 89.60 a \scriptsize{\citeauthor{sun2020adversarial}(F1)} \phantom{\tiny{a}}
        89.60 88.70 b \phantom{\tiny{a}} \footnotesize{\citeauthor{wu2018conditional}(AUC)}
        74.96 73.48 b \phantom{\tiny{a}} \footnotesize{\citeauthor{guan2019breast}(ACC)} 
        63.81 62.53 b \phantom{\tiny{a}} \footnotesize{\citeauthor{jendele2019adversarial}(F1)}  
            67.00 57.00 b \tiny{\citeauthor{lee2020simulating}(AUC)} \phantom{\tiny{a}}
        84.60 82.90 b \phantom{\tiny{a}} \footnotesize{\citeauthor{wu2020synthesizing}(AUC)}
        96.00 90.00 b \phantom{\tiny{a}} \footnotesize{\citeauthor{alyafi2020dcgans}(F1)}
        87.00 78.23 b \phantom{\tiny{a}} \footnotesize{\citeauthor{desai2020breast}(ACC)}
        81.40 79.20 b \phantom{\tiny{a}} \footnotesize{\citeauthor{muramatsu2020improving}(ACC)}
        94.10 92.50 b \phantom{\tiny{a}} \footnotesize{\citeauthor{swiderski2021deep}(AUC)}
        93.7 92.0 b \phantom{\tiny{a}} \footnotesize{\citeauthor{kansal2020generative}(ACC)} 
            17.2 15.1 b \footnotesize{\citeauthor{shen2021mass}(FROC-AUC)} \phantom{\tiny{a}}
        87.94 86.60 b \phantom{\tiny{a}} \footnotesize{\citeauthor{pang2021semi}(SEN)}
        85.70 78.60 c \phantom{\tiny{a}} \footnotesize{\citeauthor{frid2018gan}(SEN)}
            89.4 79.2 c \footnotesize{\citeauthor{zhao2020tripartite}(ACC)} \phantom{\tiny{a}}
            95.0 65.0 c \footnotesize{\citeauthor{doman2020lesion}(DR)} \phantom{\tiny{a}}
        63.2 59.6 d \phantom{\tiny{a}} \footnotesize{\citeauthor{kanayama2019gastric}(AP)}
        85.3 81.9 d \phantom{\tiny{a}} \footnotesize{\citeauthor{shin2018abnormal}(PRE)}
            33.9 24.7 d \tiny{\citeauthor{rau2019implicit}(L1)} \phantom{\tiny{a}}
        88.5 85.1 d \phantom{\tiny{a}} \footnotesize{\citeauthor{yu2020synthesis}(AUC)}
            77.7 75.7 d  \footnotesize{\citeauthor{krause2021deep}(AUC)} \phantom{\tiny{a}}
        84.7 83.4 e \phantom{\tiny{a}} \footnotesize{\citeauthor{bissoto2018skin}(AUC)}
        89.0 89.0 e \phantom{\tiny{a}} \footnotesize{\citeauthor{creswell2018denoising}(AUC)} 
        74.0 71.6 e \phantom{\tiny{a}} \footnotesize{\citeauthor{baur2018melanogans}(ACC)}
        86.1 81.5 e \phantom{\tiny{a}} \footnotesize{\citeauthor{rashid2019skin}(ACC)}
        76.3 72.1 e \phantom{\tiny{a}} \footnotesize{\citeauthor{fossen2020synthesizing}(REC)}
        76.9 71.8 e \phantom{\tiny{a}} \footnotesize{\citeauthor{qin2020gan}(PRE)} 
            63.84 66.38 f \footnotesize{\citeauthor{bi2017synthesisCTtoPET}(F1)} \phantom{\tiny{a}}
            92.1 70.87 f  \tiny{\citeauthor{salehinejad2018generalization}(ACC)} \phantom{\tiny{a}}
        95.24 92.86 f \phantom{\tiny{a}} \footnotesize{\citeauthor{zhao2018synthetic}(ACC)}
        80.5 72.5 f \phantom{\tiny{a}} \footnotesize{\citeauthor{onishi2019automated}(ACC)}
            95.0 84.8 f \footnotesize{\citeauthor{gao2019augmenting}(SEN)} \phantom{\tiny{a}}  
            55.0 51.8 f  \footnotesize{\citeauthor{han2019synthesizing}(CPM)} \phantom{\tiny{a}}
        88.12 87.56 f \phantom{\tiny{a}} \footnotesize{\citeauthor{yang2019class}(AUC)}
        89.03 85.88 f \phantom{\tiny{a}} \footnotesize{\citeauthor{wang2020class}(F1)}
        95.32 91.6 f \phantom{\tiny{a}} \footnotesize{\citeauthor{kuang2020unsupervised}(ACC)} 
        97.0 95.0 f \phantom{\tiny{a}} \footnotesize{\citeauthor{ghosal2020lung}(AUC)}
            94.5 93.8 f \scriptsize{\citeauthor{sun2020classification}(ACC)} \phantom{\tiny{a}} 
            60.5 53.2 f \scriptsize{\citeauthor{wang2020combination}(ACC)} \phantom{\tiny{a}}
            98.57 97.81 f \footnotesize{\citeauthor{bu20203d}(SEN)} \phantom{\tiny{a}} 
            85.0 85.0 f \scriptsize{\citeauthor{nishio2020attribute}(ACC)} \phantom{\tiny{a}}
        77.8 66.7 f \phantom{\tiny{a}} \footnotesize{\citeauthor{onishi2020multiplanar}(SPE)}
        85.3 81.0 f \phantom{\tiny{a}} \footnotesize{\citeauthor{teramoto2020deep}(ACC)}
            89.0 73.0 g \footnotesize{\citeauthor{schlegl2017unsupervised}(AUC)} \phantom{\tiny{a}}
        98.83 95.67 g \phantom{\tiny{a}} \footnotesize{\citeauthor{zhang2018medical}(ACC)}
        90.925 69.2575 g \phantom{\tiny{a}} \scriptsize{\citeauthor{chaudhari2019data}(ACC)}
            70.97 64.52 g \footnotesize{\citeauthor{liu2019wasserstein}(ACC)} \phantom{\tiny{a}}
        94.7 89.2 g \phantom{\tiny{a}} \footnotesize{\citeauthor{rubin2019top}(AUC)} 
    };
    \legend{\scriptsize{Head/Neck},\scriptsize{Breast},\scriptsize{Liver}, \scriptsize{Abdomen/Prostate}, 
    \scriptsize{Skin}, \scriptsize{Lung}, \scriptsize{Other}}
\end{axis}
\end{tikzpicture}
    \caption[]{\textcolor{mycorrect}{
    Scatter plot illustrating the performance improvement attributable to adversarial networks for the surveyed disease diagnosis and detection publications. The shown performance is based on the respective publication's metric reported in Table \ref{table:detection-table} and include f1-score (F1), sensitivity (SEN), accuracy (ACC), area under the receiver operating characteristic curve (AUC), and detection rate (DR). Each publication is represented by a \textcolor{mycorrect2}{marker} with a colour \textcolor{mycorrect2}{and shape} encoding depicting the publication's anatomical category.
    The black identity line indicates no change between baseline and adversarial network intervention, while dots below this line represent an improvement. The red regression line depicts the trend of improvement across publications. The author names of a few publications have been manually selected for highlighting based on the distance to the trend line.}
    } 
  	\label{fig:ScatterDia1}
\end{figure*}
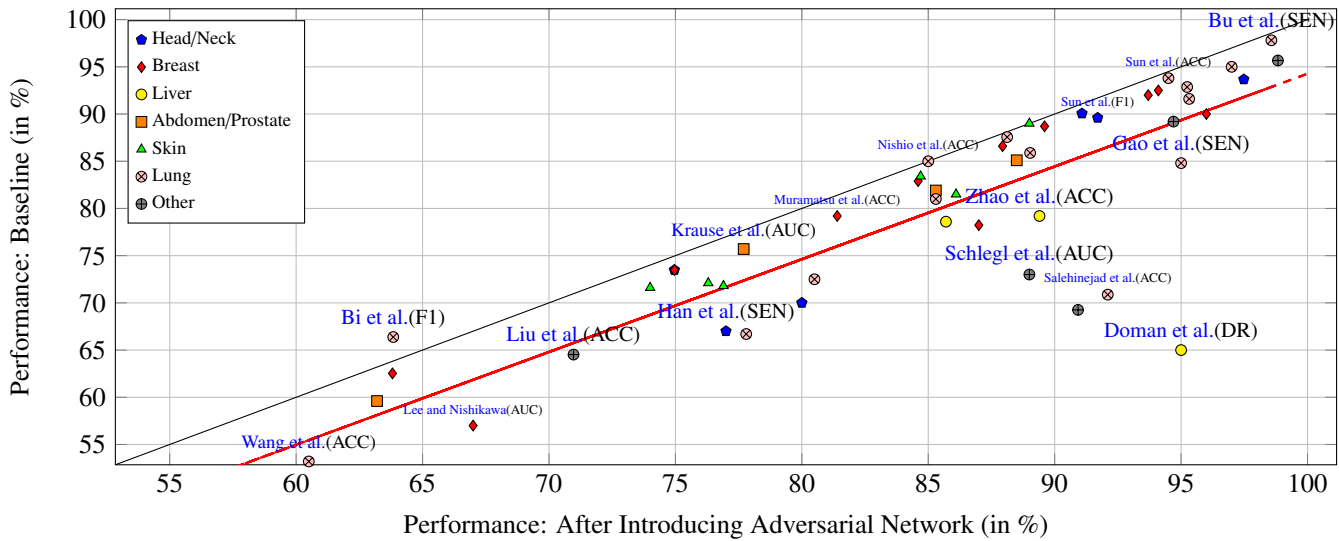

\begin{figure*}[tb]
\begin{tikzpicture}[scale=1.0,transform shape]
\centering
  \pgfplotsset{
      scale only axis,
  }
 
\begin{axis}[enlargelimits=0.03,  
        ylabel=Year of Publication,
        xlabel=Baseline+$\Delta$: Change in Performance After Introducing Adversarial Network (in \textit{\%}), 
        grid=both,
        scale only axis=true,
        width=0.91\textwidth, 
        height=4.15cm,
        ytick={2017,2018,...,2021},
        xtick={-3,-2,...,31}, 
        xmin=-3,
        xmax=31,
        ymin=2017,
        ymax=2021,
        legend style={at={(1,0)},anchor=south east},
        legend cell align={left},
        /pgf/number format/.cd,
        use comma,
        1000 sep={}]
      ]
    \addplot[
        scatter/classes={a={mark=pentagon*, mark options={line width=0.35pt}, draw=black, fill=blue}, b={mark=diamond*, mark options={line width=0.35pt}, draw=black, fill=red}, c={mark=*, mark options={line width=0.35pt}, draw=black, fill=yellow}, d={mark=square*, mark options={line width=0.35pt}, draw=black, fill=orange}, e={mark=triangle*, mark options={line width=0.35pt}, draw=black, fill=green}, f={mark=otimes*, mark options={line width=0.35pt}, draw=black, fill=pink}, 
        g={mark=oplus*, mark options={line width=0.35pt}, draw=black, fill=gray}},
        scatter, mark=*, only marks, 
        scatter src=explicit symbolic,
        nodes near coords*={\Label},
        visualization depends on={value \thisrow{label} \as \Label}
    ] 
        table [meta=class] {
        x y class label hidden
            10.00 2018 a \tiny{\citeauthor{chen2018deep}(AUC)} \phantom{\tiny{a}}
        1.02 2020 a \phantom{\tiny{a}} \footnotesize{\citeauthor{han2020infinite}(ACC)}
        3.81 2019 a \phantom{\tiny{a}} \footnotesize{\citeauthor{han2019combining}(SEN)}
        1.48 2020 a \phantom{\tiny{a}} \footnotesize{\citeauthor{benson2020gan}(ACC)}
            10.00 2019 a \scriptsize{\citeauthor{han1904h2019}(SEN)} \phantom{\tiny{a}}  
        8.70 2019 a \phantom{\tiny{a}} \footnotesize{\citeauthor{siddiquee2019learning}(IoU)}
        2.10 2020 a \phantom{\tiny{a}} \scriptsize{\citeauthor{sun2020adversarial}(F1)}
        0.9 2018 b \phantom{\tiny{a}} \footnotesize{\citeauthor{wu2018conditional}(AUC)}
        1.48 2019 b \phantom{\tiny{a}} \footnotesize{\citeauthor{guan2019breast}(ACC)} 
        1.28 2019 b \phantom{\tiny{a}} \footnotesize{\citeauthor{jendele2019adversarial}(F1)}  
        10.00 2020 b \phantom{\tiny{a}} \scriptsize{\citeauthor{lee2020simulating}(AUC)}
        1.70 2020 b \phantom{\tiny{a}} \footnotesize{\citeauthor{wu2020synthesizing}(AUC)}
            6.00 2020 b \scriptsize{\citeauthor{alyafi2020dcgans}(F1)} \phantom{\tiny{a}} 
        8.77 2020 b \phantom{\tiny{a}} \footnotesize{\citeauthor{desai2020breast}(ACC)}
        2.20 2020 b \phantom{\tiny{a}} \footnotesize{\citeauthor{muramatsu2020improving}(ACC)}
        1.6 2021 b \phantom{\tiny{a}} \footnotesize{\citeauthor{swiderski2021deep}(AUC)}
            1.70 2020 b \scriptsize{\citeauthor{kansal2020generative}(ACC)} \phantom{\tiny{a}} 
        2.10 2021 b  \phantom{\tiny{a}} \footnotesize{\citeauthor{shen2021mass}(FROC-AUC)} 
        1.34 2021 b \phantom{\tiny{a}} \footnotesize{\citeauthor{pang2021semi}(SEN)}
            7.00 2018 c \tiny{\citeauthor{frid2018gan}(SEN)} \phantom{\tiny{a}}
            10.20 2020 c \scriptsize{\citeauthor{zhao2020tripartite}(ACC)} \phantom{\tiny{a}}
            30.00 2020 c \scriptsize{\citeauthor{doman2020lesion}(DR)} \phantom{\tiny{a}}
        3.60 2019 d \phantom{\tiny{a}} \footnotesize{\citeauthor{kanayama2019gastric}(AP)}
        3.40 2018 d \phantom{\tiny{a}} \footnotesize{\citeauthor{shin2018abnormal}(PRE)}
            9.20 2019 d \phantom{\tiny{a}} \tiny{\citeauthor{rau2019implicit}(L1)}
        3.40 2020 d \phantom{\tiny{a}} \footnotesize{\citeauthor{yu2020synthesis}(AUC)}
            2.00 2021 d \scriptsize{\citeauthor{krause2021deep}(AUC)} \phantom{\tiny{a}}
        1.30 2018 e \phantom{\tiny{a}} \footnotesize{\citeauthor{bissoto2018skin}(AUC)}
            0.00 2018 e \scriptsize{\citeauthor{creswell2018denoising}(AUC)} \phantom{\tiny{a}}
        2.40 2018 e \phantom{\tiny{a}} \footnotesize{\citeauthor{baur2018melanogans}(ACC)}
            4.60 2019 e \scriptsize{\citeauthor{rashid2019skin}(ACC)} \phantom{\tiny{a}} 
        4.20 2020 e \phantom{\tiny{a}} \footnotesize{\citeauthor{fossen2020synthesizing}(REC)}
        5.10 2020 e \phantom{\tiny{a}} \footnotesize{\citeauthor{qin2020gan}(PRE)} 
            -2.54 2017 f \scriptsize{\citeauthor{bi2017synthesisCTtoPET}(F1)} \phantom{\tiny{a}}
            21.23 2018 f  \scriptsize{\citeauthor{salehinejad2018generalization}(ACC)} \phantom{\tiny{a}}
        2.38 2018 f \phantom{\tiny{a}} \footnotesize{\citeauthor{zhao2018synthetic}(ACC)}
        8.00 2019 f \phantom{\tiny{a}} \footnotesize{\citeauthor{onishi2019automated}(ACC)}
        10.20 2019 f \phantom{\tiny{a}} \footnotesize{\citeauthor{gao2019augmenting}(SEN)} 
            3.20 2019 f  \phantom{\tiny{a}} \footnotesize{\citeauthor{han2019synthesizing}(CPM)}
            0.56 2019 f \scriptsize{\citeauthor{yang2019class}(AUC)} \phantom{\tiny{a}} 
        3.15 2020 f \phantom{\tiny{a}} \footnotesize{\citeauthor{wang2020class}(F1)}
        3.72 2020 f \phantom{\tiny{a}} \footnotesize{\citeauthor{kuang2020unsupervised}(ACC)} 
        2.00 2020 f \phantom{\tiny{a}} \footnotesize{\citeauthor{ghosal2020lung}(AUC)}
        0.70 2020 f \phantom{\tiny{a}}   \scriptsize{\citeauthor{sun2020classification}(ACC)}
        7.3 2020 f \phantom{\tiny{a}} \scriptsize{\citeauthor{wang2020combination}(ACC)} 
        0.76 2020 f \phantom{\tiny{a}} \scriptsize{\citeauthor{bu20203d}(SEN)}
        0.0 2020 f \phantom{\tiny{a}} \scriptsize{\citeauthor{nishio2020attribute}(ACC)} 
        11.1 2020 f \phantom{\tiny{a}} \footnotesize{\citeauthor{onishi2020multiplanar}(SPE)}
        4.3 2020 f \phantom{\tiny{a}} \footnotesize{\citeauthor{teramoto2020deep}(ACC)}
            16.0 2017 g \scriptsize{\citeauthor{schlegl2017unsupervised}(AUC)} \phantom{\tiny{a}}
        3.16 2018 g \phantom{\tiny{a}} \footnotesize{\citeauthor{zhang2018medical}(ACC)}
            21.6675 2019 g \scriptsize{\citeauthor{chaudhari2019data}(ACC)} \phantom{\tiny{a}}
        6.45 2019 g \phantom{\tiny{a}} \scriptsize{\citeauthor{liu2019wasserstein}(ACC)}
        5.5 2019 g \phantom{\tiny{a}} \footnotesize{\citeauthor{rubin2019top}(AUC)} 
    };
    \legend{\scriptsize{Head/Neck},\scriptsize{Breast},\scriptsize{Liver}, \scriptsize{Abdomen/Prostate}, 
    \scriptsize{Skin}, \scriptsize{Lung}, \scriptsize{Other}}
\end{axis}
\end{tikzpicture}
    \caption[]{\textcolor{mycorrect}{
    Scatter plot displaying year of publication and change in performance between baseline and adversarial network method (in \%). As in Table \ref{table:detection-table}, the underlying performance metrics include f1-score (F1), sensitivity (SEN), accuracy (ACC), area under the receiver operating characteristic curve (AUC), and detection rate (DR). Each publication is represented by a \textcolor{mycorrect2}{marker} with a colour \textcolor{mycorrect2}{and shape} encoding depicting the publication's anatomical category. 
    Author names have been manually selected at random for highlighting.}
    } 
    \label{fig:ScatterDia2}
\end{figure*}
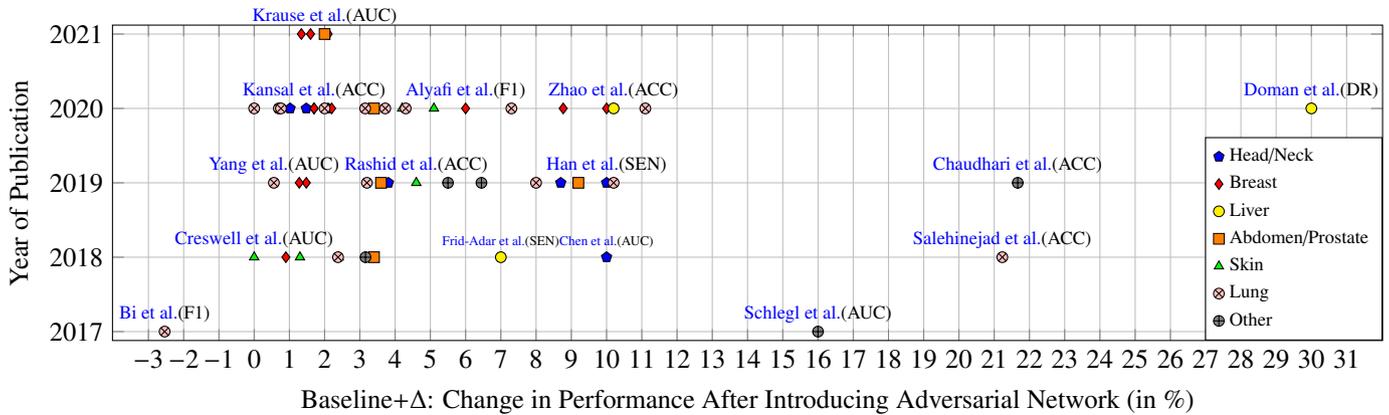

\paragraph{Adversarial Anomaly and Outlier Detection Examples} \label{sec:detection_anomaly_detection}
\citet{schlegl2017unsupervised} captured imaging markers relevant for disease prediction using a deep convolutional GAN named AnoGAN. AnoGAN learnt a manifold of normal anatomical variability, accompanying a novel anomaly scoring scheme based on the mapping from image space to a latent space. While Schlegl et al validated their model on retina optical coherence tomography images, their unsupervised anomaly detection approach is applicable to other domains including cancer detection, as indicated in Figure \ref{fig:overview}(l).~\citet{chen2018deep} used a Variational Autoencoder GAN for unsupervised outlier detection using T1 and T2 weighted brain MRI images. The scans from healthy subjects were used to train the auto-encoder model to learn the distribution of healthy images and detect pathological images as outliers.
\citet{creswell2018denoising} proposed a semi-supervised Denoising Adversarial Autoencoder (ssDAAE) to learn a representation based on unlabelled  skin lesion images. The semi-supervised part of their CNN-based architecture corresponds to malignancy classification of labelled skin lesions based on the encoded representations of the pretrained DAAE. As the amount of labelled data is smaller than the unlabelled data, the labelled data is used to fine-tune classifier and encoder. In ssDAAE, not only the adversarial autoencoder's chosen prior distribution~\citep{makhzani2015adversarial}, but also the class label distribution is discriminated by a discriminator, the latter distinguishing between predicted continuous labels and real binary (malignant/benign) labels.~\citet{kuang2020unsupervised} applied unsupervised learning to distinguish between benign and malignant lung nodules. In their multi-discriminator GAN (MDGAN) various discriminators scrutinise the realness of generated lung nodule images. After GAN pretraining, an encoder is added in front of the generator to the end-to-end architecture to learn the feature distribution of benign pulmonary nodule images and to map these features into latent space.
The benign and malignant lung nodules were scored similarly as in the f-AnoGAN framework~\citep{schlegl2019f}, computing and combining an image reconstruction loss and a feature matching loss, the latter comparing the discriminators' feature representations between real and encoded-generated images from intermediate discriminator layers. As exemplified in Figure \ref{fig:Figure3_GAN_application_pipelines}(g), the model yielded high anomaly scores on malignant images and low anomaly scores on benign images despite limited dataset size.
\citet{benson2020gan} used GANs trained from multi-modal MRI images as a 3-channel input (T1-T2 weighted, FLAIR, ADC MRI) in brain anomaly detection. The training of the generative network was performed using only healthy images together with pseudo-random irregular masks. Despite the training dataset consisting of only 20 subjects, the resulting model increased the anomaly detection rate.

\paragraph{Synthetic Detection Model Training Data}
Among the GAN publications trying to improve classification and detection performance, data augmentation is the most recurrent approach to balance, vary, and increase the detection model's training set size, as suggested in Figure \ref{fig:overview}(k).
For instance in breast imaging,~\citet{wu2018conditional} trained a class-conditional GAN to perform contextual in-filling to synthesise lesions in healthy scanned mammograms.~\citet{guan2019breast} trained a GAN on the same dataset~\citep{heath2001digital} to generate synthetic patches with benign and malignant tumours. The synthetic generated patches had clear artifacts and did not match the original dataset distribution.~\citet{jendele2019adversarial} used a CycleGAN~\citep{zhu2017unpaired} and both film scanned and digital mammograms to improve binary (malignant/benign) lesion detection using data augmentation.
Detecting mammographically-occult breast cancers is another challenging topic addressed by GANs. For instance, \citet{lee2020simulating} exploit asymmetries between mammograms of the left and right breasts as signals for finding mammography-occult cancer. They trained a  \textcolor{mycorrect}{image-conditioned} GAN (pix2pix) to generate a healthy synthetic mammogram image of the contralateral breast (e.g., left breast) given the corresponding single-sided mammogram (e.g., right breast) as input. The authors showed that there is a higher similarity (MSE, 2D-correlation) between simulated-real (SR) mammogram pairs than real-real (RR) mammogram pairs in the presence of mammography-occult cancer. Consequently, distinguishing between healthy and mammography-occult mammograms, their classifier yielded a higher performance when trained with both RR and SR similarity as input (AUC = 0.67) than when trained only with RR pair similarity as input (AUC = 0.57). 3-dimensional \textcolor{mycorrect}{conditional} image synthesis with GANs has been shown, for instance, by~\citet{han2019synthesizing}, who proposed a 3D Multi-Conditional GAN (3DMCGAN) to generate realistic and diverse nodules placed naturally on lung CT images to boost sensitivity in 3D object detection.~\citet{bu20203d} built a 3D  \textcolor{mycorrect}{image-conditioned} GAN based on pix2pix, where the input is a 3D volume of interest (VOI) that is cropped from a lung CT scan and contains a missing region in its centre. Both generator and discriminator contain squeeze-and-excitation~\citep{hu2018squeeze} residual\citep{he2016deep} neural network (SE-ResNet) modules to improve the quality of the synthesised lung nodules. Another example based on lung CT images is the method by~\citet{nishio2020attribute}, where the proposed GAN model used masked 3D CT images and nodule size information to generate images.

As to multi-modal training data synthesis,~\citet{van2015does} replaced missing sequences of a multi-sequence MRI with synthetic data. The authors illustrated that if the synthetic data generation model is more flexible than the classification model, the synthetic data can provide features that the classifier has not extracted from the original data, which can improve the performance.
During colonoscopy, depth maps can enable navigation alongside aiding detection and size measurements of polyps. For this reason,~\citet{rau2019implicit} demonstrated the synthesis of depth maps using a \textcolor{mycorrect}{image-conditioned} GAN (pix2pix) with monocular endoscopic images as input, reporting promising results on synthetic, phantom and real datasets.
In breast cancer detection,~\citet{muramatsu2020improving} translated lesions from lung CT to breast MMG using cycleGAN yielding a performance improvement in breast mass classification when training a classifier with the domain-translated generated samples.

\subsubsection{Future Prospects for Cancer Detection and Diagnosis}

\paragraph{Granular Class Distinctions for Synthetic Tumour Images}
Further research opportunity exists in exploring a more fine-grained classification of tumours that characterises different subtypes and disease grades instead of binary malignant-benign classification. Being able to robustly distinguish between different disease subtypes with similar imaging phenotypes (e.g., glioblastoma versus primary central nervous system lymphoma~\citep{kang2018diffusion}) addresses the challenge of reducing diagnostic ambiguity~\citep{bi2019artificial}. GANs can be explored to augment training data with samples of specific tumour subtypes to improve the distinction capabilities of disease detection models. This can be achieved by training a detection model on training data generated by various GANs, where each GAN is trained on a different tumour subtype distribution. Another option we estimate worth exploring is to use the tumour subtype or the disease grade (e.g., the Gleason Score for prostate cancer~\citep{hu2018prostategan}) as a conditional input into the GAN to generate additional labelled synthetic training data.

\paragraph{Cancer Image Interpretation and Risk Estimation}
Besides the detection of prospectively cancerous characteristics in medical scans, ensuring a high accuracy in the subsequent interpretation of these findings are a further challenge in cancer imaging. Improving the interpretation accuracy can reduce the number of unnecessary biopsies and harmful treatments (e.g., mastectomy, radiation therapy, chemotherapy) of indolent tumours~\citep{bi2019artificial}. For instance, the rate of overdiagnosis of non-clinically significant prostate cancer ranges widely between 1.7\% up to a noteworthy 67\%~\citep{loeb2014overdiagnosis}.
To address this, detection models can be extended to provide risk and clinical significance estimations. For example, given both an input image, and an array of risk factors (e.g., BRCA1/BRCA2 status for breast cancer~\citep{li2017deep}, comorbidity risks), a deep learning model can weight and evaluate a patient's risk based on learned associations between risk factors and input image features. The GAN framework is an example of this, where clinical, non-clinical and imaging data can be combined, either as conditional input for image generation or as prediction targets. For instance, given an input image, an AC-GAN~\citep{odena2017conditional, kapil2018deep} can classify the risk as continuous label (see Figure \ref{fig:Figure3_GAN_application_pipelines}(e)) or, alternatively, a discriminator can be used to assess whether a risk estimate provided by a generator is realistic. Also, a generator can learn a function for transforming and normalising an input image given one or several conditional input target risk factors or tumour characteristics (e.g., a specific mutation status, a present comorbidity, etc) to generate labelled synthetic training data.

\begin{table*}[htbp] 
\centering
\scriptsize
\caption{Overview of adversarially-trained models applied to \textbf{detection and diagnosis} tasks in cancer imaging. Publications are clustered by organ type and ordered by year in ascending order.}
\label{table:detection-table}
\scalebox{0.67}{
\begin{tabular}{{p{0.16\textwidth}p{0.11\textwidth}p{0.30\textwidth}p{0.10\textwidth}p{0.12\textwidth}p{0.13\textwidth}p{0.06\textwidth}p{0.32\textwidth}}}
    \hline
    \textbf{Publication} &
    \textbf{Method} &
    \textbf{Dataset} &
    \textbf{Modality} &
    \textbf{Task} &
    \textbf{Metric w/o GAN (Baseline)} &
    \textbf{Metric with GAN (Baseline+$\Delta$)} &
    \textbf{Highlights}
    \\
    \hline\hline
    \textbf{Brain}
    \\
    \hline
    \citet{chen2018deep} &
    VAE GAN & 
    Cam-CAN~\citep{shafto2014cambridge} BRATS~\citep{menze2014multimodal} ATLAS~\citep{liew2018large} &
    MRI & 
    Anomaly detection & 
    AUC(\%): 80.0 &
    70.0 & 
    Comparison of unsupervised outlier detection methods.
    \\
    \hline
    \citet{han2020infinite} &
    PGGAN & 
    BRATS~\citep{menze2014multimodal}& 
    MRI &
    Data augmentation & 
    Accuracy(\%): 90.06 & 
    91.08 & 
    PGGAN-based augmentation method of whole brain MRI.
    \\
    \hline
    \citet{han2019combining} &
    PGGAN, SimGAN & 
    BRATS~\citep{menze2014multimodal}& 
    MRI & 
    Data augmentation & 
    Sensitivity(\%): 93.67 & 
    97.48 & 
    Two-step GAN for noise-to-image and image-to-image data augmentation.
    \\
    \hline
    \citet{benson2020gan} &
    GAN & 
    TCIA~\citep{schmainda2018data} & 
    MRI & 
    Anomaly detection & 
    Accuracy(\%): 73.48 & 
    74.96 & 
    Multi-modal MRI images as input to the GAN.
    \\
    \hline
    \citet{han1904h2019} &
    GAN & 
    Private & 
    MRI & 
    Data augmentation & 
    Sensitivity(\%): 67.0 & 
    77.0 & 
    Synthesis and detection of brain metastases in MRI with bounding boxes.
    \\
    \hline
    \citet{siddiquee2019learning} & 
    Fixed-Point GAN &
    BRATS~\citep{menze2014multimodal} & 
    MRI & 
    Anomaly detection & 
    \textcolor{mycorrect}{AUC(\%): 95, IoU:0.261} & 
    \textcolor{mycorrect}{92, 0.348} & 
    Brain lesion detection and localization with fixed-point image-to-image translation concept.
    \\
    \hline
    \citet{sun2020adversarial} & 
    ANT-GAN & 
    BRATS 2013~\citep{menze2014multimodal} & 
    MRI & 
    Data augmentation, Anomaly detection & 
    F1-Score(\%): 89.6 & 
    91.7 &
    Abnormal to normal image translation in cranial MRI, VGG lesion classification.
    \\
    \hline\hline
    \textbf{Breast}
    \\
    \hline
    \citet{wu2018conditional}
    
    \href{https://github.com/ericwu09/mammo-cigan}{Mammo-ciGAN} &
    ciGAN & 
    DDSM~\citep{heath2001digital} & 
    Film MMG & 
    Data augmentation & 
    ROC AUC(\%): 88.7 &
    89.6 & 
    Synthetic lesion in-filling in healthy mammograms.
    \\
    \hline
    \citet{guan2019breast} & 
    GAN & 
    DDSM~\citep{heath2001digital} & 
    Film MMG & 
    Data augmentation & 
    Accuracy(\%): 73.48 & 
    74.96 & 
    Generate patches containing benign and malignant tumours.
    \\
    \hline
    \citet{jendele2019adversarial} 
    
    \href{https://github.com/BreastGAN}{BreastGAN} & 
    CycleGAN & 
    BCDR~\citep{lopez2012bcdr}, INbreast~\citep{moreira2012inbreast}, CBIS–DDSM~\citep{lee2016curated} & 
    Digital/Film MMG & 
    Data augmentation  & 
    ROC AUC(\%): 83.50, F1(\%): 62.53 & 
    $\Delta$-1.46, $\Delta$+1.28  & 
    Scanned \& digital mammograms evaluated together for lesion detection.
    \\
    \hline
    \citet{lee2020simulating} &
    CGAN & 
    Private & 
    Digital MMG & 
    Data augmentation & 
    ROC AUC(\%): 57 & 
    67 & 
    Synthesising contralateral mammograms.
    \\
    \hline
    \citet{wu2020synthesizing} &
    U-Net based GAN & 
    OPTIMAM~\citep{halling2020optimam} & 
    Digital MMG & 
    Data augmentation & 
    ROC AUC(\%): 82.9 & 
    84.6 & 
    Removed/added MMG lesions for malignant/benign classification.
    \\
    \hline
    \citet{alyafi2020dcgans} & 
    DCGAN &
    OPTIMAM~\citep{halling2020optimam} & 
    Digital MMG & 
    Data augmentation &  
    F1-Score(\%): \textcolor{mycorrect}{$\approx$90} & 
    \textcolor{mycorrect}{$\approx$96}& 
    Synthesise breast mass patches with high diversity.
    \\
    \hline
    \citet{desai2020breast} &
    DCGAN & 
    DDSM~\citep{heath2001digital} & 
    Film MMG & 
    Data augmentation &  
    Accuracy(\%): 78.23 & 
    87.0 & 
    Synthesise full view MMGs and used them in visual Turing test.
    \\
    \hline
    \citet{muramatsu2020improving} &
    CycleGAN & 
    DDSM~\citep{heath2001digital} & 
    CT, Film MMG & 
    Data augmentation & 
    Accuracy(\%): 79.2 & 
    81.4 & 
    CT lung nodule to MMG mass translation and vice versa.
    \\
    \hline
    \citet{swiderski2021deep} &
    AGAN & 
    DDSM~\citep{heath2001digital} & 
    Film MMG & 
    Data augmentation &  
    ROC AUC(\%): 92.50 & 
    94.10 & 
    AutoencoderGAN (AGAN) augments data in normal abnormal classification.
    \\
    \hline
    \citet{kansal2020generative} & 
    DCGAN & 
    Private & 
    OCT & 
    Data augmentation &  
    Accuracy(\%): 92.0 & 
    93.7 & 
    Synthetic Optical Coherence Tomography (OCT) images.
    \\
    \hline
    \citet{shen2021mass} &
    ciGAN & 
    DDSM~\citep{heath2001digital} & 
    Film MMG & 
    Data augmentation &  
    \textcolor{mycorrect}{FROC-AUC(\%): 15.1} &
    \textcolor{mycorrect}{17.2} &   
    Generate labelled breast mass images for precise detection.
    \\
    \hline
    \citet{pang2021semi} &
    TripleGAN-based & 
    Private  & 
    Ultrasound & 
    Data augmentation & 
    Sensitivity(\%): 86.60 & 
    87.94 & 
    Semi-supervised GAN-based Radiomics model for mass CLF.
    \\
    \hline\hline
    \textbf{Liver}
    \\
    \hline
    \citet{frid2018gan} & 
    DCGAN, ACGAN & 
    Private & 
    CT &
    Data augmentation & 
    Sensitivity(\%): 78.6 & 
    85.7  & 
    Synthesis of high quality focal liver lesions from CT for lesion CLF.
    \\
    \hline
    \citet{zhao2020tripartite} & 
    Tripartite GAN & 
    Private & 
    MRI & 
    Data augmentation & 
    Accuracy(\%): 79.2 & 
    89.4 & 
    Synthetic contrast-enhanced MRI $\rightarrow$ tumour detection without contrast agents.
    \\
    \hline
    \citet{doman2020lesion} &
    DCGAN & 
    JAMIT~\citep{jamit}, 3Dircadb~\citep{3dircadb} & 
    CT &
    Data augmentation & 
    Detection rate(\%): 65  & 
    95 & 
    Generate metastatic liver lesions in abdominal CT for improved cancer detection.
    \\
    \hline\hline
    \textbf{Stomach/Colon/Prostate}
    \\
    \hline
    \citet{kanayama2019gastric} & 
    DCGAN & 
    Private & 
    Endoscopy & 
    Data augmentation & 
    AP(\%): 59.6 & 
    63.2 & 
    Synthesise lesion images for gastric cancer detection.
    \\
    \hline
    \citet{shin2018abnormal} & 
    cGAN & 
    CVC-CLINIC~\citep{bernal2015wm}, CVC-ClinicVideoDB~\citep{angermann2017towards} &
    Colonoscopy & 
    Data augmentation & 
    Precision(\%): 81.9 & 
    85.3 & 
    Synthesise polyp images from normal colonoscopy images for  polyp detection.
    \\
    \hline
    \citet{rau2019implicit} \href{http://cmic.cs.ucl.ac.uk/ColonoscopyDepth/}{ColonoscopyDepth} & 
    pix2pix-based & 
    Private & 
    Colonoscopy & 
    Data augmentation & 
    Mean L1(\%): 33.9 & 
    24.7 & 
    Transform monocular endoscopic images from two domains to depth maps.
    \\
    \hline
    \citet{yu2020synthesis} & 
    CapGAN & 
    BrainWeb phantom~\citep{cocosco1997brainweb}, Prostate MRI~\citep{choyke2016data} & 
    MRI & 
    Data augmentation &  
    ROC AUC(\%): 85.1 & 
    88.5 & 
    Synthesise prostate MRI using Capsule Network-Based DCGAN instead of CNN. 
    \\
    \hline
    \citet{krause2021deep} & 
    CGAN & 
    \href{https://www.cancer.gov/tcga.}{TCGA}, NLCS~\citep{van1990large} & 
    Histopathology & 
    Data augmentation &  
    ROC AUC(\%): 75.7 & 
    77.7 & 
    GANs to enhance genetic alteration detection in colorectal cancer histology.
    \\
    \hline\hline
    \textbf{Skin}
    \\
    \hline
    \citet{bissoto2018skin}
    
    \href{https://github.com/alceubissoto/gan-skin-lesion}{gan-skin-lesion} & 
    PGAN, DCGAN, pix2pix & 
    Dermofit~\citep{ballerini2013color}, ISIC 2017~\citep{codella2018skin}, IAD~\citep{argenziano2002dermoscopy} & 
    Dermoscopy & 
    Data augmentation & 
    ROC AUC(\%): 83.4 & 
    84.7 & 
    Comparative study of GANs for skin lesions generation
    \\
    \hline
    \citet{creswell2018denoising} & 
    ssDAAE & 
    ISIC 2017~\citep{codella2018skin} & 
    Dermoscopy & 
    Representation learning, classification & 
    ROC AUC(\%): 89.0 & 
    89.0 & 
    Adversarial autoencoder fine-tuned on few labelled lesion classification samples.
    \\
    \hline
    \citet{baur2018melanogans} & 
    DCGAN, LAPGAN & 
    ISIC 2017~\citep{codella2018skin} & 
    Dermoscopy & 
    Data augmentation & 
    Accuracy (\%): \textcolor{mycorrect}{71.6} & 
    74.0 & 
    Comparative study, 256x256px skin lesions synthesis. LAPGAN acc=74.0\%
    \\
    \hline
    \citet{rashid2019skin} & 
    GAN & 
    ISIC 2017~\citep{codella2018skin} & 
    Dermoscopy &  
    Data augmentation &  
    Accuracy(\%): 81.5 & 
    86.1 & 
    Boost CLF performance with GAN-based skin lesions augmentation.
    \\
    \hline
    \citet{fossen2020synthesizing} & 
    AC-GAN, CycleGAN & 
    HAM10000 \& BCN20000 \citep{tschandl2018ham10000, combalia2019bcn20000}, ISIC 2017~\citep{codella2018skin} & 
    Dermoscopy &  
    Data augmentation &  
    Recall(\%): 72.1 & 
    76.3 & 
    Realistic-looking, class-specific synthetic skin lesions.
    \\
    \hline
    \citet{qin2020gan} & 
    Style-based GAN & 
    ISIC 2017~\citep{codella2018skin} & 
    Dermoscopy &  
    Data augmentation & 
    Precision(\%): 71.8 & 
    76.9 & 
    Style control \& noise input tuning in G to synthesise high quality lesions for CLF.
    \\
    \hline\hline
    \textbf{Lung}
    \\
    \hline
    \citet{bi2017synthesisCTtoPET} & 
    M-GAN & 
    Private & 
    PET & 
    Data augmentation & 
    F1-Score(\%): 66.38 & 
    63.84 & 
    Synthesise PET data via multi-channel GAN for tumour detection.
    \\
    \hline
    \citet{salehinejad2018generalization} &
    DCGAN & 
    Private & 
    Chest X-Rays & 
    Data augmentation & 
    Accuracy(\%): 70.87 & 
    92.10 & 
    Chest pathology CLF using synthetic data. 
    \\
    \hline
    \citet{zhao2018synthetic} & 
    F(\&)BGAN & 
    LIDC-IDRI~\citep{armato2011lung} & 
    CT &
    Data augmentation & 
    Accuracy(\%): 92.86 & 
    95.24 & 
    Forward GAN generates diverse images, Backward GAN improves their quality.
    \\
    \hline
    \citet{onishi2019automated} & 
    WGAN & 
    Private & 
    CT &
    Data augmentation & 
    Accuracy(\%): 63 (Benign), 82 (Malign) & 
    67, 94 & 
    Synthesise pulmonary nodules on CT images for nodule CLF.
    \\
    \hline
    \citet{gao2019augmenting}
    
    \href{https://github.com/andy1445/3D_GAN_Lung_Nodules}{3DGANLungNodules} & 
    WGAN-GP & 
    LIDC-IDRI~\citep{armato2011lung} & 
    CT &
    Data augmentation & 
    Sensitivity(\%): 84.8 & 
    95.0 & 
    Synthesise lung nodule 3D data for nodule detection.
    \\
    \hline
    \citet{han2019synthesizing} & 
    3DMCGAN & 
    LIDC-IDRI~\citep{armato2011lung} & 
    CT & 
    Data augmentation & 
    CPM(\%): 51.8 & 
    55.0 & 
    3D multi-conditional GAN (2 Ds) for misdiagnosis prevention in nodule detection.
    \\
    \hline
    \citet{yang2019class} & 
    GAN & 
    LIDC-IDRI~\citep{armato2011lung} & 
    CT &
    Data augmentation & 
    ROC AUC(\%): 87.56 & 
    88.12 & 
    Class-aware 3D lung nodule synthesis for nodule CLF. 
    \\
    \hline
    \citet{wang2020class}
    
    \href{https://github.com/qiuliwang/CA-MW-Adversarial-Synthesis}{CA-MW-AS} &
    CGAN & 
    LIDC-IDRI~\citep{armato2011lung} & 
    CT &
    Data augmentation &  
    F1-Score(\%): 85.88 & 
    89.03 & 
    Nodule synthesis conditioned on semantic features.
    \\
    \hline
    \citet{kuang2020unsupervised} & 
    Multi-D GAN & 
    LIDC-IDRI~\citep{armato2011lung} & 
    CT &
    Anomaly detection & 
    Accuracy(\%): 91.6 & 
    95.32 & 
    High anomaly scores on malignant images, low scores on benign.
    \\
    \hline
    \citet{ghosal2020lung} & 
    WGAN-GP & 
    LIDC-IDRI~\citep{armato2011lung} & 
    CT &
    Data augmentation & 
    ROC AUC(\%): 95.0 & 
    97.0 & 
    Unsupervised AE \& clustering augmented learning method for nodule CLF.
        \\
    \hline
    \citet{sun2020classification} &
    DCGAN & 
    LIDC-IDRI~\citep{armato2011lung} & 
    CT &
    Data augmentation & 
    Accuracy(\%): 93.8 & 
    94.5 & 
    Nodule CLF: Pre-training AlexNet \citep{krizhevsky2012imagenet} on synthetic, fine-tuning on real. 
    \\
    \hline
    \citet{wang2020combination} & 
    pix2pix, PGWGAN,  WGAN-GP & 
    Private & 
    CT &
    Data augmentation & 
    Accuracy(\%): 53.2 & 
    60.5 & 
    Augmented CNN for subcentimeter pulmonary adenocarcinoma CLF.
    \\
    \hline
    \citet{bu20203d} & 
    3D CGAN & 
    LUNA16~\citep{setio2017validation} & 
    CT &
    Data augmentation & 
    Sensitivity(\%): 97.81 & 
    98.57 & 
    Squeeze-and-excitation mechanism and residual learning for nodule detection.
    \\
    \hline
    \citet{nishio2020attribute} & 
    3D pix2pix & 
    LUNA16~\citep{setio2017validation} & 
    CT &
    Data augmentation & 
    Accuracy(\%): 85 &
    85 & 
    Nodule size CLF. Masked image + mask + nodule size conditioned paired translation.
    \\
    \hline
    \citet{onishi2020multiplanar} & 
    WGAN & 
    Private & 
    CT &
    Data augmentation & 
    Specificity(\%): 66.7 & 
    77.8 & 
    AlexNet pretrained on synthetic, fine-tuned on real nodules for malign/benign CLF.
    \\
    \hline
    \citet{teramoto2020deep} &
    PGGAN & 
    Private & 
    Cytopathology & 
    Data augmentation & 
    Accuracy(\%): 81.0 & 
    85.3 & 
    Malignancy CLF: CNN pretrained on synthetic lung cytology images, fine-tuned on real.
    \\
    \hline\hline
    \textbf{Others}
    \\
    \hline
    \citet{schlegl2017unsupervised} & 
    AnoGAN & 
    Private & 
    OCT & 
    Anomaly detection &  
    ROC AUC(\%): 73 & 
    89 & 
    D representations trained on healthy retinal image patches to score abnormal patches.
    \\
    \hline
    \citet{zhang2018medical} & 
    DCGANs, WGAN, BEGANs & 
    Private & 
    OCT &  
    Data augmentation & 
    Accuracy(\%): 95.67 & 
    98.83 & 
    CLF of thyroid/non-thyroid tissue. Comparative study for GAN data augmentation. 
    \\
    \hline
    \citet{chaudhari2019data} & 
    MG-GAN & 
    NCBI~\citep{edgar2002gene} & 
    Expression microarray data & 
    Data augmentation & 
    Accuracy(\%) P:71.43 L:68.21 B:69.8 C:67.59 & 
    93.6, 88.1, 90.3, 91.7 &
    Prostate, Lung, Breast, Colon. Interesting for fusion with imaging data.
    \\
    \hline
    \citet{liu2019wasserstein} & 
    WGAN-based & 
    Private & 
    Serum sample staging data & 
    Data augmentation &
    Accuracy(\%): 64.52 & 
    70.97 & 
    Synthetic training data for CLF of stages of Hepatocellular carcinoma.
    \\
    \hline
    \citet{rubin2019top} & 
    TOP-GAN & 
    Private & 
    Holographic microscopy & 
    Data augmentation &  
    AUC(\%): 89.2 & 
    94.7 & 
    Pretrained D adapted to CLF of optical path delay maps of cancer cells (colon, skin).
    \\
    \hline
\end{tabular}
}
\end{table*}

\subsection{Treatment and Monitoring Challenges}\label{sec:treatment}

After a tumour is detected and properly described, new challenges arise related to planning and execution of medical intervention. In this section we examine these challenges, in particular: tumour profiling and prognosis; challenges related to choice, response and  discovery of treatments; as well as further disease monitoring. Table \ref{Table:treatment-table} provides an overview of the cancer imaging GANs that are applied to treatment and monitoring challenges, which are discussed in the following.

\subsubsection{Disease Prognosis and Tumour Profiling} \label{sec:treatment-profiling}
\paragraph{Challenges for Disease Prognosis}
An accurate prognosis is crucial to plan suitable treatments for cancer patients. However, in specific cases, it could be more beneficial to actively monitor the tumours instead of treating them~\citep{bi2019artificial}. Challenges in cancer prognosis include the differentiation between long-term and short term survivors~\citep{bi2019artificial}, patient risk estimation considering the complex intra-tumour heterogeneity of the tumour microenvironment (TME)~\citep{nearchou2021comparison}, or the estimation of the probability of disease stages and tumour growth patterns, which can strongly affect outcome probabilities~\citep{bi2019artificial}. In this sense, GANs~\citep{li2021normalization,kimoh2018genesbiomarkers} and AI models in general~\citep{cuocolo2020machine, dimitriou2018principled} have shown potential in prognosis and survival prediction for oncology patients.

\paragraph{GAN Disease Prognosis Examples}
\citet{li2021normalization} (in Table \ref{table:scarcity-table}) show that their GAN-based CT normalisation framework for overcoming the domain shift between images from different centres significantly improves accuracy of classification between short-term and long-term survivors. \citet{ahmed2021multi} trained omicsGAN to translate between microRNA and mRNA expression data pairs, but could be readily enhanced to also translate between cancer imaging features and genetic information. The authors evaluate omicsGAN on breast and ovarian cancer datasets and report  improved prediction signals fo synthetic data tested via cancer outcome classification.
Another non-imaging approach is provided by~\citet{kimoh2018genesbiomarkers}, who apply a GAN for patient cancer prognosis prediction based on identification of prognostic biomarker genes. They train their GAN on reconstructed human biology pathways data, which allows for highlighting genes relevant to cancer development, resulting in improvement of the prognosis prediction accuracy. 
In regard to these works on non-imaging approaches, we promote future extensions combining prognostic biomarker genes and -omics data with the phenotypic information present in cancer images into multi-modal prognosis models.
\paragraph{GAN Tumour Profiling Examples} \label{sec:treatment_tumour_profiling}
Related to Figure \ref{fig:overview}(l),~\citet{vu2020unsupervised} propose that \textcolor{mycorrect}{image-conditioned} GANs (pix2pix) can learn latent characteristics of tissues of tumours that correlate with specific tumour grade. The authors show that when inferring their proposed BenignGAN on malignant tumour tissue images after training it exclusively on benign ones, it generates less realistic results. This allows for quantitative measurement of the differences between the original and the generated image, whereby these differences can be interpreted as tumour grade.

\citet{kapil2018deep} explore AC-GAN~\citep{odena2017conditional} on digital pathology imagery for semi-supervised quantification of the Non-Small-Cell-Lung-Cancer biomarker \textit{programmed death ligand 1 (PD-L1)}. Their class-conditional generator receives a one-hot encoded PD-L1 label as input to generate a respective biopsy tissue image, while their discriminator receives the image and predicts both PD-L1 label and whether the image is fake or real. The AC-GAN method compares favourably to other supervised and non-generative semi-supervised approaches, and also systematically yields high agreement with visual\footnote{A visual estimation of pathologists of the tumour cell percentage showing PD-L1 staining.} tumour proportional scoring (TPS).

As for the analysis of the TME, \citet{quiros2019pathologygan} propose PathologyGAN, which they train on breast and colorectal cancer tissue imagery. This allows for learning the most important tissue phenotype descriptions, and provides a continuous latent representation space, enabling quantification and profiling of differences and similarities between different tumours' tissues. \citet{quiros2019pathologygan} show that lesions encoded in an GAN's latent space enable using vector distance measures to find similar lesions that are close in the latent space within large patient cohorts. We highlight the research potential in lesion latent space representations to assess inter-tumour heterogeneity. Also, the treatment strategies and successes of patients with a similar lesion can inform the decision-making process of selecting treatments for a lesion at hand, as denoted by Figure \ref{fig:overview}(m).

\paragraph{Outlook on Genotypic Tumour Profiling with Phenotypic Data}
A further challenge is that targeted oncological therapies require genomic and immunological tumour profiling~\citep{cuocolo2020machine} and effective linking of tumour genotype and phenotype. Biopsies only allow to analyse the biopsied portion of the tumour's genotype, while also increasing patient risk due to the possibility of dislodging and seeding of neoplastic altered cells~\citep{shyamala2014risk, parmar2015machine}. Therefore, a trade-off\footnote{Due to this and due to the high intra-tumour heterogeneity, available biopsy data likely only describes a subset of tumour's clonal cell population.} exists between minimising the number of biopsies and maximising the biopsy-based information about a tumour's genotype. These reasons and the fact that current methods are invasive, expensive, and time-consuming~\citep{cuocolo2020machine} make genotypic tumour profiling an important issue to be addressed by AI cancer imaging methods. In particular adversarial deep learning models are promising to generate the non-biopsied portion of a tumour's genotype after being trained on paired genotype and radiology imaging data \footnote{Imaging data on which the entire lesion is visible to allow learning correlations between phenotypic tumour manifestations and genotype signatures.}. We recommend future studies to explore this line of research, which is regarded as a key challenge for AI in cancer imaging~\citep{bi2019artificial, parmar2015machine}. 

\subsubsection{Treatment Planning and Response Prediction}

\begin{table*}[t]
\centering
\scriptsize
\caption{Overview of adversarially-trained models applied to \textbf{treatment and monitoring} challenges. Publications are clustered by section and ordered by year in ascending order.}\label{Table:treatment-table}
\scalebox{0.78}{
\begin{tabular}{{p{0.18\textwidth}p{0.11\textwidth}p{0.26\textwidth}p{0.14\textwidth}p{0.13\textwidth}p{0.33\textwidth}}}
    \hline
    \textbf{Publication} &
    \textbf{Method} &
    \textbf{Dataset} &
    \textbf{Modality} &
    \textbf{Task} &
    \textbf{Highlights}
    \\
    \hline\hline
    \textbf{Disease Prognosis}
    \\
    \hline
    \citet{kimoh2018genesbiomarkers} &
    GAN-based & 
    TCGA~\citep{tomczak2015TCGA}, Reactome~\citep{croft2014reactome, fabregat2017reactome} &
    [non-imaging] multi-omics cancer data & 
    Data synthesis & 
    Biomarker gene identification for pancreas, breast, kidney, brain, and stomach cancer with GANs and PageRank.
    \\
    \hline
    \citet{ahmed2021multi} &
    omicsGAN & 
    TCGA~\citep{cancer2011integrated, ciriello2015comprehensive} &
    [non-imaging] ovarian/ breast gene expression & 
    Paired translation & 
    microRNA to mRNA translation and vice versa. Synthetic data improves cancer outcome classification.
    \\
    \hline\hline
    \textbf{Tumour Profiling}
    \\
    \hline
    \citet{kapil2018deep} &
    AC-GAN & 
    Private &
    Lung histopathology & 
    Classification & 
    AC-GAN CLF of PD-L1 levels for lung tumour tissue images obtained via needle biopsies.
    \\
    \hline
    \citet{quiros2019pathologygan}
    
    \href{https://github.com/AdalbertoCq/Pathology-GAN}{PathologyGAN} &
    PathologyGAN & 
    VGH/NKI~\citep{beck2011nki_vgh}, NCT~\citep{kather_jakob_nikolas_2018_1214456} &
    Breast/colorectal histopathology & 
    Representation learning & 
    Learning tissue phenotype descriptions \& tumour 
    representations. \textcolor{mycorrect}{Combines BigGAN \citep{brock2018large}, StyleGAN \citep{karras2019style} \& RAD \citep{jolicoeur2018relativistic}} 
    \\
    \hline
    \citet{vu2020unsupervised} &
    BenignGAN & 
    Private &
    Colorectal histopathology & 
    Paired translation & 
    Edge map-to-image. As trained on only benign, malignant images quantifiable via lower realism.
    \\
    \hline\hline
    \textbf{Treatment Response Prediction}
    \\
    \hline
    \citet{kadurin2017cornucopia, kadurin2017drugan} &
    AAE-based druGAN &
    Pubchem BioAssay~\citep{wang2014pubchem} &
    [non-imaging] Molecular fingerprint data & 
    Representation learning & 
    AAE for anti-cancer agent drug discovery. AAE input/output: molecular fingerprints \& log concentration vectors.
    \\
    \hline
    \citet{goldsborough2017cytogan} &
    CytoGAN & 
    BBBC021~\citep{ljosa2012bbbc021} &
    Cytopathology & 
    Representation learning &
    Grouping cells with similar treatment response via cell image representations. \textcolor{mycorrect}{Based on DCGAN, LSGAN, WGAN.}
    \\
    \hline
    \citet{Yoon2018GANITEEO} &
    GANITE & 
    USA 89-91 Twins~\citep{almond2004twins} &
    [non-imaging] individualized treatment effects & 
    \textcolor{mycorrect}{Multi-class-}conditional synthesis & 
    cGANs for individual treatment effect prediction, including unseen counterfactual outcomes and confidence intervals.
    \\
    \hline
    \citet{ge2020mcgantreatmentresponse} &
    MGANITE & 
    AML clinical trial~\citep{kornblau2009aml_dataset} &
    [non-imaging] individualized treatment effects  & 
    {Multi-class-}conditional synthesis & 
    GANITE extension introducing dosage quantification and continuous and categorical treatment effect estimation.
    \\
    \hline
    \citet{bica2020estimating} &
    SCIGAN & 
    CGA~\citep{weinstein2013cancer}, MIMIC III~\citep{johnson2016mimic} &
    [non-imaging] individualized treatment effects  & 
    {Multi-class-}conditional synthesis & 
    GANITE extension introducing continuous interventions and theoretical explanation for GAN counterfactuals.
    \\
    \hline\hline
    \textbf{Radiation Dose Planning}
    \\
    \hline
    \citet{mahmood2018automatedDOSE} &
    pix2pix-based & 
    Private &
    Oropharyngeal CT & 
    Paired translation & 
    Translating CT to 3D dose distributions without requiring hand-crafted features.
    \\
    \hline
    \citet{maspero2018dose} &
    pix2pix & 
    Private &
    Prostate/rectal/cervical CT/MRI & 
    Paired translation & 
    MR-to-CT translation for MR-based radiation dose planning without CT acquisition.
    \\
    \hline
    \citet{murakami2020doseprediction} &
    pix2pix & 
    Private &
    Prostate CT &
    Paired translation &
    CT-to-radiation dose distribution image translation without time-consuming contour/organs at risk (OARs) data.
    \\
    \hline
    \citet{peng2020magnetic} &
    pix2pix, CycleGAN & 
    Private &
    Nasopharyngeal CT/MRI & 
    Unpaired/Paired translation & 
    Comparison of pix2pix \& CycleGAN-based generation of CT from MR for radiation dose planning.
    \\
    \hline
    \citet{kearney2020doseganDOSE} &
    DoseGAN  & 
    Private &
    Prostate CT/PTV/OARs & 
    Paired translation & 
    Synthesis of volumetric dosimetry from CT+PTV+OARs even in the presence of diverse patient anatomy.
    \\
    \hline\hline
    \textbf{Disease Tracking \& Monitoring}
    \\
    \hline
    \citet{kim2019hepaticcgan} &
    CycleGAN & 
    Private &
    Liver MRI/CT/dose & 
    Unpaired translation & 
    Pre-treatment MR+CT+dose translation to post-treatment MRI $\rightarrow$ predicting hepatocellular carcinoma progression.
    \\
    \hline
    \citet{elazab2020gpgan} &
    GP-GAN & 
    BRATS 2014~\citep{menze2014multimodal} & 
    Cranial MRI & 
    Paired translation & 
    3D U-Net G generating progression image from longitudinal MRI to predict glioma growth between time-step.
    \\
    \hline
    \citet{li2020dc} &
    DC-AL GAN, DCGAN & 
    Private &
    Cranial MRI & 
    \textcolor{mycorrect}{Noise-to-}image synthesis & 
    CLF uses D representations to distinguish pseudo- and true glioblastoma progression.
    \\
    \hline
    \end{tabular}
}
\end{table*}

\paragraph{Challenges for Cancer Treatment Predictions}

A considerable number of malignancies and tumour stages have various possible treatment options and almost no head-to-head evidence to compare them to. Due to that, oncologists need to subjectively select an approved therapy based on their individual experience and exposure~\citep{troyanskaya2020artificial}.

Furthermore, despite existing treatment response assessment frameworks in oncology, inter- and intra-observer variability regarding choice and measurement of target lesions exists among oncologists and radiologists~\citep{levy2008tool}. To achieve consistency and accuracy in standardised treatment response reporting frameworks~\citep{levy2008tool}, AI and GAN methods can identify quantitative biomarkers\footnote{For example, characteristics and density variations of the parenchyma patterns on breast images~\citep{bi2019artificial}} from medical images in a reproducible manner useful for risk and treatment response predicts~\citep{hosny2018artificial}.

Apart from the treatment response assessment, treatment response prediction is also challenging, particularly for cancer treatments such as immunotherapy~\citep{bi2019artificial}. In cancer immunogenomics, for instance, unsolved challenges comprise the integration of multi-modal data (e.g., radiomic and genomic biomarkers~\citep{bi2019artificial}), immunogenicity prediction for neoantigens, and the longitudinal non-invasive monitoring of the therapy response~\citep{troyanskaya2020artificial}. In regard to the sustainability of a therapy, the inter- and intra-tumour heterogeneity (e.g., in size, shape, morphology, kinetics, texture, etiology) and potential sub-clone treatment survival complicates individual treatment prediction, selection, and response interpretation~\citep{bi2019artificial}.

\paragraph{GAN Treatment Effect Estimation Examples} \label{sec:treatment_treatment_effect}
In line with Figure \ref{fig:overview}(n), \citet{Yoon2018GANITEEO} propose the conditional GAN framework "GANITE", where individual treatment effect prediction allows for accounting for unseen, counterfactual outcomes of treatment. GANITE consists of two GANs: first, a counterfactual GAN is trained on feature and treatment vectors along with the factual outcome data. Then, the trained generator's output is used for creating a dataset, on which the other GAN, called ITE (Individual Treatment Response) GAN, is being trained. GANITE provides confidence intervals along with the prediction, while being readily scalable for any number of treatments. However, it does not allow for taking time, dosage or other treatment parameters into account. MGANITE, proposed by \citet{ge2020mcgantreatmentresponse}, extends GANITE by introducing dosage quantification, and thus enables continuous and categorical treatment effect estimations. SCIGAN~\citep{bica2020estimating} also extends upon GANITE and predicts outcomes of continuous rather than one-time interventions and the authors further provide theoretical justification for GANs' success in learning counterfactual outcomes. As to the problem of individual treatment response prediction, we suggest that quantitative comparisons of GAN-generated expected post-treatment images with real post-treatment images can yield interesting insight for tumour interpretation. We encourage future work to explore generating such post-treatment tumour images given a treatment parameter and a pre-treatment tumour image as conditional inputs.
With varying treatment parameters as input, it is to be investigated whether GANs can inform treatment selection by simulating various treatment scenarios prior to treatment allocation or whether GANs can help to understand and evaluate treatment effects by generating counterfactual outcome images after treatment application.

\citet{goldsborough2017cytogan} present an approach called CytoGAN, where they synthesise fluorescence microscopy cell images using DCGAN, LSGAN, or WGAN. The discriminator's latent representations learnt during synthesis enable grouping encoded cell images together that have similar cellular reactions to treatment by chemicals of known classes (morphological profiling)\footnote{CytoGAN uses an approach comparable to the one shown in Figure \ref{fig:Figure3_GAN_application_pipelines}(g).}. Even though the authors reported that CytoGAN obtained inferior result\footnote{i.e. mechanism-of-action classification accuracy} compared to classical, widely applied methods such as CellProfiler~\citep{singh2014cellprofiler}, using GANs to group tumour cells representations to inform chemical cancer treatment allocation decisions is an interesting approach in the realm of treatment selection, development~\citep{kadurin2017cornucopia, kadurin2017drugan} and response prediction.

\paragraph{GAN Radiation Dose Planning Examples} \label{sec:treatment_treatment_dose}
As radiation therapy planning is labour-intensive and time-consuming, researchers have been spurred to pursue automated planning processes~\citep{sharpeDOSEINTRO}. As outlined in the following and suggested by Figure \ref{fig:overview}(o), the challenge of automated radiation therapy planning can be approached using GANs.

By framing radiation dose planning as an image colourisation problem, \citet{mahmood2018automatedDOSE} introduced an end-to-end GAN-based solution, which predicts 3D radiation dose distributions from CT without the requirement of hand-crafted features. They trained their model on Oropharyngeal cancer data along with three traditional ML models and a standard CNN as baselines. The authors trained a pix2pix~\citep{isola2017image} GAN on 2D CT imagery, and then fed the generated dose distributions to an inverse optimisation (IO) model~\citep{babier2018inverse}, in order to generate optimised plans. Their evaluation showed that their GAN plans outperformed the baseline methods in all clinical metrics.

\citet{kazemifar2020dosimetricMRItoCT} (in Table \ref{table:scarcity-table}) proposed a cGAN with U-Net generator for paired MRI to CT translation. Using conventional dose calculation algorithms, the authors compared the dose computed for real CT and generated CT, where the latter showed high dosimetric accuracy. The study, hence, demonstrates the feasibility of synthetic CT for intensity-modulated proton therapy planning for brain tumour cases, where only MRI scans are available.

\citet{maspero2018dose} proposed a GAN-assisted approach to quicken the process of MR-based radiation dose planning, by using a pix2pix for generating synthetic CTs (sCTs) required for this task. They show that a conditional GAN trained on prostate cancer patient data can successfully generate sCTs of the entire pelvis.

A similar task has also been addressed by \citet{peng2020magnetic}. Their work compares two GAN approaches: one is based on pix2pix and the other on a CycleGAN~\citep{zhu2017unpaired}. The main difference between these two approaches was that pix2pix was trained using registered MR-CT pairs of images, whereas CycleGAN was trained on unregistered pairs. Ultimately, the authors report pix2pix to achieve results (i.e. mean absolute error) superior to CycleGAN, and highlight difficulties in generating high-density bony tissues using CycleGAN.

\begin{figure*}
	\begin{center}
       		\includegraphics[width=0.8\textwidth]{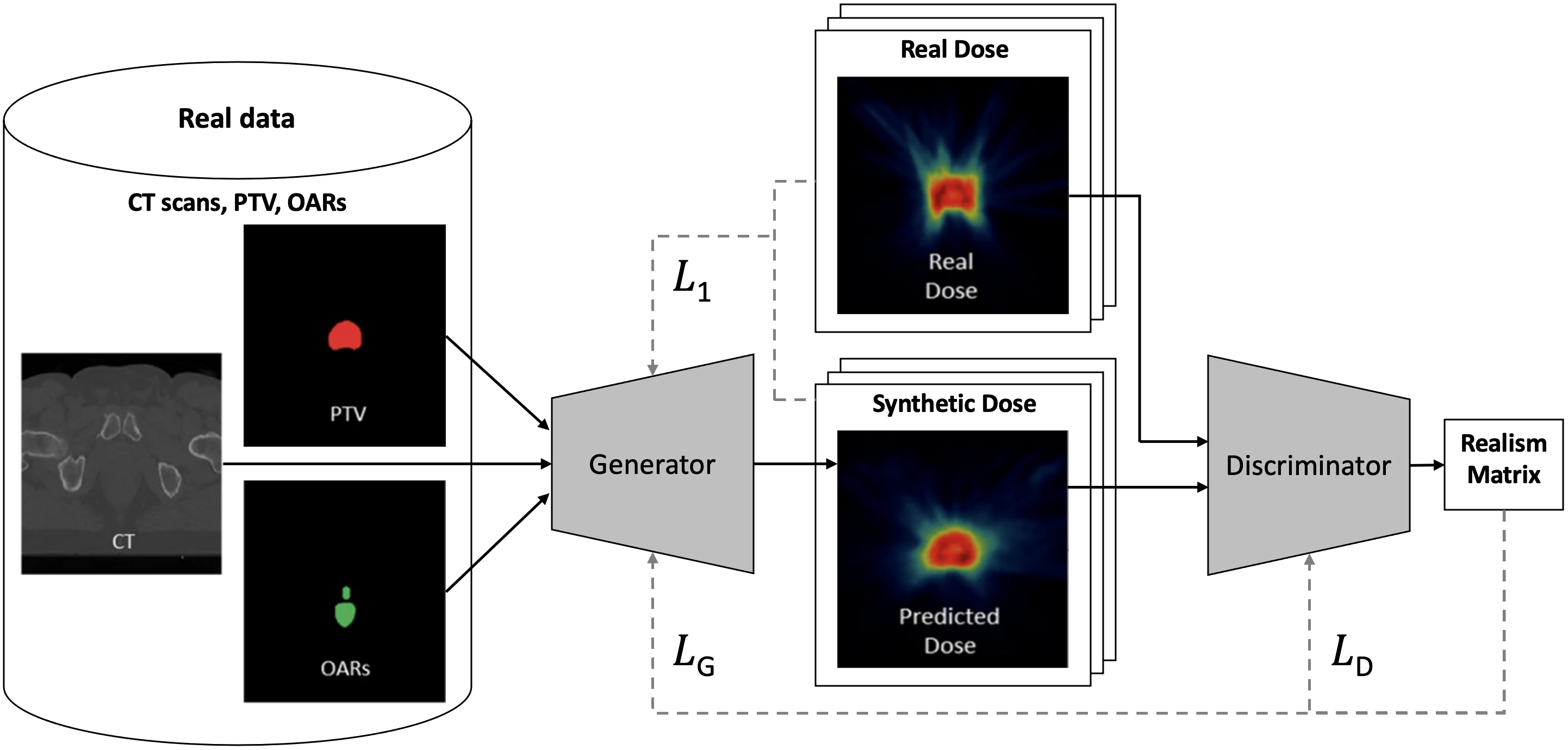}
	\end{center}
	 \caption[]{GAN architecture of DoseGAN adapted from~\citet{kearney2020doseganDOSE} and based on pix2pix~\citep{isola2017image}. Given concatenated CT scans, planning target volume (PTV) and organs at risk (OARs), the generator of DoseGAN addresses the challenge of volumetric dose prediction for prostate cancer patients.}
  	\label{fig:Figure7}
\end{figure*}

The recently introduced attention-aware DoseGAN~\citep{kearney2020doseganDOSE} overcomes the challenges of volumetric dose prediction in the presence of diverse patient anatomy. As illustrated in Figure \ref{fig:Figure7}, DoseGAN is based on a variation of the pix2pix architecture with a 3D encoder-decoder generator (L1 loss) and a patch-based patchGAN discriminator (adversarial loss). The generator was trained on concatenated CT, planning target volume (PTV) and organs at risk (OARs) data of prostate cancer patients, and the discriminator's objective was to distinguish the real dose volumes from the generated ones. Both qualitatively and quantitatively, DoseGAN was able to synthesise more realistic volumetric doses compared to current alternative state-of-the-art methods.

\citet{murakami2020doseprediction} published another GAN-based fully automated approach to dose distribution of Intensity-Modulated Radiation Therapy (IMRT) for prostate cancer. The novelty of their solution is that it does not require the tumour contour information, which is time-consuming to create, to successfully predict the dose based on the given CT dataset. Their approach consists of two pix2pix-based architectures, one trained on paired CT and radiation dose distribution images, and the other trained on paired structure images and radiation dose distribution images. From the generated radiation dose distribution images the dosimetric parameters for the PTV and OARs are computed. The generated dosimetric parameters differed on average only between 1-3\% with respect to the original ground truth dosimetric parameters.

\citet{koike2020deepARTIFACTREMOVAL} proposed a CycleGAN for dose estimation for head and neck CT images with metal artifact removal in CT-to-CT image translation as described in Table \ref{table:scarcity-table}). Providing consistent dose calculation against metal artifacts for head and neck IMRT, their approach achieves dose calculation performance similar to commercial metal artifact removal methods.

\subsubsection{Disease Tracking and Monitoring} \label{sec:4.5.3}
\paragraph{Challenges in Tracking and Modelling Tumour Progression}
Tumour progression is challenging to model~\citep{huang2020artificial} and commonly requires rich, multi-modal longitudinal data sets. As cancerous cells acquire growth advantages through genetic mutation in a process arguably analogous to Darwinian evolution~\citep{hanahan2000hallmarks}, it is difficult to predict which of the many sub-clones in the TME will outgrow the other clones. A tumour lesion is, hence, constantly evolving in phenotype and genotype~\citep{bi2019artificial} and might acquire dangerous further mutations over time, anytime. The TME's respective impact is exemplified by the stage II colorectal cancer outcome classification performance gain in~\citet{dimitriou2018principled}, which is likely attributable to the high prognostic value of the TME information in their training data.

In addition, concurrent conditions and alterations in the organ system surrounding a tumour, but also in distant organs may not only remain undetected, but could also influence patient health and progression~\citep{bi2019artificial}. GANs can generate hypothetical comorbidity data\footnote{For example from EHR~\citep{hwang2017adversarial, dashtban2020predicting}, imaging data, or a combination thereof.} to aid awareness, testing, finding, and analysis of complex disease and comorbidity patterns.
A further difficulty for tumour progression modelling is the a priori unknown effect of treatment. Treatment effects may even remain partly unknown after treatment for example in the case of radiation therapy\footnote{Radiation therapy can result in destruction of the normal tissue (e.g., radionecrosis) surrounding the tumour. Such heterogeneous normal tissue can become difficult to characterise and distinguish from the cancerous tissue~\citep{verma2013differentiating}.}~\citep{verma2013differentiating} or after surgery\footnote{It is challenging to quantify the volume of remaining tumour residuals after surgical removal~\citep{bi2019artificial}.}~\citep{bi2019artificial}.

\paragraph{GAN Tumour Progression Modelling Examples} \label{sec:treatment_tumour_progression}
Relating to Figure \ref{fig:overview}(p), GANs can not only diversify the training data, but can also be applied to simulate and explore disease progression scenarios~\citep{elazab2020gpgan}. 
For instance,~\citet{elazab2020gpgan} propose GP-GAN, which uses stacked 3D conditional GANs for growth prediction of glioma based on longitudinal MR images. The generator is based on the U-Net architecture~\citep{ronneberger2015u} and the segmented feature maps are used in the training process. 
\citet{kim2019hepaticcgan} trained a CycleGAN on concatenated pre-treatment MR, CT and dose images (i.e. resulting in one 3-channel image) of patients with hepatocellular carcinoma to generate follow-up enhanced MR images. This enables tumour image progression prediction after radiation treatment, whereby CycleGAN outperformed a vanilla GAN baseline.

The deep convolutional (DC)~\citep{radford2015unsupervised} - AlexNet (AL)~\citep{krizhevsky2012imagenet} GAN (DC-AL GAN) proposed by \citet{li2020dc} is trained on longitudinal diffusion tensor imaging (DTI) data of pseudoprogression (PsP) and true tumour progression (TTP) in glioblastoma multiforme (GBM) patients. Both of these progression types can occur after standard treatment\footnote{Pseudoprogression occurs in 20-30\% of GBM patients~\citep{li2020dc}.} and they are often difficult to differentiate due to similarities in shape and intensity. In DC-AL GAN, representations are extracted from various layers of its AlexNet discriminator that is trained on discriminating between real and generated DTI images. These representations are then used to train a support vector machine (SVM) classifier to distinguish between PsP and TTP samples achieving promising performance.

We recommend further studies to extend on these first adversarial learning disease progression modelling approaches. One potential research direction are GANs that simulate environment and tumour dependent progression patterns based on conditional input data such as the tumour's gene expression data~\citep{xu2020correlation} or the progressed time between original image and generated progression image (e.g., time passed between image acquisitions or since treatment exposure). To this end, unexpected changes of a tumour may be uncovered between time points or deviations from a tumour's biopsy proven genotypic growth expectations\footnote{For example, by comparing the original patient image  after progression with the GAN-generated predicted image (or its latent representation) after progression for time spans of interest.}.

\section{\textcolor{mycorrect}{Meta-Analysis: Towards a Framework for the Assessment of Trustworthiness and Validation}}\label{sec:trustworthiness}

\subsection{\textcolor{mycorrect}{Trustworthiness of Medical Image Synthesis Studies}}

\textcolor{mycorrect}{Section \ref{sec:currentchallenges} presented an extensive analysis of the challenges, existing publications, and state-of-the-art data synthesis and adversarial network methods in cancer imaging. While the methodologies, experiments, and results of these studies were elaborated, their validity and trustworthiness was not specifically addressed. The validity and trustworthiness varies between studies and depends on the breadth and depth of the methodological evaluation and the analysis of potential limitations. In the absence of a rigorous evaluation indicating otherwise, the methodology and experimental results of a study cannot be readily assumed to be transferable across domains, settings, tasks, datasets and modalities. Hence, while a study reports promising results for a particular task and seemingly solves the task's underlying (cancer imaging) challenge, modest changes in the dataset, evaluation method or evaluation metrics can lead to different results and conclusions. 
This points to the need of a principled assessment of trustworthiness and validity of studies in the cancer and medical imaging domains, in particular, the ones contributing and evaluating synthetic data and data generation methodology. }

\textcolor{mycorrect}{Some frameworks have proposed guidelines and best practices for the development of trustworthy artificial intelligence solutions in medical imaging \citep{lekadir2021future, hasani2022trustworthy}. However, to the best of our knowledge, no framework has been proposed for trustworthiness assessment of studies focused on medical image synthesis solutions. Building upon the FUTURE-AI consensus guidelines \citep{lekadir2021future} and the lesson's learned from the extensive analysis of the \textcolor{mycorrect}{164} publications presented in Section \ref{sec:currentchallenges}, we propose the Synthesis Study Trustworthiness Test (\textit{SynTRUST}) as a principled framework to evaluate medical image synthesis studies.}

\begin{table*}[t]
\centering
\scriptsize
\caption{\textcolor{mycorrect}{Illustration of the \textit{\textbf{SynTRUST} Framework} for evaluation of trustworthiness and validation vigour of studies that propose generative models, synthetic data, or adversarial training in medical and cancer imaging. \textit{\textbf{SynTRUST}} is based on the 5 core principles \textit{\textbf{T}horoughness, \textbf{R}eproducibility, \textbf{U}sefulness, \textbf{S}calability, and \textbf{T}enability}. For each overarching principle, a set of concrete corresponding measures is defined. Each of the 26 \textit{\textbf{SynTRUST}} measures is associated to an ID for reference and is assigned an importance rating, where 1 stands for \textit{Essential}, 2 for \textit{Desirable}, and 3 for \textit{Recommended}.}}\label{Table:SynTRUST-framework}
\scalebox{0.94}{
\begin{tabular}{{p{0.25\textwidth}p{0.02\textwidth}p{0.04\textwidth}p{0.65\textwidth}}}
    \hline
    \textbf{Principle} &
    \textbf{ID} &
    \textbf{Rating} &
    \textbf{Definition} 
    \\
    \hline\hline
    \textbf{Thoroughness: Validity of experiments}
    \\
    \hline
    Minimum test set size &
    Th1 &
    1 &
    Representative test set size should allow confident conclusions (e.g., $>30$ cases, $>100$ images, $>=20\%$ of training data).
    \\
    \hline
    Multi-metric reporting &
    Th2 &
    1 &
    Multiple standardised metrics (e.g., FID, SSIM, downstream task metric) evaluate the synthesis method (e.g., $>=2$).
    \\
    \hline
    Multiple result validation runs &
    Th3 &
    2 &
    Mean \& variance over multiple runs to be reported for all metrics (e.g., $>=3$ random seeds, or $>=5$-fold cross-validation).
    \\
    \hline
    Fair baseline comparison &
    Th4 &
    2 &
    Fairest-possible comparison with closest-possible generative/adversarial/downstream task model baseline.  
    \\
    \hline
    Principled benchmark definition &
    Th5 &
    2 &
    Systematic construction of heterogeneous (patients, pathologies, acquisitions) test set(s), without data leaking.
    \\
    \hline
    Statistical significance testing &
    Th6 &
    3 &
    Validation that reported performance variation  \& improvements are statistically significant. 
    \\
    \hline
    Ablation study of key components &
    Th7 &
    3 &
    Testing removal of both downstream \& generative model parts for insight on impact.
    \\
    \hline
    Effect of varying training set sizes &
    Th8 &
    3 &
    Testing data scarcity impact by systematically reducing downstream and generative model training data.
    \\
    \hline\hline
    \textbf{Reproducibility: Transparency of study}
    \\
    \hline
    Detailed reporting of design decisions  &
    R1 &
    1 &
    Study design \& experiments are defined \& reported with rationales and attention to detail.
    \\
    \hline
    Public availability of dataset &
    R2 &
    2 &
    At least one reported evaluation dataset is publicly accessible allowing repetition of experiments.
    \\
    \hline
    Public availability of software code &
    R3 &
    2 &
    Source code is shared in publicly accessible repository providing method implementation, ideally with documentation.
    \\
    \hline
    Public availability of model weights &
    R4 &
    3 &
    Sharing of model weights for reusing the trained model for faster and sustainable reproducing of experiments.
    \\
    \hline\hline
    \textbf{Usefulness: Versatility of synthesis method}
    \\
    \hline
    Synthesis method usability testing &
    U1 &
    1 &
    Solid generative/adversarial model usefulness evaluation on at least one community-defined (downstream) clinical task.
    \\
    \hline
    Quantitative quality measurement &
    U2 &
    2 &
    Assessment of synthetic data quality (e.g., via FID) or adversarial loss and their correlation with downstream task metrics. 
    \\
    \hline
    Qualitative quality measurement &
    U3 &
    3 &
    Observer study with clinicians to assess synthesis model output on realism, utility, quality, and diversity. 
    \\
    \hline
    Mode collapse analysis &
    U4 &
    3 &
    Diversity of generative model modes is analysed (e.g., via visual inspection and t-sne of synthetic \& real distributions).
    \\
    \hline\hline
    \textbf{Scalability: Transferability of methodology}
    \\
    \hline
    Real-world representing data &
    S1 &
    1 &
    Evaluation on cases \& samples highly representative of medically-relevant real-world clinical data.
    \\
    \hline
    Multi-dataset evaluation &
    S2 &
    2 &
    Evaluation of generative model on multiple datasets/modalities demonstrating scalability, ideally for different organs.
    \\
    \hline
    Multi-centre evaluation &
    S3 &
    3 &
    Evaluation of generative model per centre showing generalisability across centre-specific variations.
    \\
    \hline
    Multi-downstream task evaluation &
    S4 &
    3 &
    Evaluate generative model versatility via test with multiple downstream models \& tasks (e.g. segmentation, classification).
    \\
    \hline
    Downstream task robustness evaluation &
    S5 &
    3 &
    Performance variation test for simulated train \& test acquisition, manifestation, population, annotation, prevalence shifts.
    \\
    \hline\hline
    \textbf{Tenability: Acceptability of trained model}
    \\
    \hline
    Condition adherence testing &
    Te1 &
    1 &
    Test of preciseness \& reliability of presence of (input) conditions in the (synthetic) data (e.g., via classification).
    \\
    \hline
    Bias awareness analysis &
    Te2 &
    2 &
    Discussion \& analysis how bias from dataset (e.g., age, gender, ethnicity, in/exclusion criteria) transfers into model.
    \\
    \hline
    Model hallucination tendency analysis  &
    Te3 &
    3 &
    Analysis of undesired removal/addition of features such as artifacts or tumours (e.g., via inspection or classification).
    \\
    \hline
    Fairness variation testing &
    Te4 &
    3 &
    Change in fairness is measured for generative/adversarial model intervention (e.g., via equalised odds in downstream task). 
    \\
    \hline
    Privacy preservation testing &
    Te5 &
    3 &
    Investigation of patient-identifying feature leakage and training data reconstruction risk given generative model (output). 
    \\
    \hline
    \end{tabular}
}
\end{table*}

\subsection{\textcolor{mycorrect}{Proposing the \textit{SynTRUST} Framework}}
\textcolor{mycorrect}{The Synthesis Study
Trustworthiness Test (\textit{SynTRUST}) framework consists of a principled set of measures to assess the trustworthiness and validity of studies proposing generative models, synthetic data, or adversarial training methods in medical and cancer imaging. It is based on five core principles, namely,
\begin{enumerate}[label=(\roman*)]
    \item \textit{\textbf{T}horoughness} of experimental design and validation.
    \item \textit{\textbf{R}eproducibility} and transparency of results, data, models, and implementation.  
    \item \textit{\textbf{U}sefulness} and versatility of synthesis method, model, and generated data.
    \item \textit{\textbf{S}calability} and transferability of the methodology and the results across clinical domains.
    \item \textit{\textbf{T}enability}, acceptability, and reliability of the properties of the model and respective synthetic data.
\end{enumerate}}

\textcolor{mycorrect}{The methodology applied to derive the SynTRUST framework is composed of several consecutive steps, outlined as follows.
\begin{enumerate}
  \item Observation of experimental evaluation methods in the surveyed cancer imaging papers.
  \item Questioning to which extent an observed study concludes with a generally-applicable, scientifically-sound finding.
  \item Definition of causes as to why the results of the study are limited in general-applicability and trustworthiness. 
  \item Suggestion of additional validation methods that can increase the study's general-applicability.
  \item Grouping and formalisation of suggestions into 26 concrete validation measures.
  \item Definition of an overarching principle for each group of measures resulting in the 5 core principles: \textit{\textbf{T}horoughness}, \textit{\textbf{R}eproducibility}, \textit{\textbf{U}sefulness}, \textit{\textbf{S}calability}, and \textit{\textbf{T}enability} .
  \item Refinement of the measures to complement with and extend on expert consensus on best practices for the application of artificial intelligence in medical imaging \citep{lekadir2021future}.
  \item Importance rating of each measure from 1 to 3 based on their estimated impact on trustworthiness. A rating of 1 indicates \textit{essential} measures with the highest importance, a rating of 2 characterises \textit{desirable} measures, and a rating of 3 depicts measures that are \textit{recommended} additions to a study.
\end{enumerate}
}
\textcolor{mycorrect}{The resulting \textit{SynTRUST} framework is illustrated in Table \ref{Table:SynTRUST-framework}. Table \ref{Table:SynTRUST-framework} contains the title, the definition, the importance rating, and an ID for reference for each of the 26 measures, grouped by the 5 \textit{SynTRUST} principles.}

\subsection{\textcolor{mycorrect}{Analysis of Cancer Imaging Challenges using \textit{SynTRUST}}}

\begin{table*}[t]
\centering
\caption{\textcolor{mycorrect}{Selection of studies that employ data synthesis and adversarial networks methodology curated based on their promising potential towards solving the cancer imaging challenges surveyed in sections \ref{sec:scarcity}-\ref{sec:treatment}. Each of the \textit{studies} represents one concrete \textit{proposed solution} to one of the \textit{challenges}.}}
\label{tab:SynTRUST-studies}
\scalebox{0.90}{
\begin{tabular}{|lll|} 
\cline{1-3}
 \textbf{Cancer Imaging Challenge} & \textbf{Proposed Solution} & \textbf{Representative Study} \\ 
 \hline
 \hline
 Imbalanced/biased data (\ref{sec:imbalance_fairness}) &  Adversarially-trained bias-free representations & \citet{li2021estimating} \\ 
 Dataset shifts (\ref{sec:cross_modal}) & Multi-modal image translation & 
 \citet{yurt2019mustgan}  \\ 
  Uncertain synthetic data usability (\ref{sec:feature-hallucation}) & Feature hallucination evaluation metric & \citet{cohen2018cure, cohen2018distribution} \\ 
  Uncurated data (\ref{sec:curation}) & Generative image correction \& denoising model  & \citet{armanious2020medgan}      \\ 
 Privacy risks in data sharing (\ref{sec:federated}) & Federated (differentially-private) image synthesis & 
 \citet{chang2020multi,chang2020synthetic}  \\
 Adversarial attacks and defences (\ref{sec:adversarial_attacks}) & Adversarial example-based augmentation & \citet{liu2020no}         \\
   Costly human annotation (\ref{sec:annotation_issues}) & Uncertainty-aware annotation generation & \citet{hu2020coarse} \\ 
 Weak domain generalisation (\ref{sec:annotation_domain_discriminator}) & Adversarially-trained cross-domain segmentation & \citet{kamnitsas2017unsupervised} \\ 
 Extracted feature variation (\ref{sec:feature_extraction}) & Discriminator learning radiomics correlations & \citet{xiao2019radiomics} \\ 
 Intra/inter-observer variability (\ref{sec:annotation_mask_discriminator}) & Observer averaging via mask discriminator & 
 \citet{sarker2019mobilegan} \\ 
 Radiologists' high error rate (\ref{sec:detection_radiologist_error}) & Detection improving synthetic data augmentation & \citet{zhao2020tripartite} \\ 
 Intra/inter-tumour heterogeneity (\ref{sec:detection_anomaly_detection}) & Adversarially-trained anomaly detection & \citet{kuang2020unsupervised} \\ 
 Uncertain tumour profiles (\ref{sec:treatment_tumour_profiling}) & Adversarially-trained representation comparison & 
 \citet{quiros2019pathologygan} \\ 
  Unknown treatment response (\ref{sec:treatment_treatment_effect}) & Semi-supervised treatment biomarker quantification &  
 \citet{kapil2018deep} \\ 
 Unknown treatment dose (\ref{sec:treatment_treatment_dose}) & Synthesis of volumetric dosimetry images &  
 \citet{kearney2020doseganDOSE} \\ 
 Uncertain disease progression (\ref{sec:treatment_tumour_progression}) & Tumour progression image generation & 
 \citet{elazab2020gpgan} \\ 

\hline
\end{tabular}
}
\end{table*}

\begin{table*}[t]
\centering
\caption{\textcolor{mycorrect}{Results of the in-depth analysis of all \textit{essential} and \textit{desirable} measures of the \textit{SynTRUST} framework for studies proposing adversarial network methodology. The analysed studies are selected in Table \ref{tab:SynTRUST-studies} and represent solutions to key cancer imaging challenges. The \textit{SynTRUST} measures are referenced by ID from Table \ref{Table:SynTRUST-framework}. The blue check mark symbol  indicates a positive evaluation, while the red and orange cross symbols respectively indicate a negative evaluation of an essential or desirable measure. The evaluated \textit{essential} measures are \textit{minimum test set size} (Th1), \textit{multi-metric reporting} (Th2), \textit{detailed reporting of design decisions} (R1), \textit{synthesis method usability testing} (U1), \textit{real-world representing data} (S1), \textit{condition adherence testing} (Te1). The evaluated \textit{desirable} measures are \textit{multiple result validation runs} (Th3), \textit{fair baseline comparison} (Th4), \textit{principled benchmark definition} (Th5), \textit{public availability of dataset} (R2), \textit{public availability of software code} (R3), \textit{quantitative quality measurement} (U2), \textit{multi-dataset evaluation} (S2), \textit{bias awareness analysis} (Te2).}}
\label{tab:SynTRUST-results}
\scalebox{0.94}{
\begin{tabular}{|l|c|c|c|c|c|c||c|c|c|c|c|c|c|c|}
\cline{2-15}
 \multicolumn{1}{l|}{} & \multicolumn{14}{c|}{\textbf{SynTRUST Framework}} \\ 
\cline{2-15}
 \multicolumn{1}{l|}{} & \multicolumn{6}{c||}{\textbf{1: Essential Measures}} & \multicolumn{8}{c|}{\textbf{2: Desirable Measures}} \\ 
\cline{1-15}
 \textbf{Representative Study} & 
  \textbf{Th1} & \textbf{Th2} & \textbf{R1} & \textbf{U1} & \textbf{S1} & \textbf{Te1} &
  \textbf{Th3} & \textbf{Th4} & \textbf{Th5} & \textbf{R2} & \textbf{R3} & \textbf{U2} & \textbf{S2} & \textbf{Te2} 
 \\ 
 \hline
 \hline
 \citet{li2021estimating} 
 & \textcolor{blue}{\ding{52}} &  \textcolor{blue}{\ding{52}} & \textcolor{blue}{\ding{52}} & \textcolor{blue}{\ding{52}} & \textcolor{blue}{\ding{52}} & \textcolor{blue}{\ding{52}} &
 \textcolor{blue}{\ding{52}} & \textcolor{blue}{\ding{52}} &  \textcolor{orange}{\ding{56}} & \textcolor{blue}{\ding{52}} &  \textcolor{orange}{\ding{56}} & 
 \textcolor{blue}{\ding{52}} &  
 \textcolor{orange}{\ding{56}} & \textcolor{blue}{\ding{52}} 
 \\ 
 \cline{1-15}

 \citet{yurt2019mustgan}  
 & \textcolor{blue}{\ding{52}} &  \textcolor{blue}{\ding{52}} & \textcolor{blue}{\ding{52}} & \textcolor{blue}{\ding{52}} & \textcolor{blue}{\ding{52}} & \textcolor{blue}{\ding{52}} &
 \textcolor{orange}{\ding{56}} & \textcolor{blue}{\ding{52}} &  \textcolor{blue}{\ding{52}} & 
 \textcolor{blue}{\ding{52}} &  \textcolor{blue}{\ding{52}} & \textcolor{orange}{\ding{56}} &  \textcolor{blue}{\ding{52}} & \textcolor{orange}{\ding{56}} 
 \\
 \cline{1-15}
  
  \citet{cohen2018cure, cohen2018distribution}  
 & \textcolor{blue}{\ding{52}} &  \textcolor{blue}{\ding{52}} & \textcolor{blue}{\ding{52}} & \textcolor{blue}{\ding{52}} & \textcolor{blue}{\ding{52}} & \textcolor{blue}{\ding{52}} &
 \textcolor{orange}{\ding{56}} & \textcolor{blue}{\ding{52}} &  \textcolor{orange}{\ding{56}} & \textcolor{blue}{\ding{52}} &  \textcolor{blue}{\ding{52}}
 & \textcolor{blue}{\ding{52}} &  \textcolor{orange}{\ding{56}} & \textcolor{orange}{\ding{56}}
 \\
 \cline{1-15}

  \citet{armanious2020medgan}  
 & \textcolor{red}{\ding{56}} &  \textcolor{blue}{\ding{52}} & \textcolor{blue}{\ding{52}} & \textcolor{blue}{\ding{52}} & \textcolor{blue}{\ding{52}} & \textcolor{blue}{\ding{52}} &
 \textcolor{orange}{\ding{56}} & \textcolor{blue}{\ding{52}} &  \textcolor{orange}{\ding{56}} & \textcolor{orange}{\ding{56}} &  \textcolor{orange}{\ding{56}} & \textcolor{blue}{\ding{52}} &  \textcolor{blue}{\ding{52}} & \textcolor{orange}{\ding{56}} 
 \\
 \cline{1-15}
 \citet{chang2020multi,chang2020synthetic}   
 & \textcolor{blue}{\ding{52}} &  \textcolor{blue}{\ding{52}} & \textcolor{blue}{\ding{52}} & \textcolor{blue}{\ding{52}} & \textcolor{blue}{\ding{52}} & \textcolor{blue}{\ding{52}} &
 \textcolor{orange}{\ding{56}} & \textcolor{blue}{\ding{52}} &  \textcolor{orange}{\ding{56}} & \textcolor{blue}{\ding{52}} &  \textcolor{blue}{\ding{52}} & \textcolor{orange}{\ding{56}} &  \textcolor{blue}{\ding{52}} & \textcolor{orange}{\ding{56}} 
 \\
 \cline{1-15}
 \citet{liu2020no}   
 & \textcolor{blue}{\ding{52}} &  \textcolor{red}{\ding{56}} & \textcolor{blue}{\ding{52}} & \textcolor{blue}{\ding{52}} & \textcolor{blue}{\ding{52}} & \textcolor{blue}{\ding{52}} &
 \textcolor{orange}{\ding{56}} & \textcolor{blue}{\ding{52}} &  \textcolor{blue}{\ding{52}} & \textcolor{blue}{\ding{52}} &  \textcolor{orange}{\ding{56}} & \textcolor{orange}{\ding{56}} &  \textcolor{blue}{\ding{52}} & \textcolor{orange}{\ding{56}} 
 \\
 \cline{1-15}
  \citet{hu2020coarse}   
 & \textcolor{blue}{\ding{52}} &  \textcolor{blue}{\ding{52}} & \textcolor{blue}{\ding{52}} & \textcolor{blue}{\ding{52}} & \textcolor{blue}{\ding{52}} & \textcolor{blue}{\ding{52}} &
 \textcolor{blue}{\ding{52}} & \textcolor{blue}{\ding{52}} &  \textcolor{orange}{\ding{56}} & \textcolor{orange}{\ding{56}} &  \textcolor{orange}{\ding{56}} & \textcolor{orange}{\ding{56}} &  \textcolor{blue}{\ding{52}} & \textcolor{orange}{\ding{56}} 
 \\
 \cline{1-15}
 \citet{kamnitsas2017unsupervised}   
 & \textcolor{blue}{\ding{52}} &  \textcolor{blue}{\ding{52}} & \textcolor{blue}{\ding{52}} & \textcolor{blue}{\ding{52}} & \textcolor{blue}{\ding{52}} & \textcolor{blue}{\ding{52}} &
 \textcolor{blue}{\ding{52}} & 
 \textcolor{blue}{\ding{52}} &  \textcolor{blue}{\ding{52}} & \textcolor{orange}{\ding{56}} &  \textcolor{blue}{\ding{52}} & 
 \textcolor{blue}{\ding{52}} &  \textcolor{blue}{\ding{52}} & \textcolor{orange}{\ding{56}} 
 \\
 \cline{1-15}
 \citet{xiao2019radiomics}   
 & \textcolor{blue}{\ding{52}} &  \textcolor{blue}{\ding{52}} & \textcolor{blue}{\ding{52}} & \textcolor{blue}{\ding{52}} & \textcolor{blue}{\ding{52}} & \textcolor{blue}{\ding{52}} &
 \textcolor{blue}{\ding{52}} & \textcolor{blue}{\ding{52}} &  \textcolor{orange}{\ding{56}} & \textcolor{orange}{\ding{56}} &  \textcolor{orange}{\ding{56}} & \textcolor{orange}{\ding{56}} &  \textcolor{orange}{\ding{56}} & \textcolor{orange}{\ding{56}} 
 \\
 \cline{1-15}
 \citet{sarker2019mobilegan}   
 & \textcolor{blue}{\ding{52}} &  \textcolor{blue}{\ding{52}} & \textcolor{blue}{\ding{52}} & \textcolor{blue}{\ding{52}} & \textcolor{blue}{\ding{52}} & \textcolor{blue}{\ding{52}} &
 \textcolor{orange}{\ding{56}} & \textcolor{blue}{\ding{52}} &  \textcolor{orange}{\ding{56}} & \textcolor{blue}{\ding{52}} &  \textcolor{orange}{\ding{56}} & \textcolor{orange}{\ding{56}} &  \textcolor{blue}{\ding{52}} & \textcolor{orange}{\ding{56}} 
 \\
 \cline{1-15}
 \citet{zhao2020tripartite}   
 & \textcolor{blue}{\ding{52}} &  \textcolor{blue}{\ding{52}} & \textcolor{blue}{\ding{52}} & \textcolor{blue}{\ding{52}} & \textcolor{blue}{\ding{52}} & \textcolor{blue}{\ding{52}} &
 \textcolor{blue}{\ding{52}} & \textcolor{blue}{\ding{52}} &  \textcolor{blue}{\ding{52}} & \textcolor{orange}{\ding{56}} &  \textcolor{orange}{\ding{56}} & \textcolor{orange}{\ding{56}} &  \textcolor{orange}{\ding{56}} & \textcolor{orange}{\ding{56}} 
 \\
 \cline{1-15}
 \citet{kuang2020unsupervised}   
 & \textcolor{blue}{\ding{52}} &  \textcolor{blue}{\ding{52}} & \textcolor{blue}{\ding{52}} & \textcolor{blue}{\ding{52}} & \textcolor{blue}{\ding{52}} & \textcolor{blue}{\ding{52}} &
 \textcolor{blue}{\ding{52}} & 
 \textcolor{blue}{\ding{52}} &  \textcolor{blue}{\ding{52}} & 
 \textcolor{blue}{\ding{52}} &  \textcolor{orange}{\ding{56}} & \textcolor{orange}{\ding{56}} &  \textcolor{orange}{\ding{56}} & \textcolor{orange}{\ding{56}} 
 \\
 \cline{1-15}
 \citet{quiros2019pathologygan}   
 & \textcolor{blue}{\ding{52}} &  \textcolor{blue}{\ding{52}} & \textcolor{blue}{\ding{52}} & \textcolor{red}{\ding{56}} & \textcolor{blue}{\ding{52}} & \textcolor{blue}{\ding{52}} &
 \textcolor{blue}{\ding{52}} & \textcolor{orange}{\ding{56}} &  \textcolor{blue}{\ding{52}} & 
 \textcolor{blue}{\ding{52}} &  \textcolor{blue}{\ding{52}} & \textcolor{orange}{\ding{56}} &  \textcolor{blue}{\ding{52}} & \textcolor{orange}{\ding{56}} 
 \\
 \cline{1-15}
 \citet{kapil2018deep}   
 & \textcolor{blue}{\ding{52}} &  \textcolor{blue}{\ding{52}} & \textcolor{blue}{\ding{52}} & \textcolor{blue}{\ding{52}} & \textcolor{blue}{\ding{52}} & \textcolor{blue}{\ding{52}} &
 \textcolor{orange}{\ding{56}} & \textcolor{blue}{\ding{52}} &  \textcolor{orange}{\ding{56}} & \textcolor{orange}{\ding{56}} &  \textcolor{orange}{\ding{56}} & \textcolor{orange}{\ding{56}} &  \textcolor{orange}{\ding{56}} & \textcolor{orange}{\ding{56}} 
 \\
  \cline{1-15}
 \citet{kearney2020doseganDOSE}   
 & \textcolor{red}{\ding{56}} &  \textcolor{blue}{\ding{52}} & \textcolor{blue}{\ding{52}} & \textcolor{blue}{\ding{52}} & \textcolor{blue}{\ding{52}} & \textcolor{blue}{\ding{52}} &
 \textcolor{orange}{\ding{56}} & \textcolor{blue}{\ding{52}} &  \textcolor{orange}{\ding{56}} & \textcolor{orange}{\ding{56}} &  \textcolor{orange}{\ding{56}} & \textcolor{blue}{\ding{52}} &  \textcolor{orange}{\ding{56}} & \textcolor{orange}{\ding{56}} 
 \\
 \cline{1-15}
 \citet{elazab2020gpgan}   
 & \textcolor{red}{\ding{56}} &  \textcolor{blue}{\ding{52}} & \textcolor{blue}{\ding{52}} & \textcolor{blue}{\ding{52}} & \textcolor{blue}{\ding{52}} & \textcolor{blue}{\ding{52}} & 
 \textcolor{blue}{\ding{52}} & \textcolor{blue}{\ding{52}} &  \textcolor{blue}{\ding{52}} & \textcolor{blue}{\ding{52}} &  \textcolor{orange}{\ding{56}} & \textcolor{orange}{\ding{56}} &  \textcolor{blue}{\ding{52}} & \textcolor{orange}{\ding{56}} 
 \\
 \cline{1-15}
 \hline
\end{tabular}
}
\end{table*}

\subsubsection{\textcolor{mycorrect}{SynTRUST Study Curation}}
\textcolor{mycorrect}{Towards the objective of evaluating the trustworthiness of cancer imaging solutions, we demonstrate in the following how the \textit{SynTRUST} framework can be used to analyse medical imaging publications. This not only shows the practicability of the \textit{SynTRUST} framework, but also estimates the trustworthiness of current results in the field. The latter allows to corroborate concrete quality-controlled conclusions about the progress and state-of-the-art in adversarial networks in cancer imaging.}

\textcolor{mycorrect}{In our analysis we first sample the present-day challenges in cancer imaging that were surveyed in Section \ref{sec:currentchallenges} and summarised in Figure \ref{fig:overview}. Next, we carefully select representative adversarial network publications to represent a particular challenge and its solution. This selection is based on the criteria that the publication (a) proposes a particularly promising solution to its respective challenge, (b) contributes a methodology that is generally-applicable across domains and (c) report promising results. Most of the sampled publications further (d) have shown more impact and were referenced in other relevant studies. The selected studies are displayed in Table \ref{tab:SynTRUST-studies} together with their representative solution and associated cancer imaging challenge.}

\subsubsection{\textcolor{mycorrect}{SynTRUST Study Assessment}}
\textcolor{mycorrect}{Next, we analyse each of the selected publications independently based on the \textit{SynTRUST} framework.
We choose to base our analysis on the most important measures of the \textit{SynTRUST} framework that, as shown in Table \ref{Table:SynTRUST-framework}, have received either a rating of \textit{1} as \textit{essential} or a rating of 2 as \textit{desirable}. For the sake of conciseness, we leave the analysis of less critical measures rated as 3 (\textit{recommended}) to further studies. The results of our analysis of each of the selected publications are summarised in Table \ref{tab:SynTRUST-results}.}

\paragraph{\textcolor{mycorrect}{Essential SynTRUST measures}}
\textcolor{mycorrect}{We observe that the analysed studies overall show strong trustworthiness and validity considering the \textit{essential} measures: 11 out of 16 studies fulfil all of the essential criteria, while the remaining 5 studies fulfil all but one essential measures. For 3 out of these 5 studies, the only essential measure that is not fulfilled is Th1 (\textit{minimum test set size}). For instance, studies that pioneer methodologies on promising new clinical applications, such as generative tumour progression modelling \citep{elazab2020gpgan}, it is particularly challenging to encounter datasets suitable for the clinical task at hand. Even though the number of test images exceeds the defined minimum of 100 in these studies, the number of different patients (cases) is lower than 30. 30 was defined as the indicative minimum of cases to allow for conclusions for the larger patient population\footnote{Based on the central limit theorem, 30 is a popular choice and rule-of-thumb for the minimum sample size of a population.}.
All 16 studies have a detailed reporting of design decisions (R1), train and test on real-world representing clinical data (S1), and test the conditions of their adversarial network (Te1). Also, 15 out of 16 studies report multiple standardised performance metrics to evaluate the adversarial network (Th2) and demonstrate their method's usefulness on a clinically relevant downstream task (U1). In sum, the result for the essential measures demonstrates that the reported performance and progress of the analysed studies are considerably reliable and trustworthy.}

\paragraph{\textcolor{mycorrect}{Desirable SynTRUST measures}}
\textcolor{mycorrect}{While the 6 \textit{essential} basic trustworthiness requirements are mostly fulfilled, the result for the 8 \textit{desirable} measures is more varied. This highlights that the studies have a general high level of trustworthiness, but a lower level of trustworthiness for the more specific and nuanced aspects of their reported results and validations. For instance, while 15 out of 16 studies included a comparison with a suitable baseline (Th4), multiple studies did not accomplish a positive evaluation of Th3 (8), Th5 (9), R2 (7), R3 (11), U2 (11), S2 (7), and Te2 (15).}

\begin{itemize}
\item \textcolor{mycorrect}{Regarding Th3, often studies defined a static train and test set without running experiments multiple times. For example, multiple different random seed network weight initialisations or k-fold cross-validation are options to corroborate results by demonstrating stable performance with reported mean and standard deviation across runs/folds.}

\item \textcolor{mycorrect}{Regarding Th5, in general the train-test split ensured no data leaking between training and testing sets, e.g., with images from the same patient not being in both sets. However, the benchmark test sets were often not defined systematically to ensure validating the methods on a varied distribution of, e.g., cases, patients, pathologies, and acquisition parameters.}

\item \textcolor{mycorrect}{For R2, we observe that often the studies' datasets are not public available, which limits the reproducibility of the results. Often, this is due to the collection and usage of private patient data from hospitals. Further limiting factors are the high effort to repeat the study on public datasets or the specificity of the clinical task rendering its evaluation non-viable on the available public datasets.}

\item \textcolor{mycorrect}{Analysing R3 shows that the software implementing the studies' methods and experiments is often not shared publicly in code repositories, which reduces reproducibility and impedes rerunning experiments with exactly the same code base used in the respective study. } 

\item \textcolor{mycorrect}{Regarding U2, often the correlation between (a) the downstream tasks and (b) either the synthetic data quality (e.g., in the case of generative models) or the adversarial loss (e.g. in the case of adversarial training) is not analysed. Such an analysis informs on the usefulness of the quality of the respective model and on its contribution to the results on the clinical task.}

\item \textcolor{mycorrect}{As to S2, often the method is validated on, both, (a) a single dataset and (b) a single modality, while a desirable evaluation would use multiple datasets, modalities, ideally further demonstrating the method's transferability across organs, clinical domains and acquisition protocols.}

\item \textcolor{mycorrect}{For Te2, we note the general absence of an analysis of the bias that is transferred from the training dataset into the models. For instance, a model trained on a  homogeneous patient population sample, e.g., in terms of gender, sex, ethnicity, geography, likely is biased towards this subset of the overall population and can result in unequal treatment of patients from other subsets. Model biases can be detected by reviewing (a) the dataset statistics, (b) the model performance shifts on carefully subset patient samples, and (c) the exclusion and inclusion criteria applied in the data acquisition and curation processes. This enables to report and potentially mitigate otherwise unknown model biases, which increases the knowledge and reliability of a model's properties.}
\end{itemize}

\textcolor{mycorrect}{In concluding our meta-analysis, we highlight the high general level of trustworthiness of the selected adversarial network publications based on our assessment of the \textit{essential} \textit{SynTRUST} measures. This demonstrates technical maturity of adversarial training and image synthesis methods in cancer imaging. As described in the sections \ref{sec:scarcity}-\ref{sec:treatment}, many approaches towards solving the challenges in cancer imaging are not yet fully explored. Nonetheless, the solutions that have been pioneered and validated are shown to be relatively trustworthy and solid.}

\textcolor{mycorrect}{However, our meta-analysis also revealed that specific \textit{desirable} trustworthiness criteria that go beyond basic \textit{essential} validation are often not fulfilled, even by the most promising and in-depth studies in the field. For instance, a wider practice of data and code sharing is desirable. Closing this gap will not only increase reproducibility, but also accelerate adoption of existing methods and further innovation. part from that, the validation of biases and fairness criteria in datasets and models is largely overlooked despite its importance to ensure a model's acceptability and trust in the clinical setting.}

\textcolor{mycorrect}{We motivate further studies to address and build upon the gaps our analysis has revealed regarding the trustworthiness of existing cancer imaging studies. In this regard, we highlight the \textit{SynTRUST} framework not only as a means for study evaluation, but also as a guideline guiding the design of future image synthesis studies. 
}

\section{Discussion and Future Perspectives} \label{sec:discussion}


\begin{figure*}
    \centering
       		\includegraphics[width=0.95\textwidth]{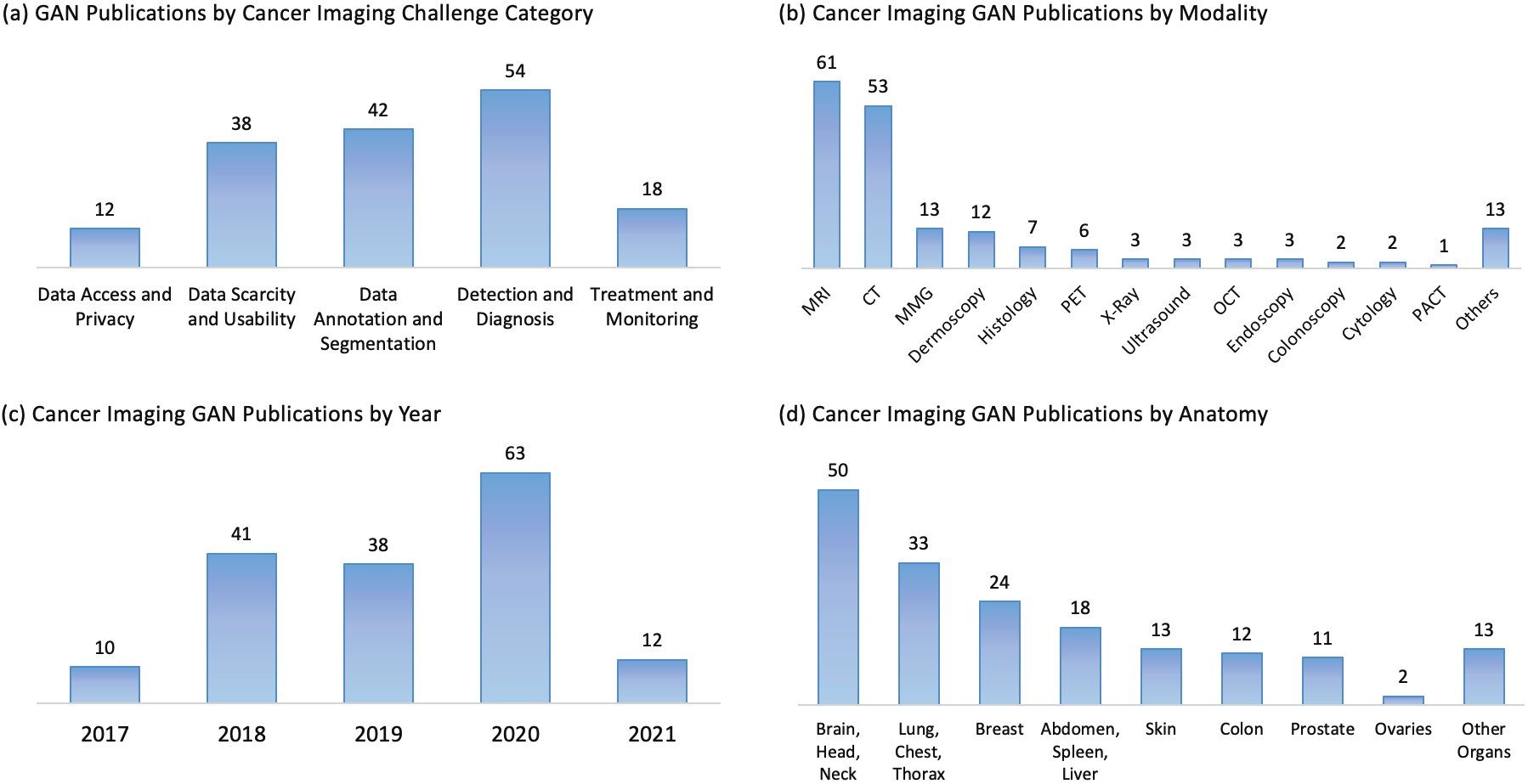}
	\caption[]{Histograms showing the distribution of the \textcolor{mycorrect}{164} analysed GAN publications in this paper by (a) cancer imaging challenge category, (b) imaging modality, (c) year of publication, and (d) anatomy/organ. These numbers are retrieved exclusively from the information in tables \ref{table:scarcity-table} - \ref{Table:treatment-table} of the respective sections \ref{sec:scarcity} - \ref{sec:treatment}. Note that (b) and (d) contain more publications in total than (a) and (c), which is caused by GAN publications that evaluate on (and are assigned to) more than one modality (b) and/or anatomy (d) due to multiple experiments or cross-domain translation. In (c), the count for 2021 is not final, as the GAN papers herein analysed have been published on or before $7^{th}$ March 2021.
	}
  	\label{fig:histogram}
\end{figure*}

\subsection{\textcolor{mycorrect}{Adversarial Methods in Cancer Imaging over the Years}}

As presented in Figure \ref{fig:histogram}(c), we have included \textcolor{mycorrect}{164} of the surveyed  GAN\textcolor{mycorrect}{-based data synthesis and adversarial training} publications in the timeframe from 2017 until March 7$^{th}$ 2021. We observe that the numbers of these cancer imaging GAN publications has been increasing from 2017 to 2020 from 10 to 63 with a surprising slight drop between 2018 to 2019 (41 to 38). The final number of respective publications for 2021 is still pending. The trend towards publications that propose GANs \textcolor{mycorrect}{and adversarial training} to solve cancer imaging challenges demonstrates the considerable research attention that \textcolor{mycorrect}{the adversarial learning scheme} has been receiving in this field. Following our literature review in Section \ref{sec:currentchallenges}, the need for further research in \textcolor{mycorrect}{ adversarial networks} seems not yet to be met. We were able to highlight various lines of research for GANs \textcolor{mycorrect}{and adversarial training} in oncology, radiology, and pathology that have received limited research attention or are untapped research potentials. These potentials indicate a continuation of the trend towards more \textcolor{mycorrect}{data synthesis and adversarial training} applications and standardised integration of GAN-generated synthetic data into medical image analysis pipelines and software solutions.

\subsection{\textcolor{mycorrect}{Modality Biases}}
In regard to imaging modalities, we analyse in Figure \ref{fig:histogram}(b) how much research attention each modality has received in terms of the number of corresponding publications. By far, MRI and CT are the most dominant modalities with 61, and 53 publications, respectively, followed by MMG (13), dermoscopy (12) and PET (6). The wide spread between MRI and CT and less investigated domains such as endoscopy (3), ultrasound (3), and digital tomosynthesis (0) is to be critically remarked. Due to variations in the imaging data between these modalities (e.g., spatial resolutions, pixel dimensions, domain shifts), it cannot be readily assumed that a GAN application with desirable results in one modality will produce equally desirable results in another. Due to that and with awareness of the clinical importance of MRI and CT, we suggest a more balanced application of GANs \textcolor{mycorrect}{and adversarial training} across modalities including experiments on rare modalities to demonstrate the clinical versatility and applicability of GAN-based solutions. \textcolor{mycorrect}{Alongside the open-access datasets described by \citet{diaz2021data}, we highlight the following additional recent open datasets to facilitate experiments on some of the cancer imaging modalities that we found to be less explored:
\begin{itemize}
  \item Breast tomosynthesis: BCS-DBT \citep{buda2021data}
  \item PET-CT: Lung-PET-CT-Dx \citep{li2020large}
  \item Endoscopy: HyperKvasir \citep{borgli2020hyperkvasir}
  \item Dermatology: HAM10000 \citep{tschandl2018ham10000}
  \item Cytology: CERVIX93 \citep{phoulady2018new}
  \item Thoracic x-ray: Node21 \citep{ecem_sogancioglu_2021_5548363}
 \end{itemize}
}
\subsection{\textcolor{mycorrect}{Anatomy Biases}}
In comparison, the GAN-based solutions per anatomy are more evenly spread, but still show a clear trend towards brain, head, neck (50), lung, chest, thorax (33) and breast (24). We suspect these spreads are due to the availability of few well-known widely-used curated benchmark datasets~\citep{menze2014multimodal, armato2011lung, heath2001digital, moreira2012inbreast} resulting in underexposure of organs and modalities with less publicly available data resources. Where possible, we recommend evaluating GAN\textcolor{mycorrect}{-based data synthesis and adversarial training} on a range of different tasks and organs. This can avoid iterating towards non-transferable solutions tuned for specific datasets with limited generalisation capabilities. Said generalisation capabilities are critical for beneficial usage in clinical environments where dynamic data processing requirements and dataset shifts (e.g., multi-vendor, multi-scanner, multi-modal, multi-organ, multi-centre) commonly exist.

\subsection{\textcolor{mycorrect}{Cancer Imaging Challenge Category Biases}}
Figure \ref{fig:histogram}(a) displays the distribution of GAN publications across cancer imaging challenge categories that correspond to the subsections of Section \ref{sec:currentchallenges}.
While the sections \ref{sec:detection} detection and diagnosis (54) and \ref{sec:detection} data annotation and segmentation (42), and \ref{sec:scarcity} data scarcity and usability (38) have received much research attention, the sections \ref{sec:treatment} treatment and monitoring (18) and \ref{sec:access} data access and privacy (12) contain substantially less GAN-related publications. This spread can be anticipated considering that classification and segmentation are popular computer vision problems and common objectives in publicly available medical imaging benchmark datasets. Early detected cancerous cells likely have had less time to acquire malignant genetic mutations~\citep{hanahan2000hallmarks, hanahan2011hallmarks} than their latter detected counterparts, which, by then, might have acquired more treatment-resistant alterations and subclone cell populations. Hence, automated early detection, location and diagnosis can provide high clinical impact via improved cancer treatment prospects, which likely influences the trend towards detection and segmentation-related GAN publications. 

\subsection{\textcolor{mycorrect}{Well-Validated Adversarial Network Solutions}}
\textcolor{mycorrect}{Our survey uncovers in Section \ref{sec:scarcity}, \ref{sec:annotation}, and \ref{sec:detection} that a vast amount of cancer imaging literature exists around a few common adversarial network solutions.}

\textcolor{mycorrect}{The most common application of GANs is data augmentation, where synthetic data is added to the training dataset to yield an improved downstream task performance. Such data augmentation can be further used to balance imbalanced datasets, which, for instance, often include much more benign tumour images than malignant ones.}

\textcolor{mycorrect}{A further well-explored application of GANs is domain adaptation via adversarial training, where a domain-adversarial loss is backpropagated into a downstream task model. Domain mapping is a related application, where images are translated from one domain to another. In general, GANs learn to translate between one source and one target domain. However, promising work has extended this technique to cross-modal synthesis between multiple domains \citep{yurt2019mustgan, li2019diamondgan, zhou2020hi}, which remains an area with much clinically-relevant research potential. Similarly, GANs for super resolution and data curation including artifact removal and image denoising achieve desirable performance and real-world applicability.}

\textcolor{mycorrect}{Image-to-image translating GANs can remove or hallucinate features such as tumours \citep{cohen2018cure, cohen2018distribution} into generated images. While this can be a major concern for clinical adoption, it also opens an avenue for future research into automated detection and assessment of removed or hallucinated features and sheds light on the need for additional metrics for GAN condition-adherence and synthetic data evaluation.}

\textcolor{mycorrect}{Furthermore, we observe that the discriminator and its associated adversarial loss can be flexibly used to classify any type of model output without necessarily following the purpose of data generation. For example, discriminator can predict whether a segmentation mask is real or created by a segmentation model, which enables the model to learn to output more globally coherent segmentation masks.}

\subsection{\textcolor{mycorrect}{New Solutions for Unexploited Areas}}

\paragraph{\textcolor{mycorrect}{Patient Privacy}}
We promote future work on the less researched open challenges in Section \ref{sec:access}, where we describe the promising research potential of \textcolor{mycorrect}{adversarial networks in patient data privacy and security. We note that secure patient data is required for legal and ethical patient data sharing and usage, which, on the other hand, is required for successful training of state-of-the-art downstream task models. For instance, sharing GANs instead of private patient data can reduce data sharing constraints, while maintaining data utility \citep{szafranowska2022sharing}. Furthermore, GANs can be trained both in a federated learning setup as well as in a differential-privacy setup. Both of these techniques can be combined to further reduce privacy risks such as the risk of generating synthetic imaging data attributable to a specific patient. Further unexploited research potential lies in adversarial identity obfuscation both on image level, as well as on latent feature representation level. In particular, devising privacy preservation testing methods to evaluate the success of adversarial identity obfuscation and related methods is a needed and not fully addressed research problem in cancer imaging and AI in healthcare at large. }

\paragraph{\textcolor{mycorrect}{Patient Security}}
\textcolor{mycorrect}{With the projected increase in clinical AI applications, adversarial learning based cybersecurity methodology becomes increasingly important to protect patients against the vulnerabilities inherent in clinically deployed deep learning solutions. Attacks can alter diagnostic markers on cancer imaging data, which can potentially result in diagnostic errors with dangerous consequences for the targeted patients. For instance, defences against adversarial examples \citep{liu2020no, samangouei2018defense} or detection of imaging data that has been tampered with \citep{mirsky2019ct} are areas where solutions based on adversarial methods will increasingly gain practical importance.}

\paragraph{\textcolor{mycorrect}{Model Debiasing}}
\textcolor{mycorrect}{The versatile ability of adversarial training to curate a model's latent space is likely to continue to increase in popularity due to the need to remove certain features in clinical AI models. For example, it is desirable to minimise a model's learned biases to increase the fairness of clinical models across patient populations \citep{lekadir2021future}. Such bias removal has been shown to be achievable via adversarial loss backpropagation \cite{zhang2018mitigating, li2021estimating}. As \citet{elazar2018adversarial} point out, some residual biases may remain in a model's latent space after converged adversarial bias removal training. Therefore, research potential lies in automated test and evaluation methodology to assess the quantity of residual bias remaining in an adversarial networks after debiasing, particularly if applied to data unseen during training.}

\paragraph{\textcolor{mycorrect}{Generative Model Evaluation}}
\textcolor{mycorrect}{A key aspect this survey observes is the absence of interpretable, standardised and exact evaluation methodology for synthetic data and generative models in the medical and cancer imaging domains. This is particularly noticeable for models without a narrow downstream task performance objective that can be used as surrogate evaluation metric nor a reconstruction objective that informs the evaluation technique. Generative models that generate a synthetic image with a clear reference value (i.e., a real image) can be evaluated based on the difference between reference and generated sample, e.g., via perceptual and reconstruction losses and metrics such as SSIM, PSNR, MSE, as discussed in Section \ref{sec:scarcity}. In the absence of such reference images, remaining methods at hand are image inspection techniques and real versus synthetic distribution comparisons, the latter including the Fréchet Inception Distance (FID) score \citep{heusel2017gans}. The popularity of the FID metric for fidelity and diversity evaluation of synthetic data has largely translated from computer vision into medical imaging. The applicability of FID in the medical domain, nonetheless, is questionable, as it internally relies on an inception classifier pretrained on the ImageNet dataset consisting of 3-channel natural images as opposed to, for instance, grayscale images from radiological domains. This demonstrates a clear need for research on further evaluation methodologies of synthetic medical images. FID extensions that pretrain the internal classifier on medical imaging datasets are potential directions, but limited by the acquisition techniques, scope, modalities, and, importantly, the size of these medical imaging datasets. Recent promising work proposed the automated generation of segmentation mask from GANs based on latent space exploration \citep{melas2021finding}. Such latent space inspection approaches can offer further potential for generative model evaluation, e.g., by helping to measure the number and difference between modes or by providing quality and diversity estimates of the segmentation masks (or other extractable pieces of information) that the model produces.}

\paragraph{\textcolor{mycorrect}{Patient Treatment}}
\textcolor{mycorrect}{Sections \ref{sec:annotation} and \ref{sec:detection} have shown that adversarial models for cancer detection, classification, and localisation are, at least for particular organs and modalities, well explored research areas. These applications are mostly relevant in diagnostic activities, which comprise only one part of the clinical workflow. We  encourage more research on GAN-based solutions in less explored subsequent clinical workflow steps such as oncological treatment planning and disease monitoring as elaborated in Section \ref{sec:treatment}. 
For example, adversarial learning offers research potential in tumour profiling and intra- and inter-tumour heterogeneity assessment via anomaly detection within the latent space of adversarial models \citep{schlegl2019f, quiros2019pathologygan}. 
The high intra- and inter-tumour heterogeneity increases the difficulty of assessing and selecting targeted treatment options. Research potential exists in precisely encoding a tumour based on imaging and/or non-imaging patient and tumour data in an adversarial model's multi-dimensional latent space. For example, this can unlock vector search applications to find similarly encoded tumours in databases to inform on therapy selection, success probabilities, and progression patterns. 
Tumour progression modelling on image-level based on generative models such as GANs remains largely unexplored. Even though not strictly necessary \citep{xia2021learning}, longitudinal and time-series cancer imaging datasets will likely trigger increased exploration of this research area once such data becomes available. For instance, given a tumour image at timepoint \textit{t1}, a GAN can learn to simulate the tumour image at timepoint \textit{t2}. 
To this end, generation of image-level counterfactuals \citep{pawlowski2020deep} as a clinically impactful solution for probing interventions. For instance, GANs can generate a tumour at \textit{t2} given the tumour image at \textit{t1} alongside multiple input conditions such as tumour growth rate, tumour type, and applied treatments.}

\subsection{\textcolor{mycorrect}{Future Perspectives and Technology Trends}}

\subsubsection{\textcolor{mycorrect}{Towards State-of-the-Art GAN Innovations in Cancer Imaging}}

\textcolor{mycorrect}{In recent years, multiple novel adversarial networks have been introduced in the field of computer vision. A lesson learned from our survey is that many of these techniques are yet to be applied thoroughly to cancer imaging. These innovations open avenues in cancer imaging that extend upon the currently used methods shown in Figure \ref{fig:chronology}, for instance, enabling improved high-resolution image generation and input-conditioned image synthesis.}

\paragraph{\textcolor{mycorrect}{Overcoming Dataset and Computation Limitations}}
\textcolor{mycorrect}{For instance, the recent VQGAN \citep{esser2021taming} combines the efficiency of convolutional networks with the expressiveness of transformers, which model the composition of a reusable codebook of context-rich visual parts. This approach is particularly relevant to medical and cancer imaging, as it allows high-resolution image synthesis despite limited computing resources.}
\textcolor{mycorrect}{Apart from containing high resolution images, cancer imaging datasets are often limited in the number of image, which may not suffice to train a GAN. In these cases, the potential issues are that during training there is convergence-failure, where the synthetic image quality is low and does not improve any further during training. While the adversarial loss often is non-interpretable not corresponding to synthetic image fidelity, diversity or condition adherence, also mode collapse may occur, where the generator has learned a particular mode to fool the discriminator instead of generating a high diversity of samples.}
\textcolor{mycorrect}{These issues are not only a function of the GAN architecture and loss function, but also of the size of the training dataset. FastGAN \citep{liu2020towards} and SinGAN \citep{shaham2019singan} shows great promise to overcome this data scarcity cancer imaging problem. FastGAN  \citep{liu2020towards} uses self-supervised training of discriminator as encoder for regularisation and generates high-resolution images despite limited computing resources and dataset size. SinGAN \citep{shaham2019singan} generates multiple synthetic images based on only a single training image. This has wide applicability and can substantially increase the usefulness of even very small cancer imaging datasets via SinGAN-based data augmentation. 
A first successful applications of SinGAN and FastGAN to cancer imaging for polyp segmentation by \citet{thambawita2022singan} shows the potential of using these models to generate not only a synthetic images, but also a corresponding segmentation mask by outputting an additional channel. This type of methodology enables training data generation for tumour detection, localisation and segmentation models without the need of conditioning the GAN on input segmentation masks.}

\paragraph{\textcolor{mycorrect}{Best Practice Combining GAN Frameworks}}
\textcolor{mycorrect}{As a vast amount of novel additions to the GANs framework has been suggested, some work \citep{brock2018large} has focused on collecting the best working practices and combining them into novel architectures, which are promising and not yet widely applied to challenges in cancer imaging. For example, BigGAN \citep{brock2018large} (a) scales model parameters by increasing the size of the feature maps, (b) applies large batch sizes, (c) uses self-attention based on SAGAN \citep{zhang2019self}, (d) provides information about the class via class-conditional batch normalization, and (e) uses hinge-loss. BigGAN and extensions thereof (e.g., \citet{zhang2019consistency, casanova2021instance, schonfeld2020u}) achieve state-of-the-art performance on class-conditional image generation.}

\textcolor{mycorrect}{Extending on PGGAN \citep{karras2017progressive} as shown in Figure \ref{fig:chronology}(m), another such example is StyleGAN \citep{karras2019style} and its variants \citep{karras2021alias, karras2020analyzing, sauer2022stylegan}, which accomplish state-of-the-art performance in conditional and unconditional computer vision image generation benchmarks. Yielding strong results, multiple architectural innovations have been introduced by the StyleGAN family, such as a style vector generating fully connected mapping network, adaptive instance normalization, and, instead of sampling from a noise vector, moving the noise input to intermediate activation maps. These innovations can inform cancer image generation models and improve their latent space exploration capabilities e.g. allowing to compare different tumour types and manifestations.} 

\paragraph{\textcolor{mycorrect}{Image-to-Image translation}}
\textcolor{mycorrect}{Image-to-image translation problems in cancer imaging are widely approached using commonly pix2pix \citep{isola2017image} (paired) and cycleGAN \citep{zhu2017unpaired} (unpaired). Nonetheless, more recent models such as OASIS \citep{sushko2020you}, ResVit \citep{dalmaz2021resvit}, and StarGAN V2 \citep{choi2020stargan} have been proposed, which are not only applicable to cancer imagery, but also have shown superior performance on computer vision benchmarks. ResVit \citep{dalmaz2021resvit}, for instance, diverges away from common CNN architectures with inductive biases by using a vision transformer architecture \citep{dosovitskiy2020image} alongside an adversarial loss \citep{goodfellow2014generative} and the common L1 losses between source and target\citep{isola2017image} and between source and reconstructed source \citep{zhu2017unpaired}. StarGAN V2 \citep{choi2020stargan} employs besides the adversarial and cycle consistency losses also a style reconstruction loss and a style diversification loss, while OASIS \citep{sushko2020you} shows that a perceptual loss is not necessary given an adversarial loss and a segmentation-based discriminator.}

\subsubsection{\textcolor{mycorrect}{GAN Alternatives and Complementary Methods}}

\paragraph{\textcolor{mycorrect}{Diffusion Models}}
\textcolor{mycorrect}{In image inpainting \citep{saharia2021palette}  and super resolution \citep{saharia2021image}, the recently proposed and increasingly popular diffusion models \citep{sohl2015deep, song2019generative, ho2020denoising} have been shown to achieve state-of-the-art and competitive performances for computer vision benchmarks and, thus, are an alternative to GANs. Diffusion models iteratively add noise to an image in a Markov chain of diffusion steps. Reversing this process, a noise vector \textit{z} is gradually denoised and transformed into an image. While achieving promising generative modelling capabilities, it still takes longer to sample from diffusion models than from GANs due to multiple denoising steps, while also further work is needed to explore the interpretability of latent representations of diffusion models \citep{NEURIPS2021_49ad23d1}. A promising line of research suggests the combination of GANs with diffusion models to increase the stability and data efficiency of GAN training \citep{Wang2022Diffusion}.} 

\paragraph{\textcolor{mycorrect}{Variational Autoencoders}}
\textcolor{mycorrect}{GANs are commonly considered to achieve higher quality outputs than variational autoencoders (VAEs) \citep{kingma2013auto} at the cost of a training process more prone towards requiring manual intervention and tuning. A promising line of research improves upon vanilla VAE by exploring combinations of GANs and VAEs \citep{larsen2016autoencoding, makhzani2015adversarial}. Extending on VAEs, \citet{van2017neural} proposed Vector Quantised Variational AutoEncoder (VQ-VAE), which learns discrete instead of continuous latent representations to avoid the issue of 'posterior collapse' that is common in VAEs. VQ-VAE has been shown to be an effective method for diverse high-quality synthetic image generation \citep{razavi2019generating}. A promising extension combines VQ-VAE with transformers \citep{vaswani2017attention} for unsupervised anomaly detection and segmentation and demonstrates its potential for tumour segmentation in brain MRI \citep{pinaya2021unsupervised}.}

\paragraph{\textcolor{mycorrect}{Normalizing Flows}}
\textcolor{mycorrect}{
The recently proposed Normalizing Flows \citep{rezende2015variational, dinh2014nice, dinh2016density} are an alternative deep generative model gaining increasing popularity for synthetic data generation tasks. As opposed to GANs and VAEs (implicit), Normalizing Flows explicitely learn the probability density function $p(x)$ and are trained via maximum likelihood estimation.
Knowing $p(x)$, unobserved but realistic new data points can be sampled with exact likelihood estimates. Normalizing Flows have been shown to be combinable with GANs and the adversarial loss function, e.g., by being the building block of the generator network \cite{grover2018flow}, and for image-to-image translation \citep{grover2020alignflow}. To date, Normalizing Flows have seen less adoption in medical and cancer imaging than GANs, but promising initial applications exist. For example, Normalizing Flows have been proposed for uncertainty estimation of lung lesion segmentation \citep{selvan2020uncertainty}, counterfactual inference on brain MRI \citep{pawlowski2020deep}, and low-dose CT image reconstruction \citep{denker2020conditional}.}

\paragraph{\textcolor{mycorrect}{Unsupervised Domain Adaptation}}
\textcolor{mycorrect}{In unsupervised domain adaptation, self-training approaches are described as an alternative to domain adversarial losses. For example, state-of-the-art methods like HRDA \citep{hoyer2022hrda} and DaFormer \citep{hoyer2022daformer} show the effectiveness of self-training in domain-adaptive semantic segmentation. DaFormer uses a transformer encoder \citep{vaswani2017attention, dosovitskiy2020image} and transfers knowledge from source to target domain via a teacher network that generates pseudo-labels for the data from the target domain. A promising avenue of research combines self-training approaches and adversarial losses \citep{li2019bidirectional,kim2020learning, wang2020classes}.}

\paragraph{\textcolor{mycorrect}{Self-Supervised Learning}}
\textcolor{mycorrect}{Given successes in learning useful representations from unlabelled data, self-supervised learning (SSL) approaches, such as BYOL \citep{grill2020bootstrap}, have become a common technique in the toolkit of deep learning researchers. Particularly when working with datasets limited in size or annotations, additional GAN-generated data can improve the learning of representations, upon which a downstream task model produces its predictions. SSL can provide an alternative, often computationally less expensive, means towards representation learning given a training task with objective function, where labels $y$ and inputs $x$ are extracted from an unlabelled dataset.
A popular and powerful SSL method is contrastive learning, where a model's latent space is learned by minimising the distance of similar samples and maximising the distance between dissimilar ones. Effective model pretraining methods such as SimCLR \citep{chen2020simple} rely on such contrastive loss functions, which, e.g., maximise agreement between differently augmented views of the same image.
Multiple recent studies propose the combination of GANs and self-supervised \citep{patel2021lt} and contrastive learning with promising results reporting improved performance and sample diversity, as well as reduced discriminator overfitting \citep{jeong2021training, kang2020contragan, liu2021divco}. In cancer imaging, for instance,
this combination has been applied to address the problem of mode collapse while retaining phenotypic tumour features for the task of colour normalisation in histopathology images \citep{ke2021contrastive}.}

\section{Conclusion}

In closing, we emphasise the versatility and the resulting modality-independent wide applicability of the adversarial learning scheme of GANs. In this survey, we strive to consider and communicate this versatility by describing the wide variety of problems in the cancer imaging domain that can be approached with \textcolor{mycorrect}{adversarial networks}. For example, we highlight GAN \textcolor{mycorrect}{and adverarial training} solutions that range from \textcolor{mycorrect}{unsupervised} domain adaptation to patient privacy preserving distributed data synthesis, to adversarial segmentation mask discrimination, to multi-modal radiation dose estimation, amongst others.\\
Before reviewing and describing GAN \textcolor{mycorrect}{and adversarial training} solutions, we surveyed the literature to understand the current challenges in the field of cancer imaging with a focus on radiology, but without excluding non-radiology modalities common to cancer imaging. After screening and analysing the cancer imaging challenges, we grouped them into the challenge categories Data Scarcity and Usability, Data Access and Privacy, Data Annotation and Segmentation, Detection and Diagnosis, and Treatment and Monitoring. After categorisation, we surveyed the literature for \textcolor{mycorrect}{adversarial networks} applied to the field of cancer imaging and found \textcolor{mycorrect}{164} relevant publications, each of which we assigned to its respective cancer imaging challenge category. Finally, we provide a comprehensive analysis for each challenge and its assigned GAN-related publications to determine to what extent it has and can be solved using GANs and \textcolor{mycorrect}{adversarial training}. \textcolor{mycorrect}{We further establish the \textit{SynTRUST} framework for assessing the trustworthiness of medical image synthesis studies. Based on \textit{SynTRUST}, we analyse 16 carefully selected cancer imaging challenge solutions. Notwithstanding the overall high level of rigour and validity of these studies, we are able to recommend a set of unaddressed trustworthiness improvements in order to guide future studies.} To this end, we also highlight research potential for challenges where we were able to propose \textcolor{mycorrect}{data synthesis or adversarial training} solutions that have not yet been fully explored by the literature.\\
With our work, we strive to uncover and motivate promising lines of research in \textcolor{mycorrect}{data synthesis and adversarial networks} that we envision to ultimately benefit the field of cancer imaging in clinical practice.

\section*{Acknowledgments}
This project has received funding from the European Union’s Horizon 2020 research and innovation programme under grant agreement No 952103.



\bibliography{mybibfile}

\end{document}